\documentstyle[12pt,aaspp4]{article}

\def\lsim{\mathrel{\lower .85ex\hbox{\rlap{$\sim$}\raise
.95ex\hbox{$<$} }}}
\def\gsim{\mathrel{\lower .80ex\hbox{\rlap{$\sim$}\raise
.90ex\hbox{$>$} }}}

\def\and{\&\ }

\def\Msol{M$_\odot$}

\begin{document}

%
%
%
%
%
%
%
%

\hsize=6.5truein
\vsize=9.0truein
\hoffset=0.0truein
\voffset=0.0truein

\pagestyle{plain}
\setcounter{page}{1}

\noindent{\Large {\bf DWARF GALAXIES OF THE LOCAL GROUP}}

\vskip1em

{\large \noindent {Mario Mateo}}

\vskip0.2em

\noindent{KEY WORDS: Stellar Populations, Local Group Galaxies, Photometry, 
Galaxy Formation, Spectroscopy, Dark Matter, Interstellar Medium}

\vspace{0.2 truein}

\noindent{Shortened Title: LOCAL GROUP DWARFS}
\vspace{0.2 truein}

\noindent{Send Proofs To: Mario Mateo\\
\noindent{\phantom{Send Proofs To: }Department of Astronomy; University of Michigan\\
\noindent{\phantom{Send Proofs To: }Ann Arbor, MI \ \ \ 48109-1090\\
\noindent{\phantom{Send Proofs To: }Phone: 313 936-1742;\ \ \ Fax: 313 763-6317\\
\noindent{\phantom{Send Proofs To: }Email: mateo@astro.lsa.umich.edu}

\begin{abstract}

The Local Group (LG) dwarf galaxies offer a unique window to the
detailed properties of the most common type of galaxy in the Universe.
In this review, I update the census of LG dwarfs based on the most
recent distance and radial velocity determinations.  I then discuss
the detailed properties of this sample, including (a) the integrated
photometric parameters and optical structures of these galaxies, (b)
the content, nature and distribution of their ISM, (c) their
heavy-element abundances derived from both stars and nebulae, (d) the
complex and varied star-formation histories of these dwarfs, (e) their
internal kinematics, stressing the relevance of these galaxies to the
dark-matter problem and to alternative interpretations, and (f)
evidence for past, ongoing and future interactions of these dwarfs
with other galaxies in the Local Group and beyond.  To complement the
discussion and to serve as a foundation for future work, I present an
extensive set of basic observational data in tables that summarize
much of what we know, and what we still do not know, about these
nearby dwarfs.  Our understanding of these galaxies has grown
impressively in the past decade, but fundamental puzzles remain that
will keep the Local Group at the forefront of galaxy evolution studies
for some time.

\end{abstract}

\clearpage

\parindent10em \hang{\it While observing the Andromeda Nebula with a
fine 18-ft telescope $\ldots$ I saw another small nebula about one
minute in diameter which appeared to throw out two small rays; one to
the right and the other to the left.}

\parindent1.5em

\vskip1em
\hfill G.-J.-H.-J.-B. Le Gentil de la Galazi\`ere,  October 29, 1749

\hfill{\it Remarks on the Nebulous Stars (1759)}

\vskip2em

\noindent{\Large {\bf1. INTRODUCTION}}

\vskip1em

\noindent To the rest of the Universe, the Local Group (LG) is an
ordinary collection of dwarf galaxies dominated by two giant spirals.
But to Earthbound astronomers interested in galaxy evolution, the
Local Group is particularly special.  The dwarfs of the Local Group
provide a uniquely well-studied and statistically useful sample of
low-luminosity galaxies.  Indeed, virtually all currently known dwarfs less
luminous than $M_V \sim -11.0$ are found in the Local Group (Whiting
et al 1997).  Dwarf galaxies represent the dominant population, by
number, of the present-day Universe (Marzke and Da~Costa 1997), and
they were almost certainly much more numerous at past epochs (Ellis
1997).  Studies of nearby clusters (Phillips et al 1998 and references
therein) suggest that the summed optical luminosity of all dwarfs may
rival that of the giant, high-surface brightness galaxies in these
environments. If low-luminosity galaxies are universally dominated by
dark matter (DM) to the extent LG dwarfs may be (Section 6), they
could account for a large fraction of the mass of galaxy clusters, and
perhaps of the entire Universe.  The dwarf galaxies of the Local Group
offer the best opportunity to study a representative sample of these
important, but by nature, inconspicuous galaxies in detail.

Dwarf galaxies are known to exist in large numbers in other
environments, particularly nearby groups (Karachentseva et al 1985,
Miller 1996, C\^ot\'e et al 1997), and clusters (Sandage \&\ Binggeli
1984, Sandage et al 1985b, Ferguson \&\ Sandage 1991, Phillips et al
1998).  But there are fundamental reasons why the dwarfs of the Local
Group remain especially important:

\begin{itemize}

\item What is the relationship, if any, between dwarf irregular
(dIrr) and dwarf spheroidal/dwarf elliptical (dSph/dE)
galaxies?  The Local Group contains a mixture of low luminosity
galaxies of both types that provide some unique ways to address this
question (e.g. Bothun et al 1986, Binggeli 1994, Skillman
\&\ Bender 1995).

\item Low-luminosity dwarfs tend to be metal poor (Section 5); thus,
the low-luminosity dwarfs in the Local Group represent a sample of
galaxies that is still largely composed of nearly primordial
material.  Dwarf galaxy abundances are typically determined from their
H~II regions, but only in the Local Group can we also obtain reliable
abundances from resolved stellar populations.  The large luminosity
range of LG dwarfs makes them excellent labs to study how other
fundamental parameters vary with luminosity, such as DM content, the
interstellar medium (ISM) properties, and star-formation history.

\item Dwarfs are the simplest galactic systems known.  However, LG
dwarfs are plainly telling us that `simple' is a relative term. The
star-formation and chemical enrichment histories of these galaxies are
complex, varied, and in most cases triggered and sustained by as-yet
unknown mechanisms (Section 6).  HST is capable of reaching the
main-sequence turnoff of the oldest stars in galaxies throughout the
Local Group (Gregg and Minniti 1997), so these galaxies will likely
remain for some time the only well-defined sample for which we can
derive complete star-formation histories.

\item Dwarf galaxies may be among the darkest single galaxies known
(Section 7).  They play an important role in addressing the
dark-matter (DM) problem, having already placed interesting
constraints on the nature and distribution of DM, and even whether the
DM paradigm is valid for these systems (Ashman 1992, Mateo 1997).  The
Local Group currently provides our only opportunity to measure the
internal kinematics of ultra-low surface brightness dwarfs.

\item There is considerable evidence of ongoing, past and future
interactions between Local Group dwarfs and larger galaxies (Section
8), which may have helped assemble the larger Local Group galaxies
over time.  Recent measurements of the star-formation histories, space
motions, three-dimensional shapes, and detailed internal kinematics of
some of these interacting dwarfs in the Local Group offer new
constraints on the dwarf/giant relationship and interaction models
(Unavane et al 1996, Mateo 1996).

\end{itemize}

I have two principle goals for this review.  First, I present here a
set of tables that aim to provide a summary of basic observational
data for all of the currently known Local Group dwarf galaxies.  These
tables also highlight areas where fundamental observations are lacking
or remain of poor quality.  My second goal is to illustrate ways in
which studies of Local Group dwarfs offer unique opportunities to
understand galaxy evolution, DM, galaxy interactions, and the relation
between stellar populations and the ISM.  At the start of most
sections, I cite specialized reviews pertinent to the topics that
follow.  Some recent general reviews relevant to LG dwarf
galaxies include those by Hodge (1989), Gallagher \&\ Wyse
(1994), Ferguson \&\ Binggeli (1994), Binggeli (1994), Skillman\&\
Bender (1995), Da~Costa (1994a,b, 1998), and Grebel (1997).  I have
also found the proceedings of three recent meetings to be particularly
helpful: the CTIO/ESO workshop on the Local Group held in La Serena,
Chile in 1994 (Layden et al 1994); the ESO/OHP workshop on dwarf
galaxies in France in 1993 (Meylan \&\ Prugniel 1994); and the Tucson
workshop on the Galactic halo and in honor of George Preston (Morrison
\&\ Sarajedini 1996).

\vskip0.25em

\noindent SOME WORDS ABOUT THE TABLES AND NOMENCLATURE \ \ The tables
are meant to be self-contained with complete notes and references.
Although the data are presented in a uniform format, the tables are
based on a large, inhomogeneous set of independent studies.  Many
entries are subjectively-weighted mean values of the results from
independent sources.  Galaxies are listed in order of increasing right
ascension in all the tables.  I try to provide realistic estimates of
the 1$\sigma$ errors throughout.  Where possible, the errors are taken
from the original source, or they reflect the scatter of independent
results.  In several tables and one figure, errors were omitted for
photometric results if $\leq$ 0.04 mag.  {\tt ASCII} files containing
most of the information in the tables are available via anonymous ftp
at {\tt ftp://ra.astro.lsa.umich.edu/pub/mateo/get/LGDtables.dat}.

`Transition' galaxies are the five objects listed in Table~1 as
dIrr/dSph systems.  Sandage \&\ Hoffman (1991) referred to such
galaxies as `mixed-morphology' systems.  The non-dIrr galaxies of the
Local Group may or may not belong to a single family; hence, when I
wish the discussion to include M32 (a dE system) and the five
transition galaxies, I use the term `early-type' galaxies.

\vfill\eject

\noindent{\Large {\bf 2. THE CENSUS OF LOCAL GROUP DWARFS}}

\vskip1em

\noindent In 1971, Paul Hodge wrote an influential review in these
volumes devoted to dwarf galaxies.  The total dwarf population of the
Local Group at that time was 14 galaxies, including the Magellanic
Clouds, with six additional uncertain cases.  As we see below, the
current census of likely dwarf members of the Local Group now stands
at $38_{-2}^{+6}$.  More dwarfs have been confirmed or identified
as LG members in the past 27 years than during the previous
222 years beginning with Le Gentil's discovery of M32!  And it is
almost certain that the full census of LG dwarfs is not yet
complete (Section 2.2).

Merely making a list of LG dwarf galaxies demands reliable distances
and radial velocities, as well as agreement on the definition of a
dwarf galaxy.  Table~1 lists all the galaxies that I consider
potential LG members, without regard yet to whether they may be dwarf
or `giant' galaxies.  Also listed are their Hubble types (van den
Bergh 1994a), with some modifications and additions for galaxies
discovered since that study.  M31, the Milky Way, and M33 are normal
`giant' spirals, and I do not consider them in any detail in this
review.  The Magellanic Clouds -- particularly the Small Magellanic
Cloud (SMC) -- have a more valid claim to dwarfhood, but I do not
discuss them here either.  Both have been the subject of recent
comprehensive reviews (Olszewski et al 1996b, Westerlund 1990, 1997).
All remaining galaxies listed in Table~1 will be considered `dwarfs'
for the purposes of this review.

\vskip1em
\noindent {\large {\it 2.1  Membership in the Local Group}}
\vskip0.5em

\noindent RESOLVABILITY AND THE BRIGHTEST STARS\ \ A common method of
flagging possible Local Group members is to identify `resolved'
galaxies with low heliocentric velocities.  This has been useful to
confirm the proximity of candidate LG members; some recent examples
include galaxies studied by Hoessel et al (1988; EGB~0427+63), Lavery
\&\ Mighell (1992; Tucana), Whiting et al (1997; Antlia).  A
quantitative variation of this technique is to identify and measure
the brightest red and blue stars in a resolved system (Sandage 1986b,
Sandage \&\ Carlson 1982, 1985a,b, 1988).  Piotto \&\ Capaccioli
(1992) and Rozanski \&\ Rowan-Robinson (1994) re-assessed this
technique, and both concluded that it is a crude 
distance indicator ($\sigma > 0.5$ mag in the
distance modulus) in the optical and the infrared (IR).  In contrast,
Karachentsev \&\ Tikhonov (1993) and Lyo \&\ Lee (1997) conclude that
the systematic uncertainties are considerably smaller: $\sim 0.4$ mag in
the optical, and even smaller in the K band for the brightest red stars
(Lyo \&\ Lee 1997).  In view of the wide range of star-formation
histories exhibited by LG dwarfs (Section 6) and these contradictory
conclusions, the brightest blue and red stars should be presently used
with care as distance indicators.  However, this approach
clearly remains suitable to identify possible LG members worthy of
further study.

\vskip0.25em

\noindent DISTANCES \ \ Table~2 lists the latest information on
distances for all the dwarf members and candidates of the Local Group.
It is satisfying that nearly all of the galaxies listed now have
reasonable distance determinations based on one or more high-precision
distance indicator, including (a) Cepheid variables (e.g. Madore \&\
Freedman 1991, Capaccioli et al 1992, Piotto et al 1994, Saha et al
1996, Wilson et al 1996), (b) the I-band tip of the red giant branch,
or TRGB (e.g. Da Costa \&\ Armandroff 1990, Lee et al 1993b, Aparicio
1994, Aparicio et al 1997b), (c) RR~Lyr stars (Saha \&\ Hoessel 1990,
Saha et al 1990, Saha et al 1992a,b), (d) SX~Phe stars (McNamara 1995,
Mateo et al 1998b), and (e) the luminosity of the horizontal branch
(HB) (Smecker-Hane et al 1994, Ibata et al 1994, Da Costa et al 1996,
Caputo et al 1995).  Other indicators that serve for more luminous
galaxies such as supernovae, surface-brightness fluctuations,
Tully-Fisher (TF) or $D_n$-$\sigma$ relations are either inapplicable
or remain uncalibrated for very low luminosity dwarfs.  Since many
individual LG galaxies have distances determined with independent
techniques, an analysis similar to that of Huterer et al (1995) --
which seeks simultaneous consistency of all distance indicators in all
galaxies considered -- would be of great interest.  Generally, when
two or more groups have studied the same galaxy, the agreement of
CCD-based results is excellent.

There are exceptions.  Aparicio (1994) used the TRGB method to argue
that the distance of the Pegasus dwarf irregular galaxy is
considerably closer than the Cepheid-based distance of Hoessel et al
(1990).  Aparicio noted that many of the putative Cepheids are located
on or near the red giant branch (RGB).  Wilson (1994a) identified
variables in IC~10 that she tentatively claimed might be Cepheids and
proposed a surprisingly small distance for the galaxy.  Soon
afterwards, Wilson et al (1996) and Saha et al (1996) obtained IR and
optical photometry of {\it bona fide} Cepheids in IC~10 to settle the
issue, finding good consistency for $D \sim 830$ kpc.  Hoessel et al
(1994) identified five possible Cepheids in Leo~A, deriving a distance
of 2.2 Mpc.  This result is inconsistent with the low velocity of
Leo~A relative to the LG barycenter (Figure~1), implies a
pathologically large red-to-blue supergiant ratio (Wilson 1992a),
forces the H~I mass of Leo~A to exceed the kinematic mass of the
entire galaxy (Section 7), and results in an abnormally low M/L ratio
for the galaxy (Young \&\ Lo 1996).  Tolstoy et al (1998) have
re-addressed this problem using new HST and ground-based photometry,
and conclude that Leo~A is $\sim 0.8\pm0.2$ Mpc away.  This result
resolves all of the problems noted above for Leo~A.

Evidently, some dwarf galaxies contain red variables (possibly
classical long-period variables or semi-regular variables) or other
sorts of variables (W~Vir stars?) that can mimic classical Cepheids.
This underscores the crucial need for color information during
searches for variable stars in nearby galaxies, the importance of
careful windowing to minimize aliases, and the utility of IR
photometry of suspected variables in nearby systems.  Given the recent
improvements in large optical and IR detectors, this
may be a good time to reconsider the Cepheid populations in LG dwarfs.

\vskip0.25em

\noindent DYNAMICAL CONSIDERATIONS. Yahil et al (1977) introduced a
dynamical approach that uses only the observed radial velocities to
determine the motions of candidate group members relative to the LG
barycenter.  Sandage (1986a) extended this approach by defining the
zero-velocity surface separating the Local Group from the local Hubble
expansion field.  This method acknowledges that the size and internal
dispersion of the Local Group (1.5-2 Mpc in radius and $\sim$60 km~s$^{-1}$,
respectively) imply a crossing time comparable to or longer than a
Hubble time.  Thus, some bound LG members may not yet have reached the
outer limit of their first orbit and consequently are still
receding from the LG center of mass. Sandage's approach requires an
estimate of the LG mass.  The current consensus for the mass of the
Galaxy has risen since 1986 to about $1.4 \pm 0.7 \times 10^{12}$
\Msol\ within 200 kpc from the Galactic center (Zaritsky et al 1989,
Fich \&\ Tremaine 1991, Peebles 1995, Lin et al 1995, Kochanek 1996).
If we assume that M31 is 30\%\ more massive than the Milky Way
(Peebles 1989), the total mass of the Local Group is $M_{LG} = 3.3 \pm
1.6 \times 10^{12}$ \Msol; the zero-velocity surface is about 1.8 Mpc
from the LG barycenter (Sandage 1986a).

A simple variant of this approach which I use here recognizes that for
a bound system traveling at velocity $v$ along a radial orbit, the
enclosed mass is $M_{enc} \leq v^2 R / G$ for a distance $R$ from the
barycenter.  For radial orbits, heliocentric velocities can
be used to estimate the mass of the Local Group for each dwarf galaxy
with or without corrections for the sun's offset from the LG
barycenter.  Galaxies that imply large values of $M_{enc}$ are
probably not bound to the Local Group and are well into the Hubble
flow.  Using the radial velocities compiled in Table~2, I have
calculated $M_{enc}$ for all LG dwarf-galaxy candidates in Table~1,
Karachentsev (1996) developed a related method that employs estimates
of both the energy and orbital timescales of individual dwarfs to
establish membership.

The results of this analysis are shown in Figure~1.  Following Yahil
et al (1977), I plot the heliocentric velocity, $V_\odot$, vs $\cos
\lambda$ for individual LG candidates, where $\lambda$ is the angle
between the galaxy and the apex of the solar motion relative to the
Local Group barycenter.  The two panels correspond to the solar-motion
solutions of Sandage (1986a) and Karachentsev \&\ Makharov (1996;
their equation 4).  For the most distant galaxies the correction from
heliocentric to barycentric velocity is small.  Galaxies for which
$M_{enc}$ exceeds the observed LG mass are denoted as {\it open
squares}.  For both solar-motion solutions, NGC~55 -- usually
considered a member of the Sculptor Group -- is consistent with LG
membership, as are NGC~3109 and the other galaxies near it (see
Section 2.3).  EGB~0427+63 and GR~8 are marginal LG members; both
galaxies imply $M_{enc}$ is slightly larger the adopted LG mass.

Of course, galaxies with large tangential velocities might still be
unbound even if their radial velocities relative to the LG barycenter
are low (Dunn \&\ Laflamme 1993).  Proper motions have now been
measured (impressively) for some Milky Way satellites: Sculptor
(Schweitzer et al 1995), Ursa Minor (Schweitzer 1998), the Large
Magellanic Cloud (LMC; Jones et al 1994), and Sagittarius (Ibata et al
1997).  We can expect to measure the tangential motions of some LG
galaxies out to $\sim$1 Mpc over four- to five-year baselines using
HST.  For now it is probably not possible to unambiguously determine
membership for galaxies near the outer fringe of the Local Group armed
with only radial velocities and distances.  This is further
complicated by the effects that nearby groups must have had on the
orbits of galaxies in the outer Local Group (Sandage 1986a, Peebles
1989,1995).

\vskip1em
\noindent{\large {\it 2.2 What's Missing?}}
\vskip0.5em

\noindent van den Bergh (1995) suggests that about 98\%\ of the
luminous mass of the Local Group has already been identified.  But
have we identified all of the individual galaxies?  Since 50\%\ of the
members listed in Table~2 have been found since 1971, the era of
discovery within the Local Group probably is not yet over.  If we
assume that the LG galaxy distribution is uniform on the sky (though
see Section 2.3), the cumulative number of galaxies should increase
from the Galactic poles as $1-\sin |b|$, where $b$ is Galactic
latitude.  Figure 2 shows the cumulative number of all LG members (top
panel) and MW satellites (bottom panel) as a function of $1 - \sin
|b|$.  For comparison, the dotted lines show how a uniformly
distributed sample would appear in these diagrams.  Note that in both
samples the observed number of galaxies exceeds the predicted
distribution at high galactic latitude (small values of $1 - \sin
|b|$). This suggests that many LG galaxies remain to be found at low
latitudes.  If I extrapolate from the observed number of galaxies at
the 50th and 67th percentile values of $1 - \sin |b|$ (the vertical
lines in Figure 2), as many as 15-20 more LG galaxies may be hidden at
low latitudes, up to half of which may be satellites of the Milky Way.
Using the technique of Yahil et al (1977), Grebel (1997) has
identified 3-4 additional LG candidates from the catalog of Schmidt
\&\ Boller (1992); none of these have preliminary distance estimates
yet.  Karachentsev \&\ Karachentseva (1998) have compiled a list of
nearby dwarfs which may contain additional candidate LG members (see als
Karachentsev \&\ Makarov 1998).

The recent discoveries of Sextans (Irwin et al 1990) and Sagittarius
(Ibata et al 1994) illustrate that nearby galaxies can also hide if
they are too diffuse.  Caldwell et al (1998) recently found a
low-surface brightness, but relatively luminous, dwarf in the M81
group with $\Sigma_{0,V} = 25.4$ mag~arcsec$^{-2}$, and $M_V = -14.3$.
If this galaxy -- which has a core radius of 1.6 kpc -- were located 1
Mpc from us, it would contribute 22 stars~arcmin$^{-2}$ brighter than
I = 22 (corresponding to the more luminous giants), or only 0.8
stars~arcmin$^{-2}$ brighter than I = 18.5 if located 200 kpc away.
The apparent core radius would be 5.4 and 27 arcmin, respectively.  An
object like this would be difficult to identify even at intermediate
Galactic latitude (Irwin 1994), though methods used to find stars far
from the centers of known dwarfs might succeed (e.g. Gould et al 1992,
Mateo et al 1996, Kuhn et al 1996).

Henning (1997) summarizes prospects for finding `hidden' galaxies from
HI observations at low latitudes.  Although the Galaxy is transparent
at 21-cm, confusion with Galactic emission at low velocities would
present a major obstacle to finding LG galaxies in this manner.  Of
course, only gas-rich objects would be found, effectively ruling out
detection of gas-poor early-type dwarfs at low latitudes.  Ongoing IR
surveys such as DENIS (Epchtein et al 1997) and 2MASS (Kirkpatrick et
al 1997) may successfully penetrate much of the foreground dust in
some regions of the Galactic Plane.  Although the stellar density of
Galactic field stars would be quite high, the signature of a nearby
dwarf galaxy might be possible to detect as a concentrated excess of
faint stars at low latitudes.

\vskip1em

\noindent {\large {\it 2.3 The Structure of the Local Group}}

\vskip0.5em

\noindent Because they are so numerous, the Local Group dwarfs can be
used effectively as tracers of substructure.  Gurzadyan et al (1993)
and Karachentsev (1996) have previously carried out substructure
analyses of the Local Group; both found the well-known concentration
of galaxies towards M31 and the Milky Way (see also van den Bergh 1995
and Grebel 1997).  A powerful -- but more subjective -- way to
visualize local substructure is via stereoscopic images of the
distribution of LG galaxies. Three stereo views are shown in
Figure~3a, corresponding to observers located well outside the Local
Group along the orthogonal axes $(l,b) = (0^\circ,0^\circ)$,
$(90^\circ,0^\circ)$, and $b = +90^\circ$, centered on the Galactic
Center.  Figure~3b identifies some of the more distant individual
galaxies for each of the three viewing positions of Figure 3a.

Four subgroups are evident in Figure~3.  The prominent group located
near the center is the Milky Way and its satellites (the origin of the
coordinate system is the center of our Galaxy).  The second surrounds
M31.  The third forms an extended `cloud' populated only by dwarfs,
mostly dIrr systems.  This `Local Group Cloud', or LG Cloud, is best seen
in the {\it middle panel} of Figure~3a.  The fourth subgroup is relatively
isolated and contains NGC~3109 as its most luminous member.  The
individual members of these four subgroups are identified in Table~1.
Remarkably few galaxies have ambiguous subgroup assignments: IC~1613
and Phoenix could both be plausibly placed in the MW or M31 subgroups;
Leo~A's new distance (Tolstoy 1998) places it in an isolated location
between the MW and N3109 subgroups.  Only GR~8 -- a marginal LG member
-- cannot be clearly assigned to any of these four subgroups.

It is interesting that the N3109 subgroup `points' towards the nearby
Maffei/IC~342 group (Karachentsev et al 1996), while the LG Cloud very
roughly points towards the Sculptor group (C\^ot\'e et al 1997).
NGC~55 is located on the far side of the LG Cloud, closest to the
Sculptor group.  Perhaps the substructure of the Local Group reflects
a dynamical history in which the Maffei and Sculptor groups have had
an important role and have `stretched-out' the Local Group in these
directions (Byrd et al 1994).  Alternatively, this may reflect
a false excess of candidates in these directions where the surface
densities of galaxies just beyond the Local Group is relatively high.
NGC~55, in particular, may be an example of this effect.

\vfill\eject

\noindent{\Large {\bf 3. OPTICAL PHOTOMETRIC AND STRUCTURAL PROPERTIES OF
LOCAL GROUP DWARFS}}

\vskip1em

\noindent Measurement of the integrated photometric and structural
properties of LG dwarfs is challenging.  Because they are so close,
many LG dwarfs are quite extended, ranging from under 10 arcmin in
diameter to over 40$^\circ$!  Few telescope/detector combinations can
survey the entire extent of the larger systems in one or even several
exposures (though see Kent 1987, Bothun \&\ Thompson 1988).  Nearly
all LG dwarfs have very low surface brightnesses, which not only makes
it difficult to discover these galaxies, but greatly complicates
obtaining reliable follow-up photometry.  Nonetheless, there have been
many attempts over the past 35 years to study the integrated
properties and structural parameters of LG dwarfs.  Hodge,
de~Vaucouleurs, and Ables pioneered these studies, and in many cases
their results remain the only ones available (e.g. Hodge 1963a,b,
1973, de Vaucouleurs \&\ Ables 1965, 1968, 1970, Ables 1971, Ables \&\
Ables 1977).

\vskip1em

\noindent{\large {\it 3.1 Integrated Photometry}}
\vskip0.5em

\noindent Table~3 lists the integrated V-band magnitudes and, when
available, the integrated colors of LG dwarfs.  In some cases, these
values are based on observations of only a small fraction of the
galaxy. For example, less than 1\%\ of the surface area of Sagittarius
has been measured photometrically (Mateo et al 1995c, Fahlman et al
1996, Mateo et al 1996), though a large fraction has been mapped
photographically (Ibata et al 1997).  Combined with the distance and
reddening values in Table~2, the photometry in Table~3 can be used to
derive integrated absolute magnitudes and luminosities (Table~4) and
the luminosity function (LF) of the Local Group (Figure~4).  For $M_B
\lsim -14$, the LG luminosity function matches that of the `poor'
groups studied by Ferguson \&\ Sandage (1991).  The best-fitting
Schechter (1976) LF for the poor groups of Ferguson \&\ Sandage (1991)
is also shown.  Note that the analytic expression, if extrapolated as
in Figure~ 4, implies that the Local Group contains many galaxies
less luminous than $M_B \sim -12$ that have yet to be discovered.

Figure~5 is the color-magnitude diagram of the Local Group based on
the integrated V-band absolute magnitudes ($M_{V_0}$) and $(B-V)_0$
colors.  The giant/dIrr galaxies are segregated from the 
early-type galaxies (denoted as a {\it dotted line} in Figure~5).
EGB~0427+63 is the only dIrr that lies redward of this boundary, but
its photometric properties and reddening are poorly known
(Karachentseva et al 1996).  The transition galaxies -- so-named in
Table~1 solely on the basis of their morphological properties -- are
located close to but on both sides of the dIrr-early type dividing
line.  NGC~205 is a luminous dSph system that contains a bright
central region of recent star formation and has a luminous blue
nucleus (Hodge 1973, Price 1985, Table~8); it is located on the dIrr
side of the dividing line in Figure~5.  The smaller region of
young stars in NGC~185 (Hodge 1963b) has a negligible effect on that
galaxy's integrated colors (Price 1985).

\vskip1em

\noindent{\large {\it 3.2 Structural Properties}}
\vskip0.5em

\noindent In the optical, the structure of LG dIrr galaxies is
dominated by star-forming complexes and OB associations with typical
diameters of 200-300 pc (Fisher \&\ Tully 1979, Hodge et al 1991a,b,
see the `Images' references in Table~1).  These clumps are usually not 
found near the optical center of symmetry of the galaxies.  NGC~3109
-- one of the most luminous dIrr galaxies in the sample -- shows clear
evidence for spiral structure underlying a similar patchy morphology
(Demers et al 1985, Sandage \&\ Carlson 1988).  In all
suitably-studied LG dIrr systems, the clumpy young stellar populations
are superimposed on a more extended, smoother and symmetric
distribution of older stars (Hodge et al 1991a,b, Minniti \&\ Zijlstra
1996).  Either dynamical effects smooth out the structures with time,
or else the star-formation regions migrate through individual
galaxies, eventually forming a more symmetric sheet of old stars
(Skillman \&\ Bender 1995, Hunter \&\ Plummer 1996, Dohm-Palmer et al
1997).

The early-type dwarfs of the Local Group are dominated by a symmetric
spheroidal component (Hodge 1971, Irwin \&\ Hatzidimitriou 1995), with
occasional instances of superimposed concentrations of relatively
young stars (NGC~185 and NGC~205: Hodge 1963b, 1973, Price \&\
Grasdalen 1983, Price 1985, Lee et al 1993a); nearly the inverse of
the dIrr galaxies.  Interestingly, the star-forming regions in NGC~185
and NGC~205 are similar in size to those seen in dIrr galaxies, but
these young stars are found near, though slightly offset from, the
centers of the galaxies.  Demers et al (1994, 1995) carefully searched
for substructure in a number of dSph systems, but found weak evidence
for such structure only in Ursa~Minor (Olszewski \&\ Aaronson 1985).

Only three LG dwarfs contain nuclei: NGC~205, Sagittarius and M32.
The latter is now widely believed to contain a massive central black
hole (Kormendy \&\ Richstone 1995).  The nucleus of NGC~205 is
extremely blue (Price \&\ Grasdalen 1983, Lee 1996), dynamically
colder than the surrounding galaxy envelope (Carter \&\ Sadler 1990),
and has a spectrum dominated by young stars (Bica et al 1990, Jones et
al 1996).  The existence of a nucleus of Sagittarius -- the globular
cluster M54 -- is somewhat controversial.  Da~Costa \&\ Armandroff
(1995) argue that the velocity dispersion and metallicity of M54 are
incompatible with its identification as a normal dSph nucleus.
However, M54 is the second most luminous globular cluster in the
entire Milky Way, nearly as luminous as the nucleus of NGC~205
(Peterson 1993), exhibits an internal abundance dispersion (Sarajedini
\&\ Layden 1995), and is located close to the center of symmetry of
Sagittarius (Ibata et al 1994, 1997).  Such an unusual object seems to
have been merely an isolated globular cluster in a dSph galaxy such as
Sagittarius.

The observed structural parameters of LG dwarfs are listed in Table~3,
including ellipticity, major-axis position angle, King core and tidal
radii, Holmberg radii, and exponential scale lengths.  The
corresponding derived structural parameters are listed in Table 4.
Historically, the surface-brightness profiles for the dIrr galaxies
are fit with exponential profiles, while for early-type systems King
profiles are preferred.  Many authors have noted that both profiles
produce acceptable fits to the red populations of dIrr and dSph
systems (Eskridge 1988a,c, Hodge et al 1991a,b, Irwin \&\
Hatzidimitriou 1995).  Aparicio et al (1997c) find the best fit to the
surface brightness profile of Antlia requires two exponential profiles
(Table~3).  S\'ersic profiles may be better suited to describing these
varied types of surface brightness profiles with only a single
additional parameter (Prugniel \&\ Simien 1997).  Sagittarius and
NGC~205 show highly elongated or otherwise disturbed outer structures,
indicative of strong interactions with the Milky Way and M31,
respectively.  All galaxies with exponential scale lengths $>$ 500 pc
are dIrr systems, while 90\%\ with smaller scale lengths are
early-type systems (James 1991).

\vfill\eject

\noindent{\Large {\bf 4. THE ISM OF LOCAL GROUP DWARFS}}

\vskip1em

\noindent Some aspects of the relationship of the stellar populations
and the interstellar medium (ISM) in LG dwarfs are acutely puzzling.
This partly reflects the great detail with which we can now study the
ISM in these nearby systems, but also reflects some fundamental
deficiencies in our understanding of dwarf galaxy evolution.  In this
section I discuss the basic properties of the ISM in LG dwarfs and
comment on some of these puzzles.  The chemical and kinematic
properties of the ISM are discussed in Sections 7 and 8, respectively.
Good reviews on the ISM in LG dwarfs have been written by Wilson
(1994a), Kennicutt (1994) and Skillman (1998); a more general review of
the ISM in dwarf galaxies can be found by Brinks \&\ Taylor (1994).

\vskip1em

\noindent{\large {\it 4.1 HI Content and Distribution}}
\vskip0.5em

\noindent DIRR GALAXIES\ \ Single-dish and aperture-synthesis radio
observations provide total fluxes (see Table~5) and detailed HI maps
of many of the dwarfs in the Local Group.  Many HI properties show a
clear progression from dIrr to dSph galaxies.  For example, Table~4
shows that the ratio of HI-to-total masses of dIrr galaxies range from
about 7\%\ to over 50\%\ (SagDIG is anomalous), which is consistent
with expectations from standard closed chemical enrichment models of
galaxies with low mean abundances (see Section 7).  Four of the five
transition galaxies (denoted `dIrr/dSph' in Table~1) have HI-to-total
mass ratios between 1\%\ and 10\%.  The exception, DDO~210, has a
particularly uncertain distance (Table~2).  The LG dSph galaxies are
all comparatively devoid of neutral hydrogen; the few with detectable
emission contain $\lsim$0.1\%\ of their mass in the form of
interstellar neutral hydrogen.

The spatial distribution of HI emission in most LG dIrr
galaxies is clumpy on scales of 100-300 pc scales (Shostak \&\
Skillman 1989, Carignan et al 1990, Hodge et al 1991a,b, Lo et al 1993,
Young \&\ Lo 1996a, Young \&\ Lo 1997b).  Diffuse HI emission is
inferred for many galaxies from the large differences in integrated
flux from single-dish and synthesis observations.  Only the most
luminous systems -- NGC~3109 and NGC~55 -- have comparatively smooth
HI distributions (Jobin \&\ Carignan 1990, Puche et al 1991).

The peak emission of individual HI clouds is generally found near
regions of optically active star formation, but the clouds are often
offset by 50-200 pc from the locations of the nearest star-forming
complexes (Gottesman \&\ Weliachew 1977, Hodge et al 1990, Hodge \&\
Lee 1990, Hodge et al 1991a,b, Hodge et al 1994).  Skillman et al
(1988) and Sait\=o et al (1992), among others, have suggested that
star formation requires a minimum HI column density of about $N({\rm
HI}) \sim 10^{21}$ cm$^{-2}$ to proceed.  However, for some galaxies,
the peak HI surface density exceeds this limit and yet there is
no current or recent star formation (Shostak \&\ Skillman 1989, Young
\&\ Lo 1997b).  It seems that a trigger is needed to initiate star
formation in these cases.  There are also counter examples -- mostly
in dSph or transition galaxies -- where recent or ongoing star
formation is apparent, yet $N$(HI) $< 10^{20}$ cm$^{-2}$ (Hodge et al
1991, Lo et al 1993, Young \&\ Lo 1997a,b).

On the largest scales the HI emission is generally centered on the
optical centroids of LG dIrr galaxies even in systems with complex HI
morphology (e.g. Lo et al 1993, Young \&\ Lo 1996a, 1997b, see also
Puche \&\ Westpfahl 1994 for examples beyond the Local Group).
Transition galaxies are more complicated: the neutral hydrogen in
LGS~3 is centered on the optical galaxy, while in Phoenix the HI -- if
it is in fact associated with the galaxy -- is distinctly offset from
the optical light (Young \&\ Lo 1997b).  In most LG dIrr galaxies, the
neutral gas is more extended than the optical emission (Hewitt et al
1983, Lake \&\ Skillman 1989, Young \&\ Lo 1996a, 1997b).  However,
for the two most luminous dIrr systems in the sample (NGC~55 and
NGC~3109, Table~4), the surface brightness profile scale lengths and
shapes are similar for the HI emission and optical light (Jobin \&\
Carignan 1990, Puche et al 1991).

Young \&\ Lo (1996, 1997b) have found evidence that the atomic
component of the ISM in Leo~A has two distinct phases.  The warm
component has a velocity dispersion of 9 km~s$^{-1}$ pervades much of the
galaxy, while the cooler component ($\sigma \sim 3$ km~s$^{-1}$) is found
principally near optical HII regions and contributes 10-20\%\ of the
total HI flux.  Remarkably, though Leo~A is 400 times fainter than LMC, both
exhibit this two-phase HI structure.  The HI gas in NGC~185
and NGC~205 also seems to exhibit the same two-phase structure, even
though in these cases the HI is clearly in a non-equilibrium
configuration and has a much lower column density (Young \&\ Lo
1997a).

\vskip0.25em

\noindent EARLY-TYPE DWARFS\ \ Deep single-dish HI observations have
failed to detect most of the early-type LG dwarfs (Knapp et al 1978,
Mould et al 1990, Koribalski et al 1994, Oosterloo et al 1996).  Knapp
et al (1978) detected HI emission near Sculptor, but lacking a
precise optical velocity for the galaxy, they tentatively concluded
that it was not associated with the galaxy.  HI emission has long
been known to exist in NGC~185 and NGC~205, but no HI is
detected in NGC~147 and M32 to similar limits (Johnson \&\ Gottesman
1983, Young \&\ Lo 1997a, Huchtmeier \&\ Richter 1986, Table~5).

The unique case of Sculptor has been recently revisited by Carignan et
al (1998), who find that the emission reported earlier (Knapp et al
1978) is probably associated with Sculptor (the optical velocity is
now accurately known; Table~2).  Figure~6 is a plot of the HI map and
the optical image of Sculptor.  The systemic optical and HI velocities
agree to within their combined errors (Armandroff \&\ Da Costa 1986,
Queloz et al 1995), though there is a hint that the HI velocity may be
larger.  For NGC~185 and NGC~205 the HI emission is clearly offset
spatially and kinematically from the optical counterparts (Young \&\
Lo 1997a, Carignan et al 1998).  As in dIrr systems, the HI emission
in these two galaxies is also spatially offset from the young stars
(Johnson \&\ Gottesman 1983, Young \&\ Lo 1997a).

Because the extent of Sculptor's HI emission is comparable to the
beam size, the map in Figure~6 may be highly incomplete.  The actual
distribution could be a ring or some other more complex bimodal
geometry (Puche \&\ Westpfahl 1994, Young \&\ Lo 1997b).  What's
certain, however, is that the flux received from the small central HI
emission is much less than the flux from the extended component.
Because past HI observations of dSph galaxies were centered on the
optical image, they could conceivably have missed all of the
emission from even a relatively strong extended component.  It would
be extremely interesting to re-examine the nearby dSph at 21-cm,
taking particular care to search for extended structures.

\vskip1em

\noindent{\large {\it 4.2 Dust and Molecular Gas}}

\vskip0.5em

\noindent Evidence of interstellar dust clouds is seen optically as
compact absorption regions in some LG dIrr galaxies and near the cores
of some dSph systems (Hodge 1963b, Hodge 1973, Hodge 1978, Price
1985).  The dust is generally clumped into small clouds ($D \sim$
20-40 pc) with inferred masses of a few hundred solar masses.  In
NGC~185, Price (1985) argues that the extinction law of the dust
regions differs significantly from the standard Galactic extinction
law.  Both the total optical extinction and total mass of individual
dust clouds in low-luminosity dIrr seem to be smaller than for the
clouds -- when seen - in dSph galaxies (Ables \&\ Ables 1977, Hodge
1978, Hodge \&\ Lee 1990.  Some LG dIrr galaxies suffer variable
internal extinction, presumably from a pervasive, non-uniform dust
sheet (Gallart et al 1996a).  van Dokkum \&\ Franx (1995) found no
optical evidence of dust clouds in the core of M32 using archival HST
observations.  Bendinelli et al (1992) claim to see a central
reddening in M32, possibly due to a central dust component; however,
Peletier (1993) finds no optical color gradients to within 1 arcsec of
the center of the galaxy.

Nearly 25\%\ of the LG dwarfs have now been detected in CO emission
(Table~5).  These detections include the lowest luminosity galaxies in
which molecular gas has been observed (Roberts et al 1991).  The CO
emission is typically confined to distinct clouds with diameters of
$\lsim 50$ pc (Ohta et al 1988, 1991, Sait\=o et al 1992, Wilson 1995,
Welch et al 1996, Young \&\ Lo 1997a).  By combining spectra from
fields without direct detections, Israel (1997) has found evidence of
a diffuse CO component in NGC~6822.  Many LG dwarfs have been detected
with IRAS at 60$\mu$m and 100$\mu$m (Melisse \&\ Isreal 1994a,b, Knapp
et al 1985).  All of the LG IRAS sources either contain dust and/or a
significant population of stars younger than about 10 Myr.
Submillemeter observations have also proven useful to track dust in LG
dwarfs both from its continuum emission (e.g. Thronson et al 1990,
Fich \&\ Hodge 1991) and from Carbon line emission (Madden et al
1997).

In general, the spatial and kinematic distribution of dust, CO
emission, HI and optical star-formation regions are well correlated,
but there are some interesting exceptions.  Hodge \&\ Lee (1990),
Richer \&\ McCall (1991), and Welch et al (1996) found or inferred
small spatial offsets between HI and CO emission regions in IC~10,
NGC~3109, and NGC~185, respectively.  The locations of optical HII
regions often correlate very well with CO emission regions (Sait\=o et
al 1992, Hodge \&\ Lee 1990).  However, the CO and H$\alpha$ emission
lines are often redshifted relative to HI (Tomita et al 1993), which
is perhaps indicative of infall or collapse in the denser regions
where CO is observed.  Interestingly, in NGC~185 some regions with
strong CO emission appear devoid of optical dust (Welch et al 1996).
But when optical dust is present, the regions are usually detected in
CO (Gallagher \&\ Hunter 1981; IC~1613 may be an exception, Hodge
1978, Ohta et al 1993).

Numerous studies have used LG dwarfs to measure the conversion factor,
$X$, between CO emission and H$_2$ molecular mass as a function of
metallicity.  $X \equiv N_{H_2}/S(CO)$, where $N_{H_2}$ is the
molecular hydrogen column density, and $S(CO)$ is the integrated CO
flux density or intensity.  Low-luminosity -- hence low-metallicity
(Section 5.2) -- dwarfs should have larger H$_2$-CO conversion factors
since CO formation will be hindered at low abundances for a given
molecular mass.  Two methods are used to estimate the molecular masses
needed to calculate $X$.  First, the observed CO line width is taken
as a measure of the cloud velocity dispersion from which the virial
mass is determined (e.g. Wilson 1994b, 1995).  The second approach
combines IRAS and HI fluxes throughout a galaxy to determine the total
hydrogen column density (neutral plus molecular) where CO is observed
(Israel 1997).  Recent studies all agree that $X$ is higher for
low-luminosity dwarfs, but the precise form, slope and zero point of
the $L$-$X$ relation is still controversial (Ohta et al 1993, Wilson
1995, Verter \&\ Hodge 1995).  In NGC~6822 ([Fe/H] $\sim -0.7$;
Section 5) $X$ is about two to five times higher than in the Galaxy
(Ohta et al 1993, Wilson 1995, Israel 1997), while for GR~8 ([O/H]
$\sim -1.3$), $X \geq 10$ times the Galactic value (Ohta et al 1993,
Verter \&\ Hodge 1995). Ohta et al (1993) and Israel (1997) both note
that $X$ shows considerable scatter at a given metallicity, implying
that some other parameter affects the H$_2$-CO ratio.

\vskip1em

\noindent{\large {\it 4.3 HII Regions, SN Remnants and X-Rays}}

\vskip0.5em

\noindent The integrated H$\alpha$ fluxes of LG dwarfs are listed
in Table~5.  All of the dIrr galaxies in the Local Group contain HII
regions.  Hodge and Lee (1990) introduced a morphological
classification scheme for these HII regions; Hunter et al (1993) have
published deep H$\alpha$ images of many LG dwarfs that provide an
excellent way to appreciate this rich morphological variety.  The
distribution of morphological types of HII regions differs between
galaxies (Hodge \&\ Lee 1990, Hunter et al 1993), though the size
distribution of HII regions is generally well-fit as a power-law
truncated at a maximum HII region diameter of about 200-400 pc
(Strobel et al 1991, Hodge et al 1994, Hodge \&\ Miller 1995).

Only one dSph galaxy (NGC~185) and one transition system (Antlia) have
detected HII emission.  The H$\alpha$ emission from NGC~185 appears to
be related to an old supernova remnant (Gallagher \&\ Hunter 1984,
Young \&\ Lo 1997a); the high excitation led Ho et al (1995, 1997) to
classify the galaxy as a Seyfert~2!  In Antlia, the HII region is
extremely faint (Aparicio et al 1997a; the region is visible in the
color image of Whiting et al 1997).  As in Pegasus (Aparicio \&\
Gallart 1995, Skillman et al 1997), its presence may merely reflect
the stochastic nature of high-mass star formation in systems with
relatively low average star-formation rates (Aparicio et al 1997b).

Using radio continuum observations, Yang \&\ Skillman (1993)
identified an unusually large non-thermal source in IC~10 that they
argue is the remnant of multiple recent supernovae shells.  This
conclusions received support from the subsequently observations of
optical filaments from the radio-continuum shell (Hunter et al 1993).
Non-thermal sources have also been observed in IC~1613 and NGC~6822
(Klein \&\ Gr\"ave 1986), which are also probably from old SN
remnants.  Virtually all other sources identified in these galaxies
are thermal sources associated with optical HII regions or non-thermal
background sources.

No diffuse X-ray emission has been detected in any LG dwarf (Markert
\&\ Donahue 1985, Fabbiano 1989, Gizis et al 1993).  This is not
surprising: if hot gas was produced during periods of active star
formation in any LG dwarf, it would have been rapidly expelled from
the galaxy and faded to invisibility. The nearby dwarf NGC~1569
appears to be the closest example of a dwarf galaxy experiencing this
short-lived X-ray emitting phase (Heckman et al 1995).  In general,
LG dwarfs contain few known X-ray sources of any kind, though some
possible X-ray binaries have been detected in a few systems 
(Eskridge \&\ White 1997, Brandt et al 1997, and references above;
but see Eskridge 1995).

\vskip1em
\noindent{\large {\it 4.4 The ISM `Crisis' in dSph Galaxies}}

\noindent NGC~147 and NGC~185 are virtually twins; their luminosities,
mean masses, abundances, abundance dispersions, average star-formation
rates, sizes, core and exponential radii are extremely similar
(Tables~3-7).  NGC~147 does have a significantly fainter central
surface brightness than NGC~185 (Table 3), but the latter contains
young stars in its core (see Section 6) which probably boosts its
central luminosity density.  Their kinematic properties, however,
indicate that both galaxies have very similar central mass densities
(Tables 4 and 7).  Yet, when considering their gaseous component
(Table~5), it is immediately apparent that while NGC~185 contains a
significant ISM, NGC~147 has none.  This is extremely puzzling.  Many
authors agree that the gas replenishment timescale in galaxies such as
these would be approximately 0.1-1 Gyr from internal sources such as
planetary nebulae or red giant winds (Ford et al 1977, Mould et al
1990, Gizis et al 1993, Welch et al 1996, Young and Lo 1997a).
Paradoxically, NGC~185 contains young stars and even an old SN remnant
(Price 1985, Lee et al 1993a, Young \&\ Lo 1997a) yet this activity
has not blown out its gas.  Since NGC~147 has no stars younger than 1
Gyr (Han et al 1997), we cannot simply claim that we have caught it
just after an energetic star-formation episode that consumed or
expelled all of its gas.

Few early-type LG galaxies have been mapped at HI, but most of the
ones that have contain distinct HI clouds with masses $\sim 10^5$
\Msol\ (if at the distance of the galaxy) and diameters of $\sim 200$
pc or larger (Carignan et al 1991, Young \&\ Lo 1997a, Carignan et al
1998).  This gas is always
significantly offset from the optical centers of the galaxies.  Young
\&\ Lo (1997a) further emphasize that the configuration and kinematics
of the gas is highly unstable: These HI clouds must be short-lived
structures.  As we shall see in Section~6, most LG dwarfs -- including
the early-type systems -- have surprisingly complex and varied
star-formation histories.  In many cases, there is evidence of star
formation in the past 10$^9$ years, yet few seem to contain any gas
that could have fueled this activity (though see Section~4.1).

If the gas is of internal origin and we have not come onto the scene
just as all dSph systems used up all their gas, then these galaxies
would have had to somehow avoid accumulating any gas between
star-formation episodes (as the lack of central HI and the NGC~147/185
paradox seems to be telling us).  Could structures such as those seen
in Sculptor serve as `holding tanks' for such quasi-expelled gas?
Another option is that the gas is of external origin (Knapp et al
1985).  This superficially explains the generally asymmetric
distribution of gas in early-type systems (except LGS~3; Young \&\ Lo
1997b), the kinematic offsets of the gas and stars in these galaxies,
and the possibly complex chemical-enrichment history of at least one
dSph system (Smecker-Hane et al 1994; accreted clouds could have any
metallicity); it even provides a repository -- the halo -- for gas
expelled from these galaxies during earlier episodes. dIrr galaxies
may have less chance to accrete clouds because they are further from
M31 and the Milky Way (Figure 3).  IC~10 shows evidence of a disturbed
outer HI velocity field and has a very high current star-formation
rate (Table~5, Shostak \&\ Skillman 1989) However, these authors warn
that many other isolated dIrr systems show similarly complex
kinematics, so such characteristics do not necessarily imply a recent
encounter or merger with another galaxy or HI cloud.  Past surveys for
high-latitude HI clouds would have missed low-velocity clouds under 10
arcmin in diameter and with $M(HI) \sim 10^5$-$10^6$ \Msol if located
$> 50$ kpc away (Wakker \&\ van Woerden 1997).  The crucial dilemma
for any accretion model is to understand how systems with escape
velocities as low as 10-15 km~s$^{-1}$ can snare gas within a halo
with a velocity dispersion at least 10 times larger.

\vfill\eject

\noindent {\Large{\bf 5. CHEMICAL ABUNDANCES IN LOCAL GROUP DWARFS}}

\vskip1em

\noindent Photometric and spectroscopic techniques can be used to
estimate heavy-element abundances in LG dwarfs.  For early-type
galaxies the properties of the red giant branch (RGB) constrains
[Fe/H], and in some cases, the abundance dispersion, $\sigma_{\rm
[Fe/H]}$.  Apart from helium and some molecular species, photometry is
poorly suited to determine abundances of other, individual elements.
Photometric abundances have now been measured in some LG dIrr systems where
the old/intermediate-age RGB population can be observed directly in
deep color-magnitude diagrams (eg Sextans~A: Dohm-Palmer et al 1997,
Leo~A: Tolstoy et al 1998 Various: Lee et al 1993c; see Table~6 for
other examples).  Spectroscopy can be used to determine abundances
both of individual stars and emission nebulae such as HII regions and
planetary nebulae.  Generally, spectroscopy provides abundances for
specific elemental, ionic, or molecular species, which somewhat
complicates comparisons with photometric [Fe/H] abundance estimates.
Some good recent reviews of dwarf abundances can be found Skillman \&\
Bender (1995) and Skillman (1998).

\vskip1em
\noindent {\large {\it 5.1 The Observational Basis for LG Dwarf
Abundances}} 
\vskip0.5em

\noindent RGB ABUNDANCES \ \ Da~Costa and Armandroff (1990) observed
RGB sequences using the V and Cousins I bands in a number of globular
clusters ranging from $-0.7$ to $-2.3$ in [Fe/H].  These sequences
have helped establish the (V$-$I) TRGB method of determining distances
(Section 2.2).  They also found that the colors of the giant branches
define a monotonic sequence with respect to abundance: [Fe/H] =
$-15.16 +17.0({\rm V-I})_{-3} - 4.9({\rm V-I})_{-3}^2$, where
(V$-$I)$_{-3}$ is the reddening-corrected (V$-$I) color of the giant
branch at $M_I = -3.0$.  This relation is valid for $-0.7 > {\rm
[Fe/H]} > -2.2$.  Lee et al (1993b) re-evaluated this expression at
$M_I = -3.5$ for easier application in distant galaxies: [Fe/H] =
$-12.64 + 12.6({\rm V-I})_{-3.5} -3.3({\rm V-I})_{-3.5}^2$. This
relation has the same range of validity as the earlier one since both
were derived from the same data (Da~Costa \&\ Armandroff 1990).

Photometric abundance indicators such as the RGB color run into
complications in dwarf galaxies.  Unlike clusters, most dwarfs exhibit
significant abundance dispersions.  Because dwarf galaxies are
considerably more distant on average than Galactic globular clusters,
few galaxies have reliable HB photometry (though see Da~Costa et al
1996); those that do often reveal unusual HB morphologies, such as
bimodal (in luminosity) HB sequences (Carina: Smecker-Hane et al 1994;
Sagittarius: Sarajedini \&\ Layden 1995) or compact, super-red HBs
(Leo~I: Lee et al 1993c, Caputo et al 1995).  It is not always
feasible to use abundance indices such as (B$-$V)$_{0,g}$ (the (B$-$V)
color of the RGB at the level of the horizontal branch (Sandage \&\
Wallerstein 1960, Sandage \&\ Smith 1966, Zinn \&\ West 1984) to
estimate the metallicities of these galaxies.  Another complication is
that most LG dwarfs exhibit complex star-formation histories (Section
6).  It remains to be seen how reliably these abundance indices --
derived from globular clusters -- can measure the composite
populations of nearby dwarfs (see also Grillmair et al 1996).  It
would clearly be useful to extend the range of photometric abundance
indicators with observations of the RGBs in populous clusters in the
Galaxy and the Magellanic Clouds.

With few exceptions, every LG dwarf for which abundances have been
determined from the RGB shows evidence of a significant abundance
dispersion (Table~6).  In M32 (Grillmair et al 1996) and NGC~205
(Mould et al 1984) the color distribution of the RGB implies an
abundance distribution that is skewed towards higher abundances.
Although these color dispersions principally reflect variations in
abundances, an age dispersion can also (slightly) broaden the RGB (eg,
Bertelli et al 1994, Meynet et al 1993).  Carina has a large age
spread and populations with distinct abundances (Smecker-Hane et al
1994, Da Costa 1994a), yet it has a narrow RGB. Leo~I also exhibits a
large spread in age but has a wide giant branch (Lee et al 1993a).  In
Carina the age spread seems to compensate for the metallicity
dispersion, while in Leo~I it does not.  Does this imply radically
different chemical enrichment patterns for the two galaxies?  This
underscores an obvious, but important, point: The star-formation and
chemical-enrichment histories of dwarfs cannot be interpreted
independently.  To derive one history requires careful consideration
of the other (Hodge 1989, Aparicio et al 1997b,c).

\vskip0.25em

\noindent SPECTROSCOPIC ABUNDANCES \ \ Among the early-type dwarfs in
the Local Group, spectroscopic abundances (typically [Fe/H] or [Ca/H])
have been measured for individual stars in Draco (Lehnert et al 1992
and references therein), Sextans (Da~Costa et al 1991, Suntzeff et al
1993), Carina (Da~Costa 1994a), Sagittarius (Da~Costa \&\ Armandroff
1995, Ibata et al 1997), and Ursa~Minor (EW Olszewski \&\ NB Suntzeff,
private communication).  Oxygen abundances have been derived from
spectroscopy of planetary nebulae in Fornax (Maran et al
1984, Richer \&\ McCall 1995), Sagittarius (Walsh et al 1997),
NGC~185, and NGC~205 (Richer \&\ McCall 1995), as well as the dIrr
NGC~6822 (Dufour \&\ Talent 1980).

Spectroscopy of HII regions in LG dIrr galaxies typically target
oxygen, but abundances of other elements such as nitrogen, sulphur,
and helium have also been measured (eg Pagel et al 1980, Garnett 1989,
1990, Garnett et al 1991).  The improved blue-sensitivity of CCD
detectors in recent years has greatly aided nebular abundance studies
by making it simpler to derive reliable physical conditions in HII
regions.  This alone greatly improves the consistency and precision of
the oxygen abundances derived in this manner (see Skillman 1998 for
details).  Table~6 lists the oxygen abundances for LG dIrrs that have
adequate data.   Unlike [Fe/H] in
early-type dwarfs, there is no evidence for significant dispersion of
the oxygen abundances in any LG dIrr in which multiple HII regions
have been studied (Pagel et al 1980, Skillman et al 1989a, Moles et al
1990, Hodge \&\ Miller 1995).  Because HII regions are associated with
young populations (though the gas itself may derive from relatively
old stars), these nebular abundances nicely complement those derived from
intermediate-age planetary nebulae (Olszewski et al 1996b).

Integrated spectroscopy is impractical for most of the dwarfs of the
Local Group largely because of their extremely low surface
brightnesses (Sembach \&\ Tonry 1996 suggest one method to overcome
this problem).  M32 is a famous exception for which a number of
integrated spectroscopic studies in the optical and ultraviolet (UV)
have been carried out (see O'Connell 1992 and Grillmair et al 1996 for
reviews).  The excess UV light in its spectrum is generally taken as
evidence of a population of relatively young stars associated with the
IR-luminous asymptotic giant branch (AGB) stars found by Freedman
(1992) and Elston \&\ Silva (1992).  However, the spectrum only weakly
constrains the stellar abundance of M32 (Grillmair et al 1996).
Optical spectra reveal complex radial gradients of the Balmer lines
(these weaken outward), while the Mg lines remain constant with radius
and CH increases in strength (Davidge 1991, Gonz\'alez 1993).  Oddly,
M32 exhibits {\it no} strong radial color gradients apart from in the
UV (Michard \&\ Nieto 1991, Peletier 1993, Silva \&\ Elston 1994,
O'Connell 1992).  These conflicting tendencies have greatly
complicated spectroscopic determinations of the galaxy's abundance
even now that deep HST photometry is available (Grillmair et al 1996).
The nuclear region of NGC~205 has also been observed spectroscopically
(Bica et al 1990), revealing a prominent young population (age $\lsim
10^8$ yrs) with a mean metallicity of [Fe/H] $\sim -0.5$, along with
more older, metal-poor stars (ages $\gsim$ 5 Gyr; [Fe/H] $\lsim
-1.0$).  A lower nuclear abundance has been derived by Jones
et al (1996) for NGC~205 ([Fe/H] $\sim -1.4$) from UV spectra.

\vskip1em
\noindent {\large {\it 5.2 The Metallicity-Luminosity Relation and the
dIrr/dSph Connection}}
\vskip0.5em

\noindent The fact that the more luminous dwarf galaxies are also on
average the most metal rich has been known for some time for both dIrr
(Lequeux et al 1979, Talent 1980, Skillman 1989a) and dSph galaxies
(Aaronson 1986, Caldwell et al 1992).  Aaronson (1986) and Skillman et
al (1989a) merged the abundance data for both types into a a single
luminosity-abundance (L-Z) relation spanning 12 magnitudes in $M_B$,
and about 1.6 dex in oxygen/iron abundance.  A recent determination of
[Fe/H] of a low surface brightness but relatively luminous dSph galaxy
in the M81 group (Caldwell et al 1998) demonstrates clearly that
luminosity, not surface brightness, is the principal parameter
correlated with metallicity in dSph, and presumably, dIrr galaxies.

Figure~7 is a plot of the mean abundances for all of the galaxies in
Table~6 (except DDO~210 for which [Fe/H] is uncertain; Greggio et al
1993) vs their mean V-band absolute magnitudes (Table~4).  The data
have been corrected for external reddening; internal extinction is
probably insignificant for most of these systems (see Table~5).  In
Table~6, 10 galaxies have reliable [Fe/H] {\it and} oxygen abundances,
including some dIrrs with stellar abundance estimates, and some
early-type dwarfs with oxygen abundances from planetary nebulae.  The
mean difference between [Fe/H] and [O/H] (defined as $\log({\rm O/H})
- \log({\rm O/H})_\odot$) is 0.37$\pm$0.06 dex.  I have therefore
added $-$0.37 to the oxygen abundances before plotting them in
Figure~7.

Even if the offset to the nebular abundances is disregarded, the
stellar [Fe/H] abundances show a bimodal, or possibly discontinuous,
behavior.  The oxygen abundances only reinforce this conclusion.  The
`upper branch' in Figure~7 (denoted by the {\it dotted line}) contains
only dSph galaxies and all four transition galaxies (LGS~3, Phoenix,
Antlia, and Pegasus; see Table~1) with precise abundance estimates.
All of the galaxies fainter than $M_V \sim -13.5$ with nebular
abundances are dIrr systems and clearly fall below the dSph relation
by about 0.6-0.7 mag.  At the other extreme, nearly all of the
galaxies brighter than $M_V \sim -13.5$ are dIrr systems -- with the
important exceptions of NGC~147, NGC~185, NGC~205 and M32.  A similar
bifurcation of the luminosity-abundance (L-Z) relation was suggested
by Binggeli (1994) and Walsh et al (1997).  Caldwell et al (1992,
1998) adopted a single relation to fit all the data for early-type
galaxies in the Local Group and other groups and clusters.  However, a
single linear relation in Figure~7 ignores the strong segregation of
dIrr and dSph for $M_V$ $\gsim -13.5$.  The proposed bimodal L-Z
relation also helps remove the abundance anomaly exhibited by
Sagittarius (Ibata et al 1994, Mateo et al 1995c, Sarajedini \&\
Layden 1995, Ibata et al 1997) for which the L-Z relation of Caldwell
et al (1992, 1998) implies that $M_{V,Sgr} \lsim -16$.  This is about
2-3 magnitudes brighter than observed (Mateo et al 1995c, Ibata et al
1997).  Based on the current data, there is no correlation between the
offset (in magnitudes) from the {\it dashed line} in the {\it upper
panel} of Figure~7 and the intrinsic color of the galaxy (Skillman et
al 1997), as might be expected if the bimodal behavior simply reflects
the effects of current star-formation on the integrated luminosities
of dIrr galaxies.

Figure~7 addresses the relationship between dwarf ellipsoidal (dSph
and dE) galaxies, and dIrr systems.  Star-formation (Dekel \&\ Silk
1986, Babul \&\ Rees 1992, De Young and Heckman 1994) and ram-pressure
stripping (Faber \&\ Lin 1983, van den Bergh 1994c) have been proposed
as means of removing gas from dwarfs in the inner Galactic halo.  The
spatial segregation of dSph and dIrr (e.g. Figure~3) appears
consistent with the latter idea.  But there are other fundamental
problems if dSph galaxies are supposed to be simply `gas-free' dIrr
(see reviews by Ferguson and Binggeli 1994, Binggeli 1994, Skillman
\&\ Bender 1995). Hunter \&\ Gallagher (1984) and Bothun et al (1986)
showed that the present-day central surface brightnesses of dIrrs will
be considerably lower than in dSph galaxies after evolutionary fading,
while James (1991) found large systematic structural differences
between Virgo dIrr and dSph galaxies that seem inconsistent with a
common origin or a single evolutionary endpoint (see also Section
3.2).  Binggeli (1994), Richer \&\ McCall (1995) and Walsh et al
(1997) all noted that at a given luminosity dIrr galaxies are
generally more metal poor than dSph systems, which is precisely the
effect seen in Figure~7.

If dIrr and dSph galaxies do indeed represent fundamentally different
objects, why does a proto-dwarf galaxy choose one type rather than the
other?  It is interesting that even the lowest-luminosity dIrr systems
show evidence for rotation (eg GR~8: Carignan et al 1990; Leo~A: Young
and Lo 1996) even though $v_{rot}/\sigma < 1.0$.  The only rotating
dSph systems are NGC~147 (Bender et al 1991) and UMi (Armandroff et al
1995, Hargreaves et al 1994b). The latter's rotation may reflect
streaming motions induced by external tides (Piatek \&\ Pryor 1995, Oh
et al 1995), and both galaxies have $v_{rot}/\sigma_0 < 1.0$.  Could
angular-momentum be the factor that distinguishes dIrr (high angular
momentum) and spheroids (low angular momentum)?  Alternatively,
Skillman and Bender (1995) suggested that the strength of the first star
formation episodes dictates this distinction; galaxies
experiencing little or no early star formation become dIrr systems (see also
Aparicio et al 1997b).  Or is environment the deciding factor after
all (van den Bergh 1994c; Figure~3)?

There are serious objections to each possibility.  Three of the four
early-type dwarf satellites of M31 that lie in the proposed dIrr
branch in Figure~7 (NGC~147, NGC~185, and NGC~205) do not
significantly rotate (Bender \&\ Nieto 1990, Held et al 1990, 1992,
Bender et al 1991), although dIrr galaxies of similar luminosity do
(see Section~7). The fourth galaxy, M32, does rotate (Tonry 1984,
Dressler \&\ Richstone 1988, Carter \&\ Jenkins 1993) but its
structural parameters are not like any dIrr.  Many dIrr galaxies do
have pronounced ancient populations, while some dSph galaxies exhibit
evidence for few or no old stars (Section 6).  Thus, the amplitude or
timing of the first episodes of star formation do not appear to
differentiate dIrr and dSph galaxies.  Finally, although galaxy types
are segregated as a function of distance from M31 or the Milky Way
(Figure~3), there are glaring exceptions.  The Magellanic Clouds are
dIrr galaxies that lie close to the Milky Way, while Tucana is an
example of an inactive dSph far from any large galaxy.  Environment
also offers no easy explanation of the segregation of faint dSph and
dIrr galaxies in Figure~7.

\vfill\eject

\noindent {\Large{\bf 6. STAR FORMATION HISTORIES OF LOCAL GROUP DWARFS}}
\vskip1em

\noindent Mould and Aaronson (1983) published deep CCD photometry of
the nearby Carina dSph galaxy and showed conclusively that it is
dominated by intermediate-age (4-8 Gyr) stars.  Although there was
already compelling evidence to suggest that some of the dSph galaxies
might contain relatively young stars (Zinn 1980), it was still widely
assumed that dSph galaxies such as Carina were all ancient stellar
systems, much like globular clusters.  The balance of opinion has
nearly completely reversed, and it is widely believed that few, if
any, LG dwarf galaxies contain only ancient stars.  Some recent
reviews of the exciting developments of this field have been written
by Da~Costa (1994a,b, 1998), Hodge (1994), Stetson (1997), and Grebel
(1997).

\vskip1em

\noindent{\large {\it 6.1 Basic Techniques and Ingredients}}

\vskip0.5em

\noindent The key ingredient to deciphering the fossil record of star
formation in nearby galaxies is deep CCD photometry of individual
stars.  Methods that rely on analysis of the most luminous, young
stars, HII regions or the integrated galaxy colors at a variety of
wavelengths (Hodge 1980, Kennicutt 1983, Gallagher et al 1984)
invariably lose age resolution for populations older than about 1
Gyr.  As shown below, many LG dwarfs were very active
during that entire time interval.

Recent technical developments have greatly expanded our ability to
constrain the SFHs of individual galaxies.  Optical and IR detectors
have greatly improved in quality and size.  Improved computing
capabilities have also been essential to progress in this field.
Extensive simulations of observations are required to correct for
effects such as crowding, internal reddening, incompleteness, and
photometric errors (eg Aparicio \&\ Gallart 1995, Gallart 1996a,b,
Tolstoy 1996, Mart\'\i nez-Delgado \& Aparicio 1997, Hurley-Keller et
al 1998).  The expansion of grids of stellar-evolutionary models has
also been crucial (Schaller et al 1992, Bertelli et al 1994):
observations of local dwarf galaxies have driven us into regions of
parameter space -- low age and low metallicity -- where we have
never before had to venture.

\vskip1em
\noindent{\large {\it 6.2 Age Indicators in Local Group Dwarfs}}
\vskip0.5em

\noindent A galaxy's star-formation history (SFH) can be determined by
simultaneously comparing its photometric data with appropriate
composite models, suitably corrected for observational effects
(Bertelli et al 1994, Gallart et al 1996a,b, Aparicio \&\ Gallart
1995, Tolstoy \&\ Saha 1996, Hurley-Keller et al 1998).  This is a
powerful approach that can, in principle, constrain the entire
star-formation and chemical history of a galaxy.  Of course, the
method depends critically on the precision of the input models and on
its implicit assumptions; e.g. that the chemical enrichment is a
monotonically rising function of time.  A heuristic drawback of this
approach is that the process is not terribly intuitive; consequently,
I shall discuss here specific evolutionary phases that have proven
especially useful as age tracers and to note the age range over which
they can be used.  The theoretical Hertzsprung-Russell (HR) diagrams
of Schaller et al (1992; their Figures 1 and 2) and Figure~1 of
Gallart et al (1996b) are particularly helpful guides to the
evolutionary phases discussed here.  A good general discussion is
given by Chiosi et al (1992).

\vskip0.25em

\noindent WOLF-RAYET STARS \ \ These high mass stars signal vigorous
star formation during the past 10 Myr (Massey 1998).  The frequency of
Wolf-Rayet (WR) stars depends on mass-loss rates, metallicity, the
star-formation rate for high-mass stars, and the high-mass end of the
initial mass function (IMF; Meynet et al 1994, Massey \&\ Armandroff
1995, Massey 1998).  Among LG dwarfs, only IC~1613, NGC~6822 and IC~10
are known to contain WR stars.

\vskip0.25em

\noindent BLUE-LOOP STARS\ \ Stars of intermediate mass evolve through
prolonged `blue loops' after they ignite He in their cores.  The
luminosity at which the loops occur depend principally on the mass of
the star, though the color and extent of the loop is critically
sensitive to metallicity.  For stars ranging in age from 100-500 Myr,
the loop luminosities ($L_{BL}$) fade monotonically with age.
Dohm-Palmer et al (1997) used deep HST photometry that clearly
separates the upper main-sequence and blue-loop stars in the
color-magnitude diagram of Sextans~A.  They constructed a luminosity
function (LF) for the blue loop stars that is uncontaminated by other
evolutionary phases. Because of the nearly one-to-one correspondence
of luminosity and age for these stars, they were able to then directly
convert the LF into the SFH of Sextans~A with only an age-L$_{BL}$
relation from models.  Cepheid variables are closely associated with
blue-loop evolution (Schaller et al 1992, Chiosi et al 1992).

\vskip0.25em

\noindent RED SUPERGIANTS\ \ Mermilliod (1981) demonstrated that red
supergiants also fade monotonically with age for populations ranging
from about 10-500 Myr in age.  However, these stars exhibit a moderate
spread in luminosity at a given age.  For a composite system -- such
as a dwarf galaxy -- the red-supergiant LF will contain stars
exhibiting a range of ages at a given luminosity.  Both the blue-loop
and red supergiant phases are short-lived (Maeder \&\ Meynet 1988,
Chiosi et al 1992, Schaller et al 1992, Wilson 1992a), and therefore
subject to added uncertainties from the stochastic nature of star
formation, particularly in dwarfs (Aparicio \&\ Gallart 1995).

\vskip0.25em

\noindent ASYMPTOTIC GIANT BRANCH (AGB) STARS \ \ Gallart et al
(1996a,b) descriptively refer to the AGB as the `red tail' extending
redward from the red giant branch.  They also discuss in detail the
practical problems of using the AGB to derive quantitatively the
intermediate-age star-formation history of a galaxy.  Figure 23 from
Gallart et al (1996b) shows clearly that the details the mass-loss
prescription used to model the AGB is critical to properly describe
AGB evolution (see also Charbonnel et al 1996).  Consequently, ages
for AGB stars probably cannot be determined to better than a factor of
2-3.  Nevertheless, AGB stars often provide our only constraint on
stellar populations older than $\sim 1$ Gyr in many distant galaxies
within and beyond the Local Group.  Long-period variables (LPV) are often
found among luminous AGB stars (Olszewski et al 1996b).

\vskip0.25em

\noindent RED GIANT BRANCH (RGB) STARS \ \ The RGB plays an important
role in understanding the chemical enrichment histories of dwarf
galaxies (Section 5.1), largely because its properties are relatively
insensitive to age.  Unfortunately, for a given metallicity, stars
spanning a large age range are funneled into a very narrow corridor
within optical color-magnitude diagrams (Chiosi et al 1992, Schaller
et al 1992).  The RGB serves only as a relatively crude age indicator
for populations older than 1 Gyr (Schaller et al 1992, Ferraro et al
1995).  LPVs are found near the upper tip of the RGB
(Caldwell et al 1998, Olszewski et al 1996b).

\vskip0.25em

\noindent RED-CLUMP AND HORIZONTAL BRANCH (HB) STARS \ \ The He core
burning phase occurs in a `red clump' located at the base of the RGB
for populations with ages in the range 1-10 Gyr.  The detailed
evolution of this clump has been studied recently by Caputo et al
(1995); its empirical behavior as a function of age in Magellanic
Cloud star clusters has been determined by Hatzidimitriou (1991).  The
clump evolution in luminosity ($\lsim 1$ mag) and color ($\lsim 0.5$
mag) is small even for large age differences.

Horizontal branch (HB) stars signal the presence of ancient
populations ($\gsim$ 10 Gyr; Olszewski et al 1996b).
RR~Lyr stars are an easily identified example of HB stars; blue HB
stars (BHB) are also distinctive but are known to exist in only two LG
dwarfs (Carina and Ursa~Minor), while RR~Lyr stars have been found in
13 systems (see Section 9).  The red HB (RHB) is also indicative of an old
population, but distinguishing it from the red clump in an
intermediate-age population can be difficult (Lee et al 1993c, Caputo
et al 1995). A beautiful example of the relationship of red-clump and
BHB stars is shown by Smecker-Hane et al (1994) for Carina.

\vskip0.25em

\noindent SUBGIANT BRANCH (SGB) STARS\ \ Stars with main-sequence
lifetimes longer than about 2-4 Gyr evolve slowly towards the RGB
after they exhaust hydrogen in their cores resulting in a
well-populated sub-giant branch below the luminosity of the
HB/red-clump stars (Meynet et al 1993).  At a given metallicity, the
minimum luminosity of subgiant branch (SGB) stars fades monotonically
with increasing age.  Bertelli et al (1992) and Hurley-Keller et al
(1998) rely heavily on the SGB to determine the star-formation history
of composite populations.

\vskip0.25em

\noindent MAIN-SEQUENCE STARS \ \ The main sequence (MS) is the only
evolutionary phase present in populations of all ages.  Unlike the
SGB, the maximum luminosity of the MS (the main-sequence turnoff)
fades with increasing age.  When an age spread is present, older
populations can be hidden by the unevolved main-sequence stars of
younger populations.  However, used in conjunction with the SGB, the
MS provides the only method of determining ages for populations older
than 1-2 Gyr with $\sim 1$ Gyr resolution (Bertelli et al 1992,
Holtzman et al 1997, Hurley-Keller et al 1998).  The main sequence has
one particularly useful feature: the maximum luminosity on the main
sequence can always be related to the age of the youngest
population at the precision of sampling uncertainties.  Short-period
dwarf Cepheids are associated with metal-poor main-sequence populations
(Nemec et al 1994, McNamara 1995, Mateo et al 1998b).

\vskip1em
\noindent {\large {\it 6.3 A Compilation of Star-Formation Histories of
Local Group Dwarfs}}

\noindent Hodge (1989) introduced the concept of `population boxes' as
a way of visualizing the star formation and chemical enrichment
histories of galaxies.  These three-dimensional plots show time,
abundance and star-formation rate (SFR) on orthogonal axes.
Unfortunately, we generally lack sufficiently detailed information to
plot real galaxies in this manner, though some recent studies have
begun to do so (Gallart et al 1996b,c, Aparicio et al 1997a,b,c,
Grebel 1997).  In order to present the star-formation history of LG
dwarfs in a uniform manner, I choose here to plot only the relative
SFR vs time in Figure~8 for 29 of these galaxies (see also van den
Bergh 1994c).  I have tried to convey the uncertainties of these
results as described in the caption; however, sampling errors and
systematic effects due to the evolutionary models are extremely
difficult to estimate with any precision.  References for individual
galaxies are listed in the caption.

A number of important conclusions regarding the star-formation histories
of LG dwarf galaxies can be drawn from Figure~8.

\begin{itemize}

\item No two LG dwarfs have the same star formation history!  

\item Many dIrr galaxies appear to contain significant old populations
as indicated by their pronounced RGBs (WLM, NGC~3109, NGC~6822), or by
the presence of RR~Lyr stars (IC~1613).

\item The most recent star-formation episodes are relatively short --
ranging from 10-500 Myr in duration -- in both dIrr and
early-type systems (WLM, NGC~205, IC~1613, Fornax, Carina, Sextans~B,
Sextans~A, NGC~6822, Pegasus).  Since all age indicators quickly lose
resolution for ages exceeding $\sim$ 1 Gyr, it is probably safe to
assume that short-duration bursts are typical of the entire
evolution of these galaxies.  For example, the seemingly long
intermediate-age episode of star formation in Carina may actually have
been a set of short, but observationally unresolved, bursts.

\item The second-parameter effect is common throughout the Local
Group.  That is, many galaxies with low metallicities possess
relatively red horizontal branches (Zinn 1980, Sarajedini et al 1997).
This suggests that the most ancient populations of these galaxies are
younger than the oldest Galactic globular clusters.  However, Draco
(Grillmair et al 1998) seems to present an interesting exception.
Main-sequence photometry reveals an old population, yet the galaxy's
HB is red.  In this case, age does not seem to be the second
parameter.

\item No single galaxy is composed exclusively of stars older than
10 Gyr with the possible exception of Ursa~Minor.

\item Some galaxies may contain very few or no stars older than 10 
Gyr (M32, and possibly Leo~I).

\item van den Bergh (1994b) suggested that dSph galaxies nearest the
Milky Way are on average older than more distant systems.  But this is
not strongly supported by Figure~8.  Carina, Fornax and Leo~I are all
`young' systems and are found 100-270 kpc from the Milky Way;
Ursa~Minor, Leo~II and Tucana are predominantly old systems found
70-880 kpc from the Milky Way.  NGC~205 contains young stars yet is
located near M31 (Section 3.2; Hodge 1973).

\end{itemize}

Most of the remaining galaxies listed in Table~2 that are not represented
in Figure~8 simply have insufficient data for even an educated guess of
their entire SFHs.  One galaxy deserves special mention.  Massey \&\
Armandroff (1995) note that IC~10 has the highest surface density of
WR stars of any region in any LG galaxy.  Two H$_2$O masers -- also
considered to be tracers of high-mass star formation -- have been
found in IC~10 (Becker et al 1993).  Radio continuum and optical
H$\alpha$ imaging reveal evidence of an enormous multiple-SN driven
bubble (Hunter et al 1993, Yang \&\ Skillman 1993).  The
inferred SFR of IC~10 (Table 5) is the highest by far of any LG dwarf.
By comparison, if Carina formed its entire dominant intermediate-age
population in 10 Myr, its total SFR would only slightly have exceeded
what we see today in IC~10 (Hurley-Keller et al 1998).  

Some galaxies are plotted in Figure~8 twice because of evidence that
they exhibit significantly different star-formation histories in their
inner and outer regions.  Radial population gradients have been
detected in many early-type systems (And~I, Leo~II, Sculptor, but not
Carina: Da~Costa et al 1996; NGC~205: Jones et al 1996; Antlia:
Aparicio et al 1997c), often as a gradient in the HB morphology.
Aparicio et al (1997c) commented on the core/halo morphology now
evident in some LG dIrr systems (see also Minniti \&\ Zijlstra 1996).
Mighell (1997) argues that the strong intermediate-age burst of
star-formation in Carina started in the center of that galaxy, and
then progressed outward.

\vfill\eject

\noindent {\Large{\bf 7. INTERNAL KINEMATICS OF LOCAL GROUP DWARFS}}
\vskip1em

\noindent For a given mass-to-light ratio, the central velocity
dispersion of a self-gravitating system in equilibrium scales as $(R_c
S_0)^{1/2}$, where $R_c$ is the characteristic radial scalelength of
the system, and $S_0$ is the central surface brightness in intensity
units (Richstone and Tremaine 1986).  Globular clusters have central
velocity dispersions of 2-15 km~s$^{-1}$.  Pressure-supported dwarf
galaxies that have central scale lengths about 10 times larger, and
surface brightnesses 10$^3$ times smaller should therefore have
central velocity dispersions of $\leq 2$ km~s$^{-1}$.  They don't.
All low-luminosity dwarfs have central velocity dispersions of $\gsim
7$ km~s$^{-1}$, independent of galaxy type, and regardless of whether
the dispersion is measured from the stars or gas.  In this section I
review the observational basis for DM in the dwarf galaxies of the
Local Group, and discuss some possible alternatives to DM in these
systems.  The seminal paper of this field was written by Aaronson
(1983).  Recent reviews include Gallagher \&\ Wyse (1994), Mateo
(1994), Pryor (1994, 1996), Gerhard (1994), and Olszewski (1998).

\vskip1em

\noindent {\large{\it 7.1 The Observational Basis for Dark Matter}}

\vskip0.5em

\noindent To estimate kinematic masses for galaxies we require a
measure of the velocity dispersion for pressure-supported systems, or
the rotation velocity for rotationally-supported galaxies, and an
estimate of the relevant scale length.  To determine mass-to-light
ratios, we further need the luminosity density or total luminosity.
The scale length required depends on the details of the dynamical model used
to interpret the kinematics; for non-rotating dwarfs the King core
radius or exponential scale length is commonly used (Table~3), while
for rotating systems the relevant length scale is taken from the
rotation curve (Table~7).

\vskip0.25em

\noindent DSPH GALAXIES\ \ Because they generally lack an ISM
component (or when they do, it is not in dynamical equilibrium; see
Section 4), and because they have such low surface brightnesses, the
internal kinematics of most LG dSph galaxies are based on
high-precision spectroscopic radial velocities of individual stars
(Olszewski 1998, Mateo 1994).   There have been at least four 
persistent criticisms of the reliability of dSph kinematics derived in
this manner.  

\begin{enumerate}

\item Some early claims of large central velocity dispersions were
based on spectra with single-epoch errors exceeding $\sim$ 5-10 km~s$^{-1}$.
Most recent studies rely on spectra and reduction techniques that
deliver single-epoch errors of 1-4 km~s$^{-1}$ (Mateo et al 1991b, Hargreaves
et al 1994b, 1996b, Olszewski et al 1995).  These smaller errors have
been confirmed from comparisons of results for common stars from
independent studies (Armandroff et al 1995, Queloz et al 1995,
Hargreaves et al 1994b, 1996b).

\item The most luminous AGB stars are subject to atmospheric motions,
or `jitter', with amplitudes of 2-10 km~s$^{-1}$.  Virtually all recent
studies have pushed far below the upper tip of the RGB to luminosities
where jitter is not seen in globular cluster red giants (Mateo et al
1991b, 1993, Vogt et al 1995, Queloz et al 1995, Hargreaves et al
1994a,b, 1996b).

\item Improved simulations of the effects of binaries have
convincingly shown that, barring a pathological binary period
distribution, dispersions based even on single-epoch observations are
negligibly affected by binary motions (Hargreaves et al 1996a,
Olszewski et al 1996a).  The limited data available suggest that the
binary frequency and period distribution are not radically different
in dSph galaxies from what is seen in the solar neighborhood (Mateo et
al 1991b, Hargreaves et al 1994b, 1996b, Olszewski
et al 1995, 1996a, Queloz et al 1995).

\item In recent years, samples of over 90 stars have become available
in many dSph galaxies (Armandroff et al 1995, Olszewski et al 1996a, 
Mateo 1997, Olszewski 1998). It is now possible to meaningfully
measure departures from Gaussian distributions for such large samples
(Merrifield \&\ Kent 1990, Olszewski 1998), or construct velocity
dispersion profiles within single galaxies (Mateo et al 1991b,
Da~Costa 1994a, Hargreaves 1994a,b, 1996b, Armandroff et al 1995,
Mateo 1997).

\end{enumerate}

These observational advances strongly suggest that modern measurements
of the velocity dispersions of LG dSph galaxies are indeed reliable
estimates of the true one-dimensional dispersions of these systems.  I
conclude that no dSph galaxy has a central velocity dispersion smaller
than 6.6 km~s$^{-1}$ (Table~7).

\vskip0.25em

\noindent DIRR GALAXIES\ \ Standard procedures exist to derive
rotation curves for dIrr galaxies within and beyond the Local Group
(Jobin \&\ Carignan 1990, Puche et al 1990, 1991).  However, only in
the Local Group do we encounter dIrr galaxies whose kinematics are not
dominated by rotation at all radii (Carignan et al 1990, 1991, Lo et
al 1993, Young \&\ Lo 1996a, 1997b).  GR~8 exhibits rotation in its
inner regions but then becomes pressure supported at large
radii (Carignan et al 1990).  The typical measure of the velocity
dispersion in gas rich but non-rotating dwarfs is from the 21-cm line
width.  Lo et al (1993) address this problem in detail; they recommend
that where possible, the dispersion be based on the line-of-sight
dispersion of the mean velocities of individual clouds.  In practice,
the two approaches seem to agree to within their combined errors.

\vskip1em

\noindent {\large{\it 7.2 Constraints on Dark Matter in Local Group
Dwarfs}}

\vskip0.5em

\noindent To determine masses and central mass densities of dSph
galaxies, the King formalism is generally adopted (Richstone \&\
Tremaine 1986), along with the simplifying assumptions that mass
follows light, and that the velocity dispersion is isotropic.
Pryor \&\ Kormendy (1990) have investigated the sensitivity of the
derived masses for dSph galaxies on these and other assumptions.  In
general, the resulting M/L ratios for most dSph galaxies with good
surface photometry are robust to within a factor of two (Pryor 1994, Mateo
1994, 1997).  This same approach is also used to derive masses for
pressure-supported dIrr systems (Lo et al 1993, Young \&\ Lo 1997b).
The rotation curves of the more luminous LG dIrr galaxies are
typically fit with two-component models.  One component of fixed M/L
follows the visible-light or HI distribution; the second dark
component is usually fit as an isothermal sphere to represent an
extended dark halo.  All of the LG dwarfs require both components for an
adequate fit to the observed rotation curves.  These techniques
implicitly assume the galaxies are in dynamical equilibrium.

Figure~9 is a plot of the derived M/L ratios for all LG dwarfs with
adequate data in Table~7.  The King method (Richstone \&\ Tremaine 1986)
allows one to calculate both the central density and total mass under
the assumptions listed above.  For the pressure-supported galaxies plotting
the central M/L ratio is therefore possible, while for all systems with
kinematic data, the global M/L can be derived.  

The dIrr and
early-type galaxies are clearly separated in the {\it lower panel} of
Figure~9.  This is not surprising: The dIrr systems all have masses
constrained by rotation curves at large radii that can only be
understood if the galaxies possess extended DM halos. For the
early-type galaxies, the King formalism assumes that mass follows
light.  Hence, it is likely that the M/L ratios of all the early-type
dwarfs are underestimated.  Figure~9 also shows that the
distribution of M/L ratios for the
early-type dwarfs can be fit with the relation $\log M/L = 2.5 +
10^7/(L/L_\odot)$; these galaxies are consistent with the idea that
each is embedded in a dark halo of fixed mass of about $10^7$
\Msol and contains a luminous component with M/L$_V$ = 2.5.  
This is a lower limit to the halo mass since the dark matter may
plausibly be more extended than the luminous material.

LG dwarfs have proven useful to restrict some possible forms of dark
matter.  For example, massive neutrinos have been ruled out in dSph
galaxies from phase-space arguments (Lin \&\ Faber 1983, Lake 1989a,
Gerhard \&\ Spergel 1992a).  Massive black-hole models (Strobel \&\
Lake 1994) are also incompatible with the generally smooth central
surface brightness distributions of the cores of dSph systems or the
large global M/L ratios inferred in these systems (Demers et al 1995).

\vskip1em

\noindent {\large{\it 7.3 Alternatives to Dark Matter}}

\vskip0.5em

\noindent The current fashion is to assume that the kinematic
observations described above constitute part of a dark matter problem.
However, it may be wise to remember that this already implies a
solution to what remains a long-standing crisis in understanding the
internal kinematics of galaxies.  In this section I discuss two
possible alternatives to DM as they apply to LG dwarfs.

\vskip0.25em

\noindent TIDES\ \ Kuhn \&\ Miller (1989) and Kuhn (1993) proposed
that one way to mimic the kinematic effects of DM was through a
resonance process between the orbital period of a dwarf and its
natural radial oscillation period.  Pryor (1996) and Olszewski (1998)
reviewed recent studies that suggested that this mechanism is unlikely
to have a significant effect on real dSph galaxies and for actual 
observational samples of stars used in kinematic studies.  I add only
two points here.  First, tidal effects are indeed visible in many LG
dwarfs (Section~8), but this alone does not imply that the central
velocity dispersions, and hence the inferred mass-to-light ratios, are
significantly affected until the galaxies are nearly completely
disrupted.  Second, the M/L ratios of isolated early-type galaxies
such as LGS~3 and Leo~II are sufficiently high to require DM, yet they
are sufficiently far from any large galaxies that they cannot be
significantly affected by tides.  Kinematic studies of other isolated
dwarfs such as Antlia and Tucana would be particular helpful in
settling this issue.

\vskip0.25em

\noindent MODIFIED GRAVITY\ \ Milgrom (1983a,b) introduced Modified
Newtonian Dynamics, or MOND, to understand the rotation curves of disk
galaxies (and some other related phenomena) with a modified form of
Newton's law of gravity without resorting to DM.  
Only at very low accelerations (defined by the
parameter $a_0 \sim 2 \times 10^{-8}$ cm~sec$^{-2}$) is Newton's law
substantially altered from its standard form.

A single unambiguous example of a galaxy strictly obeying Newtonian
dynamics in the low-acceleration regime would falsify MOND instantly.
I do not address here how MOND currently fares with regard to
understanding the dynamics of large galaxies apart from noting that no
unambiguous failures have yet been reported (Sanders 1996,
McGaugh \&\ de Block 1998).  Nor will I discuss the far-reaching
implications of MOND on cosmology and other areas of astrophysics
(Bekenstein and Milgrom 1984, Felton 1984, Sivaram 1994, Qiu et al
1995, Sanders 1997).  Instead, I focus here on the use of LG dwarfs to
test MOND.

Lake and Skillman (1989) and Lake (1989b) suggested that the rotation
curves of IC~1613 and NGC~3109 could not be explained by MOND unless
$a_0 \leq 3 \times 10^{-9}$ cm~sec$^{-2}$, a value incompatible with
that needed to interpret rotation curves of giant systems.  Milgrom
(1991) noted that (a) MOND successfully fit the shapes of the
rotations curves of these and other galaxies discussed by Lake
(1989b), and (b) the mixed success that MOND had in reproducing the
amplitudes of the rotation curves could be understood given the errors
in the galaxy distances, inclinations, and asymmetric-drift
corrections.  For NGC~3109 (Jobin and Carignan 1990), Milgrom was in
fact justified in claiming the earlier rotation curve was in error,
though in the case of IC~1613 it remains unclear if Milgrom's
objection to Lake \&\ Skillman's (1989) conclusion is valid.  More
recently, Sanders (1996) found that the rotation curves of both NGC~55
and NGC~3109 are fit well by MOND.  

Gerhard and Spergel (1992b) argued that the internal kinematics of LG
dSph galaxies demanded DM even if MOND was used to estimate their
masses.  Lo et al (1993) found that the MOND masses for many nearby
dIrr galaxies were smaller than their integrated HI masses -- an
obvious failure if correct.  Milgrom (1995) responded that if the
observational errors and most recent results were considered, neither
effect claimed by Gerhard and Spergel (1992b) was observed.  In the
second case, Lo et al (1993) used an incorrect expression to determine
the MOND masses, leading to estimates that were too small by a factor
of 20 (Milgrom 1994).  When the proper relation is used ($M_{MOND} =
81 \sigma_0^4/4 a_0 G$, where $\sigma_0$ is the observed central
velocity dispersion for an isotropic system), the MOND masses are
consistent with the inferred luminous (gaseous + stellar) masses
without invoking a dark component.

The burden of proof remains squarely on MOND, but the kinematic data for
LG dwarfs does not yet refute this alternative to DM.

\vfill\eject

\noindent {\Large{\bf 8. INTERACTIONS IN THE LOCAL GROUP}}
\vskip1em

\noindent The Local Group is a dangerous place for dwarf
galaxies. NGC~205 and Sagittarius have wandered too close to their
dominant parents and exhibit clear kinematic and structural signatures
of tidal distortions (Hodge 1973, Bender et al 1991, Pryor 1996, Ibata
et al 1997).  Irwin \&\ Hatzidimitriou (1995) noted that many nearby
dSph systems show a strong correlation of tidal radius or ellipticity
with the strength of the external tidal field.  Bellazzini et al (1996)
have shown convincingly that the central surface brightness,
$\Sigma_0$, of dSph galaxies obey a bi-variate relation in $\Sigma_0$,
$L_{tot}$, and $R_{GC}$, where $R_{GC}$ is the Galactocentric distance
of the galaxy.  They show that Sagittarius in particular appears to be
unbound, even in its core (Mateo et al 1995c; but see Ibata et al
1997).

A number of models have investigated what dwarfs look like before,
during and after strong tidal encounters (Allen \&\ Richstone 1988,
Moore \&\ Davis 1994, Piatek \&\ Pryor 1995, Oh et al 1995, Johnston
et al 1995, Vel\'azquez \&\ White 1995, Kroupa 1997).  At early times
in a strong interaction or in the weak-interaction limit, stars are
lost from the dwarf into leading and trailing orbits.  These stars
quickly fill a larger volume than that of the original galaxy, and if
they were included in kinematic samples, they would reveal streaming
motions that could be interpreted as rotation.  Extra-tidal stars
might be seen at this stage, even though the majority of the galaxy's
stars remain bound and the central velocity dispersion is unaffected
(eg Gould et al 1992, Kuhn et al 1996; in both cases many of the stars
discussed are actually within recent estimates of the tidal radii of
the respective galaxies).  At later stages of strong interactions, the
dwarfs become strongly elongated, but not necessarily parallel to the
orbital path of their center of mass.  Alcock et al (1997a) claim to
see such a tilt in Sagittarius, but Ibata et al (1997) do not.  At
very late stages of a nearly complete tidal disruption event, the
dwarf becomes a long strand that is stretched along its orbit with a
small clump ($\leq$ 10\%\ of the original mass) as the only remnant of
the original galaxy.  At no time except the very end of the tidal
episode does the central velocity dispersion significantly exceed its
virial value, even for models with no initial dark component.

These disrupted dwarfs should produce relatively long-lived streams in
the halos of galaxies such as M31 and the Milky Way (1-2 Gyr;
P Harding, private communication).  Lynden-Bell \&\ Lynden-Bell
(1995) conclude that one possible stream can be traced out with the
Magellanic Stream (Wakker \&\ van Woerden 1997), Ursa~Minor, Draco,
and possibly Carina and Sculptor.  A recent determination of the
proper motion of Sculptor (Schweitzer et al 1995) suggests that this
galaxy is not part of this putative (or any other proposed) stream.
There have been many intriguing claims of halo substructure in recent
years (eg Majewski 1992, Arnold \&\ Gilmore 1992, C\^ot\'e et al 1993,
Kinman et al 1996) that could possibly be remnants of disrupted
dwarfs.

More recently, Alcock et al (1997b) and Zaritsky \&\ Lin (1997)
claimed to detect a possible signature of a foreground galaxy or
galaxy tidal remnant towards the LMC.  Gallart (1998) suggested
instead that this new `galaxy' is in fact due to the signature of
known, but subtle stellar evolutionary phases that are becoming
apparent in the large-scale photometric surveys being carried out in
the LMC.  This is not the first time that a putative new galaxy has
been detected directly in front of a known LG dwarf: Connolly (1985)
identified a number of `foreground' RR~Lyr stars towards the LMC that
he concluded are members based on their photometric properties, but
non-members kinematically. Saha et al (1986) also identified some
anomalously bright RR~Lyr-like stars apparently in front of the Carina
dSph galaxy that could either be part of an extended halo of the LMC
or possibly associated with a foreground system.  A possibly more
natural explanation may be that these are instead anomalous Cepheids
in Carina itself (Mateo et al 1995a).  In none of these cases is the
true nature of all of these `foreground' stars conclusively
established, and in the case of the LMC it is not unreasonable to
suppose that a tidal tail is present (Zaritsky \&\ Lin 1997).
Nevertheless, it seems wise to treat claims of the existence of
galaxies or tidal features directly in front of known LG systems 
with particular caution.

Mateo (1996) and Unavane et al (1996) have discussed the possibility
that a large fraction of the Galactic halo has been constructed from
disrupted dSph systems.  The latter considered Carina to be the
template of such a system, while Mateo (1996) compared the properties
of the ensemble of {\it all} the Galactic dSph satellites with the
halo.  Neither approach is strictly correct.  Carina has arguably the
most unusual stellar population of any dSph system (Section 6.3;
Figure 8); it is clearly not an appropriate choice as a template for
the halo.  On the other hand, present-day dSph systems are
``survivors'' able to form stars over a longer period than systems
that were destroyed.  They probably also follow orbits (relative to
the Galaxy) that are quite distinct from the orbits of the galaxies
that were consumed; thus, even an average of the stellar populations
of all remaining dSph galaxies should not be expected to precisely
match the current halo population.  Given these differences in
approach, the two studies nevertheless essentially agree: No more than
10\%\ of the halo could have Carina-like progenitors (Preston et al
1994), but more than 50\%\ of the halo could have formed from galaxies
similar to the entire ensemble of Galactic dSph systems (though see
van den Bergh 1994b).

\vfill\eject

\noindent {\Large{\bf 9. CONCLUSIONS}}
\vskip1em

\noindent I hope that I have been able to convey the rich detail, the
surprises, and the broad relevance of modern research on LG dwarfs.
Nevertheless, this review has only scratched the surface of this
active field.  Table 8 provides a census of various specific types of
objects (variable stars, young, intermediate-age, and old ISM and
stellar population tracers) that can and have been used to study these
systems in greater detail.  Keep in mind that these dwarfs offer our
best opportunities to study how stellar evolution proceeds in
chemically young environments; they offer our best window into the
nature of dark matter in what may be the smallest natural size-scales
of this material; and they may help us understand if and how dwarfs
help form larger galaxies via mergers.  Many targets useful to attack
these problems can be found in Table~8.  We still have plenty of work
to do with these nearby dwarfs to address these issues as well as
the many others brought up in this review.

But perhaps most exciting of all is that as we stand on the threshold
of detailed studies of dwarf galaxies in other groups (e.g. C\^ot\'e
1995, C\^ot\'e et al 1997, Caldwell et al 1998), the LG dwarfs will
serve as a benchmark against which other systems can be compared.  We
{\it know} that complex star-formation histories are common in the
Local Group.  We {\it know} that both dIrr and early-type dwarfs
contain interstellar material but which manifests itself in many
different ways.  We {\it know} that LG dwarfs are kinematically
peculiar, whether dominated by DM or not.  Do dwarfs in other groups
behave the same way?  How do they differ?  How many of these
tendencies are truly universal, and how many due to the specific
environmental circumstances of individual groups?  These questions will
lead to insights that we cannot gain from studies of only the galaxies
in the Local Group. It will be an exciting adventure and one that is
sure to uncover as many surprises as the dwarfs of the Local Group
already have.  

\vfill\eject

\vskip2em

\noindent ACKNOWLEDGEMENTS

\noindent I dedicate this paper to the memory of my father, Luis
Ernesto Mateo, who taught me -- and gave me the opportunity -- to find
my own way.  I would like to thank the people who have helped
materially with this paper: K Chiboucas, D Hurley-Keller and K von
Braun for carefully checking the references in most of the tables,
Tina Cole for typing in hundreds of references and somehow managing
to remain her cheerful self, and to A Aparicio, J Gallagher, P Hodge
and A Sandage for their comments on the original manuscript.  It is
also a pleasure to thank colleagues who have provided many stimulating
discussions over the years which (through no fault of their own!)
have led to some of the ideas in this paper: J Bregman, N Caldwell, C
Carignan, C Chiosi, G Da~Costa, S Demers, R Dohm-Palmer, K Freeman, C
Gallart, E Grebel, P Harding, D Hunter, D Hurley-Keller, KY Lo, P
Massey, H Morrison, E Olszewski, C Pryor, D Richstone, A Saha, E
Skillman, and E Tolstoy.  I am grateful to C Carignan for providing
Figure~6 to me prior to publication.  I want to particularly thank
Paul Hodge and George Preston, both for the numerous pleasant
discussions I have had with them about some of the topics in this
review, and for sharing their enthusiasm to do astronomy.  Finally, my
heartfelt thanks to Nancy, Emilio and Carmen for putting up with me
while writing this review, and for their encouragement through it all.
Some of my research described in this paper has been supported, in
part, by grants from NASA and NSF.

\vfill\eject

\noindent{\underline{LITERATURE CITED}}
\vskip1em

\def\apj{\underline {Ap.~J.}\ }
\def\apjl{\underline {Ap.~J.~Lett.}\ }
\def\apjs{\underline {Ap.~J.~Suppl.}\ }

\noindent\hang Aaronson M. 1983. ApJ 266:L11-L15

\noindent\hang Aaronson M. 1986. In {\it Star Forming Dwarf Galaxies and Related Objects}, ed. D. Kunth, TX Thuan, JTT Van, p. 125-144. Paris: Editions Fronti\`eres

\noindent\hang Aaronson M, Olszewski EW, Hodge PW. 1983. ApJ 267:271-279

\noindent\hang Ables HD. 1971. USNO 20 pt 4

\noindent\hang Ables HD, Ables PG. 1977. ApJS 34:245-258

\noindent\hang Alard C. 1996. ApJ 458:L17-L20

\noindent\hang Alcock C, Allsman RA, Alves DR, Axelrod TS, Becker AC, et al. 1997a. ApJ 474:217-222

\noindent\hang Alcock C, Allsman RA, Alves DR, Axelrod TS, Becker AC, et al. 1997b. ApJ 490:L59-L63

\noindent\hang Allen AJ, Richstone DO. 1988. ApJ 325:583-595

\noindent\hang Allsopp NJ. 1978. MN 184:397-404

\noindent\hang Anders E, Grevesse N. 1989. Geochimica et Cosmochimica Acta  53:197-214

\noindent\hang Aparicio A. 1994. ApJ 437:L27-L30

\noindent\hang Aparicio A, Dalcanton JJ, Gallart C, Mart\'\i nez-Delgado D. 1997a. AJ 114:1447-1457

\noindent\hang Aparicio A, Gallart C. 1995. AJ 110:2105-2119

\noindent\hang Aparicio A, Gallart C, Bertelli G. 1997b AJ 114:669-679

\noindent\hang Aparicio A, Gallart C, Bertelli G. 1997c AJ 114:680-693

\noindent\hang Aparicio A, Garc\'\i a-Pelayo JM, Moles M. 1988. A\&AS 74:375-384

\noindent\hang Aparicio A, Herrero A, S\'anchez F, eds. 1998. {\it Stellar Astrophysics for the Local Group}, Cambridge: Cambridge U Press

\noindent\hang Aparicio A, Rodr\'\i guez-Ulloa JA. 1992. A\&A 260:77-81

\noindent\hang Armandroff TE, Da Costa GS. 1986. AJ 92:777-786

\noindent\hang Armandroff TE, Da Costa GS, Caldwell N, Seitzer P. 1993. AJ 106:986-998

\noindent\hang Armandroff TE, Massey P. 1991. AJ 102:927-950

\noindent\hang Armandroff TE, Olszewski EW, Pryor C. 1995. AJ 110:2131-2165

\noindent\hang Arnold R, Gilmore G. 1992 MNRAS 257:225-239

\noindent\hang Ashman K. 1992. PASP 104:1109-1138

\noindent\hang Azzopardi M. 1994. See Layden et al 1994, p. 129-140

\noindent\hang Azzopardi M, Lequeux J, Westerlund BE. 1985. A\&A 144:388-394

\noindent\hang Azzopardi M, Lequeux J, Westerlund BE. 1986. A\&A 161:232-236

\noindent\hang Baade W, Swope H. 1961. AJ 66:300-347

\noindent\hang Babul A, Rees MJ. 1992. MNRAS 255:346-350

\noindent\hang Battistini P, B\`onoli F, Barccesi A, Federici L, Fusi Pecci F, Marano B, B\"orngen F. 1987. A\&AS 67:447-482

\noindent\hang Beauchamp D, Hardy E, Suntzeff NB, Zinn R. 1995. AJ 109:1628-1644

\noindent\hang Becker R, Henkel C, Wilson TL, Wouterloot JGA. 1993. A\&A 268:483-490

\noindent\hang Bekenstein J, Milgrom M. 1984. ApJ 286:7-14

\noindent\hang Bellazzini M, Fusi Pecci F, Ferraro FR. 1996. MNRAS 278:947-952

\noindent\hang Bender R, Neito JL. 1990. A\&A 239:97-112

\noindent\hang Bender R, Paquet A, Nieto JL. 1991. A\&A 246:349-353

\noindent\hang Bendinelli O, Parmeggiani G, Zavatti F, Djorgovski S. 1992. AJ 103:110-116

\noindent\hang Bertelli G, Bressan A, Chiosi C, Fagotto F, Nasi E. 1994. A\&AS 106:275-302

\noindent\hang Bertelli G, Mateo M, Chiosi C, Bressan A. 1992. ApJ 388:400-414

\noindent\hang Bica E, Alloin D, Schmidt AA. 1990. A\&A 228:23-36

\noindent\hang Binggeli B. 1994. In {\it Panchromatic View of Galaxies}, ed. G Hensler, C Theis, J Gallagher, p. 173-191. Paris: Editions Frontieres

\noindent\hang Bothun GD, Mould JR, Caldwell N, MacGillivray HT. 1986. AJ 92:1007-1019

\noindent\hang Bothun GD, Thompson IB. 1988. AJ 96:877-883

\noindent\hang Bowen DV, Tolstoy E, Ferrara A. Blades JC, Brinks E. 1997. ApJ 478:530-535

\noindent\hang Brandt WN, Ward MJ, Fabian AC, Hodge PW. 1997. MNRAS 291:709-716

\noindent\hang Bresolin F, Capaccioli M, Piotto G. 1993. AJ 105:1779-1792

\noindent\hang Brinks E, Taylor CL. 1994. See Meylan \&\ Prugniel 1994, p. 263-272

\noindent\hang Buonanno R, Corsi CE, Fusi Pecci F, Hardy E, Zinn R. 1985. A\&A 152:65-84

\noindent\hang Burstein D, Davies RL, Dressler A, Faber SM, Stone RPS, Lynden-Bell D, Terlevich RJ, Wegner G. 1987. ApJS 64:601-642

\noindent\hang Burstein D, Heiles C. 1982. AJ 87:1165-1189

\noindent\hang Buta R, Williams KL. 1995. AJ 109:543-557

\noindent\hang Byrd G, Valtonen M, McCall M, Innanen K. 1994. AJ 107:2055-2059

\noindent\hang Caldwell N, Armandroff TE, Da Costa GS, Seitzer P. 1998. AJ 115:535-558

\noindent\hang Caldwell N, Armandroff TE, Seitzer P, Da Costa GS. 1992. AJ 103:840-850

\noindent\hang Caldwell N, Schommer RA, Graham JA. 1988. BAAS 20:1084

\noindent\hang Capaccioli C, Piotto G, Bresolin F. 1992. AJ 103:1151-1158

\noindent\hang Caputo F, Castellani V, Degl'Innocenti S. 1995. A\&A 304:365-368

\noindent\hang Carignan C. 1985. ApJ 299:59-73

\noindent\hang Carignan C, Beaulieu S, C\^ot\'e S, Demers S, Mateo M 1998. AJ in press.

\noindent\hang Carignan C, Beaulieu S, Freeman KC. 1990. AJ 99:178-190

\noindent\hang Carignan C, Demers S, C\^ot\'e S. 1991. ApJ 381:L13-L16

\noindent\hang Carlson G, Sandage A. 1990. ApJ 352:587-594

\noindent\hang Carney BW, Seitzer P. 1986. AJ 92:23-42

\noindent\hang Carter D, Jenkins CR. 1993. MNRAS 263:1049-1074

\noindent\hang Carter D, Sadler EM. 1990. MNRAS 245:P12-P16

\noindent\hang Castellani M, Marconi G, Buonanno R. 1996. A\&A 310:715-721

\noindent\hang Cesarsky DA, Lausten S, Lequeux J, Schuster HE, West RM. 1977. A\&A 61:L31-L33

\noindent\hang Charbonnel C, Meynet G, Maeder A, Schaerer D. 1996, A\&AS 115:339-344

\noindent\hang Chiosi C, Bertelli G, Bressan A. 1992. ARAA 30:235-285

\noindent\hang Ciardullo R, Jacoby GH, Ford HC, Neill JD. 1989. ApJ 339:53-69

\noindent\hang Collier J, Hodge P. 1994. ApJS 92:119-123

\noindent\hang Cook KH. 1987. {\it Asymptotic Giant Branch Populations in Composite Stellar Systems}, PhD thesis, University of Arizona

\noindent\hang Cook KH, Aaronson M, Norris J. 1986. ApJ 305:634-644

\noindent\hang Connolly LP. 1985. ApJ 299:728-740

\noindent\hang Corbelli E, Schneider SE, Salpeter EE. 1989. AJ 97:390-404

\noindent\hang C\^ot\'e P, Welch DL, Fischer P, Irwin MJ. 1993. ApJ 406:L59-L62

\noindent\hang C\^ot\'e S. 1995. {\it Dwarf Galaxies in Nearby Southern Groups}, PhD thesis, Australian National University

\noindent\hang C\^ot\'e S, Freeman KC, Carignan C, Quinn PJ. 1997. AJ 114:1313-1329

\noindent\hang Da Costa GS. 1984. ApJ 285:483-494

\noindent\hang Da Costa GS. 1992. In {\it The Stellar Populations of Galaxies}, ed. B Barbuy, A Renzini, p. 191-200. Dordrecht: Kluwer

\noindent\hang Da Costa GS. 1994a. See Meylan \&\ Prugniel 1994, p. 221-230

\noindent\hang Da Costa GS. 1994b. See Layden et al 1994, p. 101-114

\noindent\hang Da Costa GS. 1998. See Aparicio et al 1998, p. 351-406

\noindent\hang Da Costa GS, Armandroff TE. 1990. AJ 100:162-181

\noindent\hang Da Costa GS, Armandroff TE. 1995. AJ 109:2533-2552

\noindent\hang Da Costa GS, Armandroff TE, Caldwell N, Seitzer P. 1996. AJ 112:2576-2595

\noindent\hang Da Costa GS, Graham JA. 1982. ApJ 261:70-76

\noindent\hang Da Costa GS, Hatzidimitriou D, Irwin MJ, McMahon RG. 1991. MNRAS 249:473-480

\noindent\hang Danzinger IJ, Webster BL, Dopita MA, Hawarden TG. 1978. ApJ 220:458-466

\noindent\hang Davidge TJ. 1991. AJ 101:884-891

\noindent\hang Davidge TJ. 1993. AJ 105:1392-1399

\noindent\hang Davidge TJ. 1994. AJ 108:2123-2127

\noindent\hang Davidge TJ, Jones JH. 1992. AJ 104:1365-1373

\noindent\hang Davidge TJ, Nieto JL. 1992. ApJ 391:L13-L16

\noindent\hang Dekel A, Silk J. 1986 ApJ 303:39-55

\noindent\hang Demers S, Battinelli P, Irwin MJ, Kunkel WE. 1995. MNRAS 274:491-498

\noindent\hang Demers S, Beland S, Kunkel WE. 1983. PASP 95:354-357

\noindent\hang Demers S, Irwin MJ. 1987. MNRAS 226:943-961

\noindent\hang Demers S, Irwin MJ. 1993. MNRAS 261:657-673

\noindent\hang Demers S, Irwin MJ, Gambu I. 1994a. MNRAS 266:7-15

\noindent\hang Demers S, Irwin MJ, Kunkel WE. 1994b. AJ 108:1648-1657

\noindent\hang Demers S, Kunkel WE, Irwin MJ. 1985. AJ 90:1967-1981

\noindent\hang Demers S, Mateo M, Kunkel WE. 1998. MNRAS in press

\noindent\hang Dettmar RJ, Heithausen A. 1989. ApJ 344:L61-L64

\noindent\hang de Vaucouleurs G, Ables H. 1965. PASP 77:272-282

\noindent\hang de Vaucouleurs G, Ables HD. 1968. ApJ 151:105-116

\noindent\hang de Vaucouleurs G, Ables HD. 1970. ApJ 159:425-433

\noindent\hang de Vaucouleurs G, de Vaucouleurs A, Buta R. 1981. AJ 86:1429-1463

\noindent\hang de Vaucouleurs G, de Vaucouleurs A, Corwin HG, Buta RJ, Paturel G, Fouqu\'e P. 1991. {\it Third Reference Catalogue of Bright Galaxies}, New York: Springer

\noindent\hang de Vaucouleurs G, Freeman KC. 1972. Vistas in Ast 14:163-294

\noindent\hang de Vaucouleurs G, Moss C. 1983. ApJ 271:123-132

\noindent\hang De Young DS, Heckman TM. 1994. ApJ 431:598-603

\noindent\hang Dohm-Palmer RC, Skillman ED, Saha A, Tolstoy E, Mateo M, et al. 1997. AJ 114:2527-2544 

\noindent\hang Dohm-Palmer RC, Skillman ED, Saha A, Tolstoy E, Mateo M, et al. 1998. AJ in press

\noindent\hang Dressler A, Richstone DO. 1988. ApJ 324:701-713

\noindent\hang Drissen L, Roy JR, Moffat AFJ. 1993. AJ 106:1460-1470

\noindent\hang Dufour RJ, Talent DL. 1980. ApJ 235:22-29

\noindent\hang Dunn AM, Laflamme R. 1993. MNRAS 264:865-874

\noindent\hang Ellis RS. 1997. ARAA 35:389-443

\noindent\hang Elston R, Silva DR. 1992. AJ 104:1360-1364

\noindent\hang Epchtein N, de Butz B, Capoani L, Chavallier L, Copet E, et al. 1997. ESO Messenger 87:27-34

\noindent\hang Eskridge PB. 1988a. AJ 95:1706-1716

\noindent\hang Eskridge PB. 1988b. AJ 96:1336-1351

\noindent\hang Eskridge PB. 1988c. AJ 96:1352-1361

\noindent\hang Eskridge PB. 1995. PASP 107:561-565

\noindent\hang Eskridge PB, White RE. 1997. BAAS 190:02.01

\noindent\hang Fabbiano G. 1989. ARAA 27:87-138

\noindent\hang Faber SM, Lin DNC. 1983. ApJ 266:L17-L20

\noindent\hang Fahlman GG, Mandushev G, Richer HB, Thompson IB, Sivaramakrishnan A. 1996. ApJ 459:L65-L68


\noindent\hang Felton JE. 1984. ApJ 286:3-6

\noindent\hang Ferguson HC, Binggeli B. 1994. Astronomy \&\ Astrophysics Reviews 6:67-122

\noindent\hang Ferguson HC, Sandage A. 1991. AJ 101:765-782

\noindent\hang Ferguson AMN, Wyse RFG, Gallagher JS. 1996. AJ 112:2567-2575

\noindent\hang Ferraro FR, Fusi Pecci F, Tosi M, Buonanno R. 1989. MNRAS 241:433-452

\noindent\hang Ferraro FR, Fusi Pecci F, Testa V, Greggio L, Corsi CE, Buonanno R, Terndrup DM, Zinnecker H. 1995. MNRAS 272:391-422

\noindent\hang Fich M, Hodge P. 1991. ApJ 374:L17-L20

\noindent\hang Fich M, Tremaine S. 1991. ARAA 29:409-445

\noindent\hang Fisher JR, Tully RB. 1975. A\&A 44:151-171

\noindent\hang Fisher JR, Tully RB. 1979. AJ 84:62-70

\noindent\hang Fitzgibbons GL. 1990. {\it Visual, Red and Infrared Photographic Surface Photometry of NGC~55 and NGC~253}, PhD thesis, University of Florida

\noindent\hang Ford HC, Ciardullo R, Jacoby GH, Hui X. 1989. in {\it Planetary Nebulae}, ed. S. Torres-Piembert, p. 335-350. Dordrecht: Reidel

\noindent\hang Ford HC, Jacoby G, Jenner DC. 1977. ApJ 213:18-26

\noindent\hang Ford HC, Jacoby G, Jenner DC. 1978. ApJ 223:94-97

\noindent\hang Ford HC, Jenner DC. 1976. ApJ 208:683-687

\noindent\hang Fouqu\'e R, Bottinelli L, Durand N, Gouguenheim L, Paturel G. 1990. A\&AS 86:473-502

\noindent\hang Freedman WL. 1988a. AJ 96:1248-1306

\noindent\hang Freedman WL. 1988b. ApJ 326:691-709

\noindent\hang Freedman WL. 1992. AJ 104:1349-1359

\noindent\hang Gallagher JS, Hunter DA. 1981. AJ 86:1312-1322

\noindent\hang Gallagher JS, Hunter DA, Tutukov AV. 1984. ApJ 284:544-556

\noindent\hang Gallagher JS, Hunter DA, Gillett FC, Rice WL. 1991. ApJ 371:142-147

\noindent\hang Gallagher JS, Hunter DA, Mould J. 1984. ApJ 281:L63-L65

\noindent\hang Gallagher JS, Wyse RFG. 1994. PASP 106:1225-1238

\noindent\hang Gallart C. 1998. ApJ 495:L43-L46

\noindent\hang Gallart C, Freedman WL, Mateo M, Chiosi C, Thompson I. et al. 1998. AJ in press

\noindent\hang Gallart C, Aparicio A, Bertelli G, Chiosi C. 1996a. AJ 112:1950-1968

\noindent\hang Gallart C, Aparicio A, Bertelli G, Chiosi C. 1996b. AJ 112:2596-2606

\noindent\hang Gallart C, Aparicio A, V\'\i lchez JM. 1996c. AJ 112:1928-1949

\noindent\hang Garnett DR. 1989. ApJ 345:282-297

\noindent\hang Garnett DR. 1990. ApJ 363:142-153

\noindent\hang Garnett DR, Kennicutt RC, Chu YH, Skillman ED. 1991. ApJ 373:458-464

\noindent\hang Gerhard OE. 1994. See Meylan \&\ Prugniel 1994, p. 335-350

\noindent\hang Gerhard OE, Spergel DN. 1992a. ApJ 389:L9-L11

\noindent\hang Gerhard OE, Spergel DN. 1992b. ApJ 397:38-43

\noindent\hang Gizis JE, Mould JR, Djorgovski S. 1993. PASP 105:871-874

\noindent\hang Godwin PJ, Lynden-Bell D. 1987. MNRAS 229:P7-P13

\noindent\hang Gonz\'alez J. 1993. {\it Line Strength Gradients and Kinematic Profiles in Elliptical Galaxies}, PhD thesis, University of California, Santa Cruz

\noindent\hang Gottesman ST, Weliachew L. 1977. A\&A 61:523-530

\noindent\hang Gould A, Guhathakurta P, Richstone D, Flynn C. 1992. ApJ 388:345-353

\noindent\hang Grebel EK. 1997. Review of Modern Astronomy 10:29-60

\noindent\hang Gregg M, Minniti D. 1997. PASP 109:1062-1067

\noindent\hang Greggio L, Marconi G, Tosi M, Focardi P. 1993. AJ 105:894-932

\noindent\hang Grillmair CJ, Lauer TR, Worthey G, Faber SM, Freedman WL, et al. 1996. AJ 112:1975-1987

\noindent\hang Grillmair CJ, Mould JR, Holtzman JA, Worthey G, Ballester GE, et al. 1998. AJ 115:144-151

\noindent\hang Gurzadyan VG, Kocharyan AA, Petrosian AR. 1993. A\&SS 201:243-248

\noindent\hang Han M, Hoessel JG, Gallagher JS, Holtzman J, Stetson PB, et al. 1997. AJ 113:1001-1010

\noindent\hang Hargreaves JC, Gilmore G, Annan JD. 1996a. MNRAS 279:108-120

\noindent\hang Hargreaves JC, Gilmore G, Irwin MJ, Carter D. 1994a. MNRAS 269:957-974

\noindent\hang Hargreaves JC, Gilmore G, Irwin MJ, Carter D. 1994b. MNRAS 271:693-705

\noindent\hang Hargreaves JC, Gilmore G, Irwin MJ, Carter D. 1996b. MNRAS 282:305-312

\noindent\hang Harris HC, Silbermann NA, Smith HA. 1997. In {\it A Half Century of Stellar Pulsation Interpretations: A Tribute to AN Cox}, ed. PA Bradley \&\ JA Guzik, p. 164-xxx, San Francisco: ASP, Vol 135

\noindent\hang Hatzidimitriou D. 1991. MNRAS 251:545-554

\noindent\hang Heckman TM, Dahlem M, Lehnert MD, Fabbiano G, Gilmore D, Waller WH. 1995. ApJ 448:98-118

\noindent\hang Heisler CA, Hill TL, McCall ML, Hunstead RW. 1997. MNRAS 285:374-386

\noindent\hang Held EV, de Zeeuw T, Mould J, Picard A. 1992. AJ 103:851-856

\noindent\hang Held EV, Mould JR, deZeeuw PT. 1990. AJ 100:415-419

\noindent\hang Henning PA. 1997. Publications of the Astronomical Society of Australia 14:21-24

\noindent\hang Hewitt JN, Haynes MP, Giovanelli R. 1983. AJ 88:272-295

\noindent\hang Ho LC, Fillipenko AV, Sargent WLW. 1995. ApJS 98:477-593

\noindent\hang Ho LC, Fillipenko A, Sargent W. 1997. ApJS 112:315-390

\noindent\hang Hodge P. 1994. See Layden et al 1994, p. 57-64

\noindent\hang Hodge P, Kennicutt RC, Strobel N. 1994. PASP 106:765-769

\noindent\hang Hodge P, Lee MG, Gurwell M. 1990. PASP 102:1245-1262

\noindent\hang Hodge P, Lee MG, Kennicutt RC. 1988. PASP 100:917-934

\noindent\hang Hodge P, Lee MG, Kennicutt RC. 1989. PASP 101:640-648

\noindent\hang Hodge P, Miller BW. 1995. ApJ 451:176-187

\noindent\hang Hodge PW. 1963a. AJ 68:470-474

\noindent\hang Hodge PW. 1963b. AJ 68:691-696

\noindent\hang Hodge PW. 1964. AJ 69:438-442

\noindent\hang Hodge PW. 1966. AJ 71:204-205

\noindent\hang Hodge PW. 1969. ApJS 18:73-84

\noindent\hang Hodge PW. 1971. ARAA 9:35-66

\noindent\hang Hodge PW. 1973. ApJ 182:671-695

\noindent\hang Hodge PW. 1974. PASP 86:289-293

\noindent\hang Hodge PW. 1976. AJ 81:25-29

\noindent\hang Hodge PW. 1977. ApJS 33:69-82

\noindent\hang Hodge PW. 1978. ApJS 37:145-167

\noindent\hang Hodge PW. 1980. ApJ 241:125-131

\noindent\hang Hodge PW. 1981. {\it Atlas of the Andromeda Galaxy}, Seattle: UW Press

\noindent\hang Hodge PW. 1982. AJ 87:1668-1670

\noindent\hang Hodge PW. 1989. ARAA 27:139-159

\noindent\hang Hodge PW, Lee MG. 1990. PASP 102:26-40

\noindent\hang Hodge PW, Smith DW. 1974. ApJ 188:19-25

\noindent\hang Hodge PW, Smith TR, Eskridge PB, MacGillivray HT, Beard SM. 1991a. ApJ 369:372-381

\noindent\hang Hodge PW, Smith T, Eskridge P, MacGillivray H, Beard S. 1991b. ApJ 379:621-630

\noindent\hang Hodge PW, Wright FW. 1978. AJ 83:228-233

\noindent\hang Hoessel JG, Abbott MJ, Saha A, Mossman AE, Danielson GE. 1990. AJ 100:1151-1158

\noindent\hang Hoessel JG, Saha A, Danielson GE. 1988. PASP 100:680-682

\noindent\hang Hoessel JG, Saha A, Krist J, Danielson GE. 1994. AJ 108:645-652

\noindent\hang Hoffman GL, Salpeter EE, Farhat B, Roos T, Williams H, Helou G. 1996. ApJS 105:269-298


\noindent\hang Holtzman JA, Mould JR, Gallagher JS, Watson AM, Grillmair CJ, et al. 1997. AJ 113:656-668

\noindent\hang Hoopes CG, Walterbos RAM, Greenawalt BE. 1996. AJ 112:1429-1437

\noindent\hang Hopp U, Schulte-Ladbeck RE. 1995. A\&AS 111:527-563

\noindent\hang Huchtmeier WK, Richter OG. 1986. A\&AS 63:323-343

\noindent\hang Huchtmeier WK, Richter OG. 1988. A\&A 203:237-249

\noindent\hang Humason ML, Mayall NU, Sandage AR. 1956. AJ 61:97-162

\noindent\hang Hummel E, Dettmar RJ, Wielebinski R. 1986. A\&A 166:97-106

\noindent\hang Hunter DA, Gallagher JS. 1985. ApJS 58:533-560

\noindent\hang Hunter DA, Gallagher JS. 1990. ApJ 362:480-490

\noindent\hang Hunter DA, Hawley WN, Gallagher JS. 1993. AJ 106:1797-1811

\noindent\hang Hunter DA, Plummer J. 1996. ApJ 462:732-739

\noindent\hang Hurley-Keller D, Mateo M, Nemec J. 1998. AJ May issue

\noindent\hang Huterer D, Sasselov DD, Schechter PL. 1995. AJ 110:2705-2714

\noindent\hang Ibata RA, Gilmore G, Irwin MJ. 1994. Nature 370:194-196

\noindent\hang Ibata RA, Wyse RFG, Gilmore G, Irwin MJ, Suntzeff NB. 1997. AJ 113:634-655

\noindent\hang Irwin M. 1994. See Meylan \&\ Prugniel 1994, p. 27-36

\noindent\hang Irwin M, Hatzidimitriou D. 1995. MNRAS 277:1354-1378

\noindent\hang Irwin MJ, Bunclark PS, Bridgeland MT, McMahon RG. 1990. MNRAS 244:P16-P19

\noindent\hang Israel FP. 1997. A\&A 317:65-72

\noindent\hang Israel FP, Tacconi LJ, Baas F. 1995. A\&A 295:599-604

\noindent\hang Jacoby GH, Lesser MP. 1981. AJ 86:185-192

\noindent\hang James P. 1991. MNRAS 250:544-554

\noindent\hang Jobin M, Carignan C. 1990. AJ 100:648-662

\noindent\hang Johnson DW, Gottesman ST. 1983. ApJ 275:549-558

\noindent\hang Johnston KV, Spergel DN, Hernquist L. 1995. ApJ 451:598-606

\noindent\hang Jones BF, Klemola AR, Lin DNC. 1994. AJ 107:1333-1337

\noindent\hang Jones DH, Mould JR, Watson AM, Grillmair C, Gallagher JS, et al. 1996. ApJ 466:742-749

\noindent\hang Kaluzny J, Kubiak M, Szymanski M, Udalski A, Krzeminski W, Mateo M. 1995. A\&AS 112:407-428

\noindent\hang Karachentsev I. 1996. A\&A 305:33-41

\noindent\hang Karachentsev ID, Makarov DA. 1996. AJ 111:794-803

\noindent\hang Karachentsev ID, Makarov DA. 1998. A\&A 331:891-893

\noindent\hang Karachentsev ID, Tikhonov NA. 1993. A\&AS 100:227-235

\noindent\hang Karachentsev ID, Tikhonov NA, Sazonova LN. 1994. Astrophysical
Letters 20:84-88

\noindent\hang Karachentseva VE, Karachentsev ID. 1998. A\&AS 127:409-419

\noindent\hang Karachentseva VE, Karachentsev ID, B\"orngen F. 1985. A\&AS 60:213-227

\noindent\hang Karachentseva VE, Prugniel P, Vennik J, Richter GM, Thuan TX, Martin JM. 1996. A\&AS 117:343-368

\noindent\hang Kayser SE. 1967. AJ 72:134-148

\noindent\hang Kennicutt RC. 1983. ApJ 272:54-67

\noindent\hang Kennicutt RC. 1994. See Layden et al 1994, p. 28-40

\noindent\hang Kent SM. 1987. AJ 94:306-314

\noindent\hang Killen RM, Dufour RJ. 1982. PASP 94:444-452

\noindent\hang Kinman TD, Pier JR, Suntzeff NB, Harmer DL, Valdes F, et al. 1996. AJ 111:1164-1168

\noindent\hang Kirkpatrick JD, Biechman CA, Skrutskie MF. 1997. ApJ 476:311-318

\noindent\hang Klein U, Gr\"ave R. 1986. A\&A 161:155-168

\noindent\hang Knapp GR, Kerr FJ, Bowers PF. 1978. AJ 83:360-362

\noindent\hang Knapp GR, Turner EL, Cunniffe PE. 1985. AJ 90:454-468

\noindent\hang Kochanek CS. 1996. ApJ 457:228-243

\noindent\hang Kodaira K, Okamura S, Ichikawa S. 1990. {\it Photometric Atlas of Northern Bright Galaxies}, Tokyo: University of Tokyo Press

\noindent\hang K\"onig CHB, Nemec JM, Mould JR, Fahlman GG. 1993. AJ 106:1819-1825

\noindent\hang Koribalski B, Johnston S, Otrupcek R. 1994. MNRAS 270:L43-L45

\noindent\hang Kormendy J, Richstone D. 1995. ARAA 33:581-624

\noindent\hang Kroupa P. 1997. New Ast 2:139-164

\noindent\hang Kuhn JR. 1993. ApJ 409:L13-L16

\noindent\hang Kuhn JR, Miller RH. 1989. ApJ 341:L41-L45

\noindent\hang Kuhn JR, Smith HA, Hawley SL. 1996. ApJ 469:L93-L96

\noindent\hang Lake G. 1989a. AJ 98:1253-1259

\noindent\hang Lake G. 1989b. ApJ 345:L17-L19

\noindent\hang Lake G, Skillman ED. 1989. AJ 98:1274-1284

\noindent\hang Lausten S, Richter W, van der Lans J, West RM, Wilson RN. 1977. A\&A 54:639-640

\noindent\hang Lavery RJ, Mighell KJ. 1992. AJ 103:81-83

\noindent\hang Lavery RH, Seitzer P, Walker AR, Suntzeff NB, Da Costa GS. 1996. BAAS 188:09.03

\noindent\hang Layden A, Smith RC, Storm J, eds. 1994. {\it The Local Group}, Garching: European Southern Observatory

\noindent\hang Lee MG. 1993. ApJ 408:409-415

\noindent\hang Lee MG, 1995a. AJ 110:1129-1140

\noindent\hang Lee MG. 1995b. AJ 110:1155-1163

\noindent\hang Lee MG. 1996. AJ 112:1438-1449

\noindent\hang Lee MG, Freedman W, Mateo M, Thompson I, Roth M, Ruiz MT. 1993c. AJ 106:1420-1435

\noindent\hang Lee MG, Freedman WL, Madore BF. 1993a. AJ 106:964-985

\noindent\hang Lee MG, Freedman WL, Madore BF. 1993b. ApJ 417:553-559

\noindent\hang Lehnert MD, Bell RA, Hesser JE, Oke JB. 1992. ApJ 395:466-474

\noindent\hang Lequeux J, Meyssonnier N, Azzopardi M. 1987. A\&AS 67:169-179

\noindent\hang Lequeux J, Rayo JF, Serrano A, Piembert M, Torres-Piembert S. 1979. A\&A 80:155-166

\noindent\hang Light RM, Armandroff TE, Zinn R. 1986. AJ 92:43-47

\noindent\hang Liller DNC, Faber SM. 1983. ApJ 266:L21-L25

\noindent\hang Lin DNC, Jones BF, Klemola AR. 1995. ApJ 439:652-671

\noindent\hang Lo KY, Sargent WLW, Young K. 1993. AJ 106:507-529

\noindent\hang Longmore AJ, Hawarden TG, Goss WM, Mebold U, Webster BL. 1982. MNRAS 200:325-346

\noindent\hang Longmore AJ, Hawarden TG, Webster BL, Goss WM, Mebold U. 1978. MNRAS 183:P97-P100

\noindent\hang Longo G, de Vaucouleurs A. 1983. University of Texas Monograph 3

\noindent\hang Longo G, de Vaucouleurs A. 1985. University of Texas Monograph 3A

\noindent\hang Lu NY, Hoffman GL, Groff T, Roos T, Lamphier C. 1993. ApJS 88:383-413

\noindent\hang Lugger PM, Cohn HN, Cederbloom SE, Lauer TR, McClure RD. 1992. AJ 104:83-91

\noindent\hang Luppino GA, Tonry JL. 1993. ApJ 410:81-86


\noindent\hang Lynden-Bell D, Lynden-Bell RM. 1995. MNRAS 275:429-442

\noindent\hang Lyo AR, Lee MG. 1997. Journal of the Korean Astronomical Society 30:27-70

\noindent\hang Madore BF, Freedman WL. 1991. PASP 103:933-957

\noindent\hang Madden SC, Poglitsch A, Geis N, Stacey GJ, Townes CH. 1997. 483:200-209

\noindent\hang Maeder A, Meynet G. 1988. A\&AS 76:411-425

\noindent\hang Majewski SR. 1992. ApJS 78:87-152

\noindent\hang Maran SP, Gull TR, Stecher TP, Aller LH, Keyes CD. 1984. ApJ 280:615-617

\noindent\hang Marconi G, Buonanno R, Castellani M, Iannicola G, Molaro P, Pasquini L, Pulone L. 1998. A\&A 330:453-463

\noindent\hang Marconi G, Focardi P, Greggio L, Tosi M. 1990. ApJ 360:L39-L41

\noindent\hang Marconi G, Tosi M, Greggio L, Focardi P. 1995. AJ 109:173-199

\noindent\hang Markert T, Donahue M. 1985. ApJ 297:564-571

\noindent\hang Mart\'\i nez-Delgado D, Aparicio A. 1997. ApJ 480:L107-L110

\noindent\hang Marzke RO, Da Costa LN. 1997. AJ 113:185-196

\noindent\hang Massey P. 1998. See Aparicio et al 1998, p. 95-147

\noindent\hang Massey P, Armandroff TE. 1995. AJ 109:2470-2479

\noindent\hang Massey P, Armandroff TE, Conti PS. 1992. AJ 103:1159-1165

\noindent\hang Mateo M. 1994. See Meylan \&\ Prugniel 1994, p. 309-322

\noindent\hang Mateo M. 1996. See Morrison \&\ Sarajedini 1996, p. 434-443

\noindent\hang Mateo M. 1997. In {\it The Nature of Elliptical Galaxies}, ed. M Arnaboldi, GS Da Costa, P Saha, p. 259-269. San Francisco: ASP, Vol 116

\noindent\hang Mateo M. 1998. See Aparicio et al 1998, p. 407-457

\noindent\hang Mateo M, Demers S, Kunkel WE. 1998a. AJ in press

\noindent\hang Mateo M, Fischer P, Krzemi\'niski W. 1995a. AJ 110:2166-2182

\noindent\hang Mateo M, Hurley-Keller DA, Nemec J. 1998b. AJ May issue

\noindent\hang Mateo M, Kubiak M, Szyma\'nski M, Kaluzny J, Krzemi\'nski W, Udalski A. 1995b. AJ 110:1141-1154

\noindent\hang Mateo M, Mirabal N, Udalski A, Szyma\'nski M, Kaluzny J, Kubiak M, Krzemi\'nski M, Stanek KZ. 1996. ApJ 458:L13-L16

\noindent\hang Mateo M, Nemec J, Irwin M, McMahon R. 1991a. AJ 101:892-910

\noindent\hang Mateo M, Olszewski E, Welch DL, Fischer P, Kunkel W. 1991b. AJ 102:914-926

\noindent\hang Mateo M, Olszewski EW, Hodge P. 1998d, AJ in press

\noindent\hang Mateo M, Olszewski EW, Pryor C, Welch DL, Fischer P. 1993. AJ 105:510-526

\noindent\hang Mateo M, Olszewski EW, Vogt SS, Keane M. 1998c. AJ in press

\noindent\hang Mateo M, Udalski A, Szyma\'nski M, Kaluzny J, Kubiak M, Krzemi\'nski W. 1995c. AJ 109:588-593

\noindent\hang Mathewson DS, Ford VL. 1984. In {\it Structure and Evolution of the Magellanic Clouds} ed. S van den Bergh, K de Boer, p. 125-136. Dordrecht: Reidel

\noindent\hang McGaugh S, de Blok E. 1998. ApJ in press

\noindent\hang McNamara DH. 1995. AJ 109:1751-1756

\noindent\hang Melisse JPM, Israel FP. 1994a. A\&A 285:51-68

\noindent\hang Melisse JPM, Israel FP. 1994b. A\&AS 103:391-412

\noindent\hang Mermilliod JC. 1981. A\&A 97:235-244

\noindent\hang Merrifield MR, Kent SM. 1990. AJ 99:1548-1557

\noindent\hang Meylan G, Prugniel P. 1994. {\it Dwarf Galaxies}, Garching: European Southern Observatory

\noindent\hang Meynet G, Maeder A, Schaller G, Schaerer D, Charbonnel C. 1994. A\&AS 103:97-105

\noindent\hang Meynet G, Mermilliod JC, Maeder A. 1993. A\&AS 76:411-425

\noindent\hang Michard R, Nieto JL. 1991. A\&A 243:L17-L20

\noindent\hang Mighell KJ. 1990. A\&AS 82:1-39

\noindent\hang Mighell KJ. 1997. AJ 114:1458-1470

\noindent\hang Mighell KJ, Rich RM. 1996. AJ 111:777-787

\noindent\hang Milgrom M. 1983a. ApJ 270:365-370

\noindent\hang Milgrom M. 1983b. ApJ 270:371-383

\noindent\hang Milgrom M. 1991. ApJ 367:490-492

\noindent\hang Milgrom M. 1994. ApJ 429:540-544

\noindent\hang Milgrom M. 1995. ApJ 455:439-442

\noindent\hang Miller BW. 1996. AJ 112:991-1008

\noindent\hang Minniti D, Zijlstra AA. 1996. ApJ 467:L13-L16

\noindent\hang Moles M, Aparicio A, Masegosa J. 1990. A\&A 228:310-314

\noindent\hang Moore B, Davis M. 1994. MNRAS 270:209-221



\noindent\hang Morrison H, Sarajedini A, eds. 1996. {\it Formation of the Galactic Halo\ .\ .\ .\ Inside and Out}, San Francisco: ASP, Vol 92

\noindent\hang Mould J. 1997. PASP 109:125-129

\noindent\hang Mould J, Aaronson M. 1983. ApJ 273:530-538

\noindent\hang Mould JR, Bothun GD, Hall PJ, Staveley-Smith L, Wright AE. 1990. ApJ 362:L55-L57

\noindent\hang Mould J, Kristian J. 1990. ApJ 354:438-445

\noindent\hang Mould J, Kristian J, Da Costa GS. 1983. ApJ 270:471-484

\noindent\hang Mould J, Kristian J, Da Costa GS. 1984. ApJ 278:575-581

\noindent\hang Musella I, Piotto G, Capaccioli M. 1997. AJ 114:976-987

\noindent\hang Nemec JM. 1985. AJ 90:204-239

\noindent\hang Nemec JM, Nemec AF, Lutz TE. 1994. AJ 108:222-246

\noindent\hang Nemec JM, Wehlau A, Mendes de Oliveira C. 1988. AJ 96:528-559



\noindent\hang Nolthenius R, Ford H. 1986. ApJ 305:600-608

\noindent\hang O'Connell RW. 1992. In {\it The Stellar Populations of Galaxies}, ed. B Barbuy, A Renzini, p. 233-243. Dordrecht: Kluwer

\noindent\hang Oh KS, Lin DNC, Aarseth SJ. 1995. ApJ 442:142-158

\noindent\hang Ohta K, Sasaki M, Sait\=o M. 1988. PASJ 40:653-664

\noindent\hang Ohta K, Tomita A, Sait\=o M, Sasaki M, Nakai N. 1993. PASJ 45:L21-L26

\noindent\hang Olszewski EW. 1998. In {\it Santa Cruz Workshop on Galaxy Halos}, ed D Zaritsky, p. 70-78, San Francisco: ASP

\noindent\hang Olszewski EW, Aaronson M. 1985. AJ 90:2221-2238

\noindent\hang Olszewski EW, Aaronson M, Hill JM. 1995. AJ 110:2120-2130

\noindent\hang Olszewski EW, Pryor C, Armandroff TE. 1996. AJ 111:750-767

\noindent\hang Olszewski EW, Suntzeff NB, Mateo M. 1996. ARAA 34:511-550

\noindent\hang Oosterloo T, Da Costa GS, Staveley-Smith L. 1996. AJ 112:1969-1974

\noindent\hang Ortolani S, Gratton RG. 1988. PASP 100:1405-1422

\noindent\hang Pagel BEJ, Edmunds MG, Smith G. 1980. MNRAS 193:219-230

\noindent\hang Paltoglou G, Freeman KC. 1987. In {\it Structure and Dynamics of Elliptical Galaxies}, ed. T. de Zeeuw, p. 447-448.  Dordrecht: Reidel

\noindent\hang Peebles PJE. 1989. ApJ 344:L53-L56

\noindent\hang Peebles PJE. 1995. ApJ 449:52-60

\noindent\hang Peletier RF. 1993. A\&A 271:51-64

\noindent\hang Peterson CJ. 1993. In {\it Structure and Dynamics of Globular Clusters}, ed. SG Djorgovski, G. Meylan, p. 337-345. ASP: San Francisco, Vol 50

\noindent\hang Peterson RC, Caldwell N. 1993. AJ 105:1411-1419

\noindent\hang Phillips S, Parker QA, Schwartzenberg JM, Jones JB. 1998. ApJ 493:L59-L62

\noindent\hang Piatek S, Pryor C. 1995. AJ 109:1071-1085

\noindent\hang Pierce MJ, Tully RB. 1992. ApJ 387:47-55

\noindent\hang Piotto G, Capaccioli M. 1992. Memoria della Scoiet\'a Astronomica di Italia 63:465-478

\noindent\hang Piotto G, Capaccioli M, Pellegrini C. 1994. A\&A 287:371-386

\noindent\hang Poulain P, Nieto JL. 1994. A\&AS 103:573-595

\noindent\hang Preston GW, Beers TC, Shectman SA. 1994. AJ 108:538-554

\noindent\hang Price JS. 1985. ApJ 297:652-661

\noindent\hang Price JS, Grasdalen GL. 1983. ApJ 275:559-570

\noindent\hang Price JS, Mason SF, Gullixson CA. 1990. AJ 100:420-424

\noindent\hang Pritchet CJ, Richer HB, Schade D, Crabtree D, Yee HKC. 1987. ApJ 323:79-90

\noindent\hang Prugniel P, Simien F. 1997. 321:111-122

\noindent\hang Pryor C. 1994. See Meylan \&\ Prugniel 1994, p. 323-334

\noindent\hang Pryor C. 1996. See Morrison \&\ Sarajedini 1996, p. 424-433

\noindent\hang Pryor C, Kormendy J. 1990. AJ 100:127-140

\noindent\hang Puche D, Carignan C, Bosma A. 1990. AJ 100:1468:1476

\noindent\hang Puche D, Carignan C, Wainscoat RJ. 1991. AJ 101:447-455

\noindent\hang Puche D, Westpfahl D. 1994. See Meylan \&\ Prugniel, p. 273-281

\noindent\hang Qiu B, Wu XP, Zou ZL. 1995. A\&A 296:264-268

\noindent\hang Queloz D, Dubath P, Pasquini L. 1995. A\&A 300:31-42

\noindent\hang Reid N, Mould J. 1991. AJ 101:1299-1303

\noindent\hang Richer HB, Crabtree DR, Pritchet CJ. 1984. ApJ 287:138-147

\noindent\hang Richer HB, Westerlund BE. 1983. ApJ 264:114-125

\noindent\hang Richer MG, McCall ML. 1992. AJ 103:54-59

\noindent\hang Richer MG, McCall ML. 1995. ApJ 445:642-659

\noindent\hang Richstone DO, Tremaine S. 1986. AJ 92:72-74

\noindent\hang Richter GM, Schmidt KH, Th\"anert W, Stavrev K, Panov K. 1991. AN 312:309-314

\noindent\hang Richter OG, Tammann GA, Huchtmeier WK. 1987. A\&A 171:33-40

\noindent\hang Roberts MS, Hogg DE, Bregman JN, Forman WR, Jones C. 1991. ApJS 75:751-799

\noindent\hang Rowan-Robinson M, Phillips TG, White G. 1980. A\&A 82:381-384

\noindent\hang Rozanski R, Rowan-Robinson M. 1994. MNRAS 271:530-552

\noindent\hang Sage LJ, Wrobel JM. 1989. ApJ 344:204-209

\noindent\hang Saha A, Freedman WL, Hoessel JG, Mossman AE. 1992a. AJ 104:1072-1085

\noindent\hang Saha A, Hoessel JG. 1987. AJ 94:1556-1563

\noindent\hang Saha A, Hoessel JG. 1990. AJ 99:97-148

\noindent\hang Saha A, Hoessel JG. 1991. AJ 101:465-468

\noindent\hang Saha A, Hoessel JG, Krist J. 1992b. AJ 103:84-103

\noindent\hang Saha A, Hoessel JG, Krist J, Danielson GE. 1996. AJ 111:197-207

\noindent\hang Saha A, Hoessel JG, Mossman AE. 1990. AJ 100:108-126

\noindent\hang Saha A, Monet DG, Seitzer P. 1986. AJ 92:302-327

\noindent\hang Sait\=o M, Sasaki M, Ohta K, Yamada T. 1992. PASJ 44:593-600

\noindent\hang Sakai S, Madore BF, Freedman WL. 1996. ApJ 461:713-723

\noindent\hang Sandage A. 1961. {\it The Hubble Atlas of Galaxies}, Washington DC: Carnegie

\noindent\hang Sandage A. 1986a. ApJ 307:1-19

\noindent\hang Sandage A. 1986b. AJ 91:496-506

\noindent\hang Sandage A, Bedke J. 1994. {\it The Carnegie Atlas of Galaxies}, Washington DC: Carnegie

\noindent\hang Sandage A, Binggeli B. 1984. AJ 89:919-931

\noindent\hang Sandage A, Binggeli B, Tammann GA. 1985. AJ 90:395-404

\noindent\hang Sandage A, Binggeli B, Tammann GA. 1985. AJ 90:1759-1771

\noindent\hang Sandage A, Carlson G. 1982. ApJ 258:439-456

\noindent\hang Sandage A, Carlson G. 1985a. AJ 90:1019-1026

\noindent\hang Sandage A, Carlson G. 1985b. AJ 90:1464-1473

\noindent\hang Sandage A, Carlson G. 1988. AJ 96:1599-1613

\noindent\hang Sandage A, Hoffman GL. 1991. ApJ 379:L45-L47

\noindent\hang Sandage A, Smith LL. 1966. ApJ 144:886-893

\noindent\hang Sandage A, Wallerstein G. 1960. ApJ 131:598-609

\noindent\hang Sanders RH. 1996. ApJ 473:117-129

\noindent\hang Sanders RH. 1997. ApJ 480:492-502

\noindent\hang Sarajedini A, Chaboyer B, Demarque P. 1997. PASP 109:1321-1339

\noindent\hang Sarajedini A, Claver CF, Ostheimer JC. 1997. AJ 114:2505-2513

\noindent\hang Sarajedini A, Layden AC. 1995. AJ 109:1086-1094

\noindent\hang Saviane I, Held EV, Piotto G. 1996. A\&A 315:40-51

\noindent\hang Schaller G, Schaerer D, Meynet G, Maeder A. 1992. A\&AS 96:269-331

\noindent\hang Schechter P. 1976. ApJ 203:297-306

\noindent\hang Schild R. 1980. ApJ 242:63-65

\noindent\hang Schmidt KH, Boller T. 1992. AN 313:189-231

\noindent\hang Schweitzer AE, Cudworth KM, Majewski SR, Suntzeff NB. 1995. AJ 110:2747-2756

\noindent\hang Sembach KR, Tonry JL. 1996. AJ 112:797-805

\noindent\hang S\'ersic JL. 1968. {\it Atlas de Galaxias Australes}, Cordoba: Observatorio Astronomico

\noindent\hang S\'ersic JL, Cerruti MA. 1979. Obs. 99:150-151

\noindent\hang Shostak GS. 1974. A\&A 31:97-101

\noindent\hang Shostak GS, Skillman ED. 1989. A\&A 214:33-42

\noindent\hang Siegel MH, Majewski SR, Reid IN, Thompson I, Landolt AU, Kunkel WE. 1997. BAAS 191:81.03

\noindent\hang Silva DR, Elston R. 1994. ApJ 428:511-543

\noindent\hang Sivaram C. 1994. ApSS 215:185-189

\noindent\hang Skillman ED. 1998. See Aparicio et al 1998, p. 457-525

\noindent\hang Skillman ED, Bender R. 1995. RevMxA\&A~Conf Series 3:25-30

\noindent\hang Skillman ED, Bomans DJ, Kobulnicky HA. 1997. ApJ 474:205-216

\noindent\hang Skillman ED, Kennicutt RC, Hodge PW. 1989a. ApJ 347:875-882

\noindent\hang Skillman ED, Terlevich R, Melnick J. 1989b. MNRAS 240:563-572

\noindent\hang Skillman ED, Terlevich R, Teuben PJ, van Woerden H. 1988. A\&A 198:33-42

\noindent\hang Smecker-Hane TA, Stetson PB, Hesser JE, Lehnert MD. 1994. AJ 108:507-513

\noindent\hang Smith EO, Neill JD, Mighell KJ, Rich RM. 1996. AJ 111:1596-1603

\noindent\hang Sofue Y, Wakamatsu K. 1993. PASJ 45:529-538

\noindent\hang Stasinska G, Comte G, Vigroux L. 1986. A\&A, 154:352-356

\noindent\hang Stetson PB. 1997. Baltic Ast 6:3-10

\noindent\hang Strobel NV, Hodge P, Kennicutt RC. 1991. ApJ 383:148-163

\noindent\hang Strobel NV, Lake G. 1994. ApJ 424:L83-L86

\noindent\hang Suntzeff NB, Aaronson M, Olszewski EW, Cook KH. 1986. AJ 91:1091-1095

\noindent\hang Suntzeff NB, Mateo M, Terndrup DM, Olszewski EW, Geisler D, Weller W. 1993. ApJ 418:208-228

\noindent\hang Swope H. 1967. PASP 79:439-440

\noindent\hang Tacconi LJ, Young JS. 1987. ApJ 322:681-687

\noindent\hang Talent DL. 1980. {\it A Spectrophotometric Study of HII Regions in Chemically Young Galaxies}, PhD thesis, Rice University

\noindent\hang Thronson HA, Hunter DA, Casey S, Harper DA. 1990. ApJ 355:94-101

\noindent\hang Thuan TX, Martin GE. 1979. ApJ 232:L11-L16

\noindent\hang Tikhonov N, Makarova L. 1996 AN 317:179-186

\noindent\hang Tolstoy E. 1996. ApJ 462:684-704

\noindent\hang Tolstoy E, Gallagher JS, Cole AA, Hoessel JG, Saha A, et al. 1998. AJ in press

\noindent\hang Tolstoy E, Saha A. 1996. ApJ 462:672-683

\noindent\hang Tolstoy E, Saha A, Hoessel JG, Danielson GE. 1995. AJ 109:579-587

\noindent\hang Tomita A, Ohta K, Sait\=o M. 1993. PASJ 45:693-705

\noindent\hang Tonry JL. 1984. ApJ 283:L27-L30

\noindent\hang Tosi M, Greggio L, Marconi G, Focardi P. 1991. AJ 102:951-974

\noindent\hang Unavane M, Wyse RFG, Gilmore G. 1996. MNRAS 278:727-736

\noindent\hang van Agt SLTJ. 1973. In {\it Variable Stars in Globular Clusters and Related Systems}, ed. JD Fernie, p. 35-48. Dordrecht: Reidel

\noindent\hang van Agt SLTJ. 1978. PubDDO 3:205-235

\noindent\hang van den Bergh S. 1994a. AJ 107:1328-1332

\noindent\hang van den Bergh S. 1994b. AJ 108:2145-2153

\noindent\hang van den Bergh S. 1994c. ApJ 428:617-619

\noindent\hang van den Bergh S. 1995. ApJ 446:39-43

\noindent\hang van de Rydt F, Demers S, Kunkel WE. 1991. AJ 102:130-136

\noindent\hang van Dokkum PG, Franx M. 1995. AJ 110:2027-2036


\noindent\hang Vel\'azquez H, White SDM. 1995. MNRAS 275:L23-L26

\noindent\hang Verter F, Hodge P. 1995. ApJ 446:616-621

\noindent\hang Vogt SS, Mateo M, Olszewski EW, Keane MJ. 1995. AJ 109:151-163


\noindent\hang Wakker BP, van Woerden H. 1997. ARAA 35:217-266

\noindent\hang Walsh JR, Dudziak G, Minniti D, Zijlstra AA. 1997. ApJ 487:651-662

\noindent\hang Webster BL, Smith MG. 1983. MNRAS 204:743-763

\noindent\hang Welch GA, Mitchell GF, Yi S. 1996. ApJ 470:781-789 

\noindent\hang Westerlund BE. 1990. A\&ARev 2:29- 

\noindent\hang Westerlund BE. 1998. {\it The Magellanic Clouds}, Cambridge: Cambridge University Press

\noindent\hang Whitelock PA, Irwin M, Catchpole RM. 1996. New Ast. 1:57-75

\noindent\hang Whiting AB, Irwin MJ, Hau GKT. 1997. AJ 114:996-1001

\noindent\hang Wiklind T, Rydbeck G. 1986. A\&A 164:L22-L24

\noindent\hang Wilson CD. 1992a. AJ 104:1374-1394

\noindent\hang Wilson CD. 1992b. ApJ 391:144-151

\noindent\hang Wilson CD. 1994a. See Layden et al 1994, p. 15-27

\noindent\hang Wilson CD. 1994b. ApJ 434:L11-L14

\noindent\hang Wilson CD. 1995. ApJ 448:L97-L100

\noindent\hang Wilson CD, Reid IN. 1991. ApJ 366:L11-L14

\noindent\hang Wilson CD, Welch DL, Reid IN, Saha A, Hoessel J. 1996. AJ 111:1106-1109

\noindent\hang Wyatt RJ, Dufour RJ. 1993. RevMxA\&A 27:213-218

\noindent\hang Yahil A, Tammann GA, Sandage A. 1977. ApJ 217:903-915

\noindent\hang Yang H, Skillman ED. 1993. AJ 106:1448-1459

\noindent\hang Young JS, Xie S, Tacconi L, Knezek P, Viscuso P, et al. 1995. ApJS 98:219-257

\noindent\hang Young LM, Lo KY. 1996a. ApJ 462:203-214

\noindent\hang Young LM, Lo KY. 1996b. ApJ 464:L59-L62

\noindent\hang Young LM, Lo KY. 1997a. ApJ 476:127-143

\noindent\hang Young LM, Lo KY. 1997b. ApJ 490:710-72

\noindent\hang Zaritsky D, Lin DNC. 1997. AJ 114:2545-2555

\noindent\hang Zaritsky D, Olszewski EW, Schommer RA, Peterson RC, Aaronson M. 1989. ApJ 345:759-769

\noindent\hang Zijlstra AA, Walsh JR. 1996. A\&A 312:L21-L24

\noindent\hang Zinn R. 1980. In {\it Globular Clusters}, ed. D Hanes, B Madore, p. 191. Cambridge: Cambridge U Press

\noindent\hang Zinn R, West MJ. 1984. ApJS 55:45-66

\vfill\eject

\noindent{\underline{FIGURE CAPTIONS}}

\vskip1.5em 

\noindent {\bf Figure~1.}\ \ The heliocentric velocities of nearby
galaxies plotted as a function of $\cos(\lambda)$, where $\lambda$ is
the angle between the galaxy and the location of the apex of the sun's
motion relative to the center-of-mass of the Local Group.  The {\it
upper plot} adopts the solution for the solar motion from Sandage
(1986a) for which the apex is located at $(l,b) =
(101^\circ,-11^\circ)$ and the sun's velocity component in this
direction is 343 km~s$^{-1}$.  The {\it lower plot} is for the
solution from Karachentsev \&\ Makarov (1996) for which $(l,b) =
(93^\circ,-4^\circ)$, and the solar velocity is 316 km~s$^{-1}$.  The
{\it filled squares} denote the galaxies adopted as Local Group members in
this paper; the {\it open squares} represent a sample of other nearby
galaxies -- SclDIG (Heisler et al 1997), NGC~300 (Puche et al 1990),
Maffei~1 (Luppino \&\ Tonry 1993), DDO~187 and UGCA~86 (van den Bergh
1994a) -- that are shown here as a representative sample of galaxies
that have at some time or other been considered to be LG members (see
van den Bergh 1994a for other examples).

\vskip1.5em

\noindent {\bf Figure~2.}\ \ Plots of the cumulative distribution of
all LG galaxies ({\it upper plot}) and of the galaxies in the MW
subgroup ({\it lower plot}) as a function of $(1 - \sin|b|)$, where
$b$ is Galactic latitude.  The {\it dotted lines} show the expected
distribution of a uniform sample of 40 and 12 objects.  The {\it
vertical lines} show where 50\%\ ({\it left}) and 67\%\ ({\it right})
of a uniform cumulative distribution would be found.  For example,
based on the 50th percentile value of $N_c = 10$ in the {\it lower
panel}, a total of 20 MW satellites would be expected if they were
uniformly distributed, and if there was no gradient in Galactic
extinction as a function of latitude.

\vskip1.5em

\noindent {\bf Figure~3.}\ \ Stereoscopic views of the Local Group
based on the data in Tables 1 and 2.  From {\it top} to {\it bottom},
these views are from $(l,b) = (0^\circ,0^\circ)$,
$(90^\circ,0^\circ)$, and the NGP, respectively.  The directions of
the principle orthogonal axes in the plane of the paper are indicated
on the {\it top} and {\it right edges} of each stereoscopic pair.  The
`viewer' is located 2.6 Mpc from the center of the Milky Way which is
represented by the {\it large cross closest to the center of each
panel}; the {\it other large cross} represents M31 and the {\it small cross}
is M33.  {\it Open squares} represent the dSph and E galaxies (see Table~1
for galaxy types); {\it closed squares} are Irr galaxies; {\it small x's} are
transition systems (denoted dIrr/dSph in Table 1).  To produce the
stereo effect, the viewer's `eyes' were set about 250 kpc apart.

The MW, M31, and NGC~3109 subgroups are easily seen in all three
stereo pairs; the Local Group Cloud is best appreciated in the middle
pair but is also evident as a distinct structure in the other two
panels.  Only GR~8 (and possibly Leo~A) is unattached to any subgroup;
this is best seen in the lower panel where the galaxy is particularly
near the observer.  The {\it set of three single panels on the opposite
page} serve to aid identification of individual galaxies in the stereo
pairs.  The galaxies closest to M31 and the Milky Way have not been
labeled to avoid excessive clutter in the diagram.  Because of the
strong perspective effects required to make a stereo pair, none of the
panels represent orthogonal projections of the galaxy positions onto the
various planes.

\vskip1.5em

\noindent {\bf Figure~4.}\ \ The differential luminosity function of
galaxies in the Local Group based on the data in Table~4.  The {\it upper}
and {\it lower panels} show the V-band and B-band LFs, respectively.  The
{\it lower panel} also shows a best-fitting Schechter (1976) function
($\alpha = -1.16$, $M_{*,B} = -21.42$) and the empirical LF that
Ferguson \&\ Sandage (1991) derived for their sample of `poor' groups.
Both LFs are scaled to match approximately the cumulative galaxy
counts for $M_B \lsim -14$.

\vskip1.5em

\noindent {\bf Figure~5.}\ \ The $M_{V,0}$-(B$-$V)$_0$ color-magnitude
diagram of Local Group galaxies based on the data from Tables 2 and 3.
The {\it diagonal dashed line} separates galaxies classified as Spirals, or
Irregular systems ({\it filled squares}; Table~1), and dSph or Elliptical
systems ({\it open circles}).  Five 'transition' objects, Antlia, LGS~3,
Phoenix, Pegasus, and DDO~210, are plotted as {\it filled triangles};
NGC~205 is plotted as a {\it filled circle}. 

\vskip1.5em

\noindent {\bf Figure~6.}\ \ A plot of the Sculptor dSph galaxy
showing the optical and 21-cm radio components (Carignan et al 1998).
The stellar and HI velocities agree to within 5 km~s$^{-1}$; the gas is
almost certainly associated with Sculptor.  Considerable HI flux in the outer
regions of Sculptor may have been missed by these VLA observations;
the actual morphology (total flux) of the HI emitting gas may be
quite different (larger) than the bimodal distribution shown (or the
flux reported in Table~5).  The key point is that what little neutral
H gas there is in Sculptor is distributed away from the galaxy's
center.  For reference, the tidal radius of Sculptor is
approximately 76 arcmin (Table~3), slightly larger than the dimensions
of the sides of the figure.  The HI {\it contours} correspond to 0.2, 0.6,
1.0, 1.4, 1.8 and $2.2 \times 10^{19}$ cm$^{-2}$.

\vskip1.5em

\noindent {\bf Figure~7.}\ \ A plot of [Fe/H] ({\it filled squares})
or [O/H] $-$ 0.37 (I have assumed $12 + \log ({\rm O/H})_\odot = 8.93$
after Anders \&\ Grevesse (1989)) vs absolute V-band magnitude.  The
{\it dotted line} is a rough fit to the [Fe/H]-$M_V$ relation for the
dSph and transition objects.  Sagittarius corresponds to the {\it
points} near ($M_V$,[Fe/H]) $\sim$ ($-$13.4,$-$1.0)).  {\it Square
symbols} refer to dSph or dE galaxies; {\it triangles} refer to
transition galaxies (denoted dIrr/dSph in Table~1); {\it circles}
refer to dIrr systems.  {\it Filled symbols} correspond to [Fe/H]
abundances determined from stars, while {\it open symbols} denote
oxygen abundance estimates from analyses of HII regions and planetary
nebulae.  See Table~6 for details.

\vskip1.5em

\noindent {\bf Figure~8.}\ \ Schematic plots of the star-formation
histories of all Local Group dwarfs with sufficient data.  The {\it
labels within the individual panels} specify the nature of the stellar
indicators used to infer the presence of a given age component: MS =
main sequence stars; AGB = asymptotic giant branch stars; RG = red
giants; RR = RR~Lyr variables; AC = anomalous Cepheids; SG = blue and
red supergiants; W = WR stars; PN = planetary nebulae.  `2P' means
that the galaxy has an anomalously red horizontal-branch population
for its (low) metallicity -- that is, the galaxy exhibits the second
parameter effect. {\it Numbers within square brackets} denote the
metallicities of specific star-forming epochs; this information is
generally quite uncertain and is available for only a few systems.

The reliability of the various star-formation episodes for a given
galaxy is denoted by the style of the lines use to plot them: {\it solid
horizontal lines} indicate that that duration of a given age component
is fairly well determined; {\it solid vertical lines} indicate that the
relative star formation rate of a given event with respect to other
star-formation episodes is reasonably well constrained; {\it dashed
horizontal and vertical lines} indicate very great uncertainties in
the duration, or relative strength of individual star-formation
periods.  The galaxies are plotted in the same order that they are
listed in the tables by increasing right ascension.  Some of the
galaxies listed as Local Group members in Table~2 are not plotted
because of insufficient data.  For a few
galaxies, {\it separate panels} show the SFHs of the inner and outer regions,
separately.

\noindent REFERENCES FOR FIGURE 8: {\bf WLM: } Sandage \&\ Carlson (1985b), 
Cook et al (1986), Ferraro et al (1989), Minniti \&\ Zijlstra (1996);
{\bf NGC~147: } Mould et al (1983), Saha et al (1990), Davidge (1994),
Han et al (1997);
{\bf And~III: } Armandroff et al (1993); 
{\bf NGC~185: } Saha \&\ Hoessel (1990), Lee et al (1993b);
{\bf NGC~205: } Mould et al (1984), Richer et al (1984), Saha et al (1992b),
Lee (1996);
{\bf M32: } Davidge \&\ Jones (1992), Freedman (1992), Elston \&\ Silva (1992),
Grillmair et al (1996);
{\bf And~I: } Da Costa et al (1996);
{\bf Sculptor: } Da Costa (1984), Azzopardi et al (1985, 1986);
{\bf LGS~3: } Lee (1995a); Aparicio et al (1997b);
{\bf IC~1613}  Freedman (1988); Saha et al (1992a);
{\bf Phoenix: } Ortolani \&\ Gratton (1988), van de Rydt et al (1991);
{\bf Fornax: } Buonanno et al (1985), Demers et al (1995), Beauchamp et
al (1995), Demers et al (1998);
{\bf Carina: } Mould \&\ Aaronson (1983),  Azzopardi et al (1985, 1986),
Mighell (1990,1997), Smecker-Hane et al (1994), Hurley-Keller et al (1998);
{\bf Leo~A: } Tolstoy et al (1998);
{\bf Sextans~B: } Tosi et al (1991), Marconi et al (1995);
{\bf NGC~3109: } Richer \&\ McCall (1992),
Greggio et al (1993), Davidge (1993), Bresolin et al (1993);
{\bf Antlia: } Aparicio et al (1997a), Whiting et al (1997), Sarajedini et al
(1997); 
{\bf Leo~I: } Azzopardi et al (1985, 1986), Reid \&\ Mould (1991), 
Caputo et al (1995), Gallart et al (1998);
{\bf Sextans~A: } Dohm-Palmer et al (1997);
{\bf Sextans: } Mateo et al (1991a), Suntzeff et al (1993),
Mateo et al (1995a);
{\bf Leo~II: } Azzopardi et al (1985), Mighell \&\ Rich (1996);
{\bf GR~8: } Dohm-Palmer et al (1998); 
{\bf Ursa~Minor: } Olszewski \&\ Aaronson (1985);
{\bf Draco: } Carney \&\ Seitzer (1986), Azzopardi et al (1986), Grillmair
et al (1998); 
{\bf Sagittarius: } Ibata et al (1994), Mateo et al (1995b), Sarajedini
\&\ Layden (1995), Mateo et al (1996), Alard (1996), Fahlman et al (1996),
Ibata et al (1997), Marconi et al (1998); 
{\bf NGC~6822: } Hodge (1980), Armandroff \&\ Massey (1991), Gallagher 
et al (1991),
Wilson (1992a), Marconi et al (1995), Gallart et al (1996a,b,c);
{\bf DDO~210: } Marconi et al (1990), Greggio et al (1993);
{\bf Tucana: } Lavery \&\ Mighell (1992), Saviane et al (1996), Castellani 
et al (1996);
{\bf Pegasus: } Aparicio \&\ Gallart (1995), Aparicio et al (1997b).

\vskip1em

\noindent {\bf Figure~9.}\ \ Kinematically-determined mass-to-light
ratios of local group dwarfs as a function of luminosity.  {\it Top
panel}: $\log (M/L)_0$ from Table~4 vs $M_V$. {\it Filled squares} are
for dSph or dSph/Irr systems for which masses were determined from the
central velocity dispersions, while the {\it open squares} represent
Irr systems which have masses derived here from HI rotation
curves. See Table~4 for details, or the original sources to obtain
definitive kinematic mass estimates for these galaxies.  Sagittarius
is denoted as an {\it open circle}.  {\it Bottom panel}: $\log
(M/L)_{tot}$ from Table~4 vs $M_V$; the {\it symbols} are the same as
in the {\it top panel}.  In each panel I have also plotted the
function $\log M/L = 2.5 + 10^7/(L/L_\odot)$ as a {\it dashed line}.

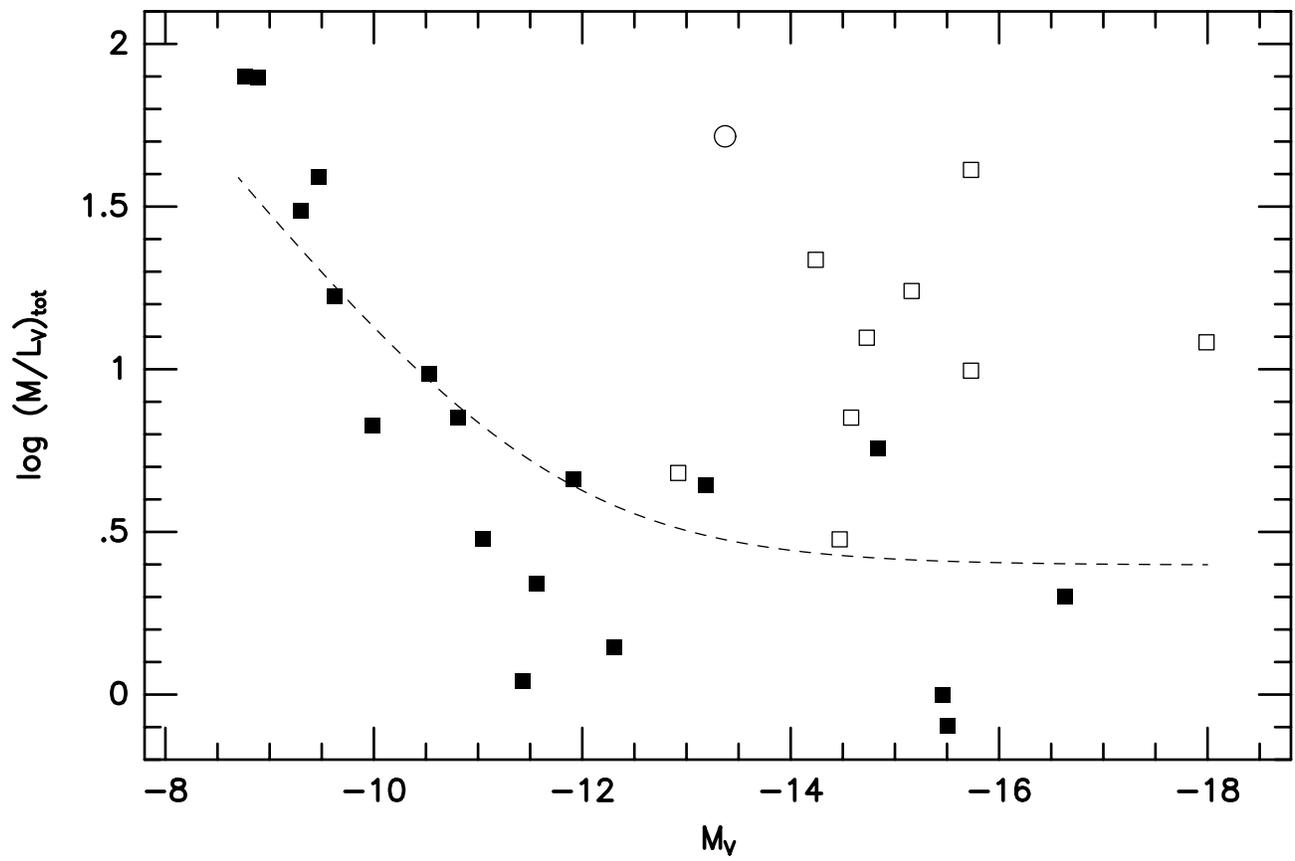
\end{document}



\setcounter{table}{0}
\begin{table}
\noindent \ \ Table 1: Local Group Galaxies
\begin{center}
\begin{tabular}{llcccclclc}
\\
\ Galaxy & Other Name & $\alpha_{2000}$ & $\delta_{2000}$ & $l$ & $b$ & \ \ Type & Subgroup & Images & Notes \\
\ \ \ (1) & \ \ \ \ (2) & (3) & (4) & (5) & (6) & \ \ \ \ (7) & (8) & \ \ (9) & (10) \\
\\
WLM        & DDO~221  & 00 01 58 & $-$15 27.8 & \pz75.9  & $-$73.6 & IrrIV-V     & LGC & 1 & \\
NGC~55       &          & 00 15 08 & $-$39 13.2 & 332.7 & $-$75.7 & IrrIV  & LGC  & 3,4,51 & \\
IC~10      & UGC~192  & 00 20 25 & +59 17.5 & 119.0 & \pz$-$3.3  & dIrr        & M31 & 5,6 & \\
NGC~147    & DDO~3    & 00 33 12 & +48 30.5 & 119.8 & $-$14.3 & dSph/dE5    &  M31 & 7,51 & \\
And III    &          & 00 35 17 & +36 30.5 & 119.3 & $-$26.2 & dSph        & M31 & 8 & \\
NGC~185    & UGC~396  & 00 38 58 & +48 20.2 & 120.8 & $-$14.5 & dSph/dE3p   & M31 & 7,51 & \\
NGC~205    & M110     & 00 40 22 & +41 41.4 & 120.7 & $-$21.1 & E5p/dSph-N & M31 & 7,51 & E \\
M32        & NGC~221  & 00 42 42 & +40 51.9 & 121.2 & $-$22.0 & E2          & M31 & 9,10,51 & \\
{\it M31}   & NGC~224  & 00 42 44 & +41 16.1 & 121.2 & $-$21.6 & SbI-II      & M31 & 10,50,51 \\
And~I      &          & 00 45 43 & +38 00.4 & 121.7 & $-$24.9 & dSph        & M31 & 8 & \\
{\it SMC}   & NGC~292  & 00 52 44 & $-$72 49.7 & 302.8 & $-$44.3 & IrrIV-V     & MW & 11,12,51 & \\
Sculptor   &          & 01 00 09 & $-$33 42.5 & 287.5 & $-$83.2 & dSph    & MW & 13,49 & \\
LGS~3      & Pisces   & 01 03 53 & +21 53.1 & 126.8 & $-$40.9 & dIrr/dSph   & M31 & 14 & \\
IC~1613    & DDO~8    & 01 04 54 & +02 08.0 & 129.8 & $-$60.6 & IrrV        & M31/LGC & 15-17 & \\
And~II     &          & 01 16 27 & +33 25.7 & 128.9 & $-$29.2 & dSph        & M31 & 8 & A \\
{\it M33}   & NGC~598  & 01 33 51 & +30 39.6 & 133.6 & $-$31.3 & ScII-III    & M31 & 17,18,51 & \\ 
Phoenix    &          & 01 51 06 & $-$44 26.7 & 272.2 & $-$68.9 & dIrr/dSph & MW/LGC & 19,20 & \\
Fornax     &          & 02 39 59 & $-$34 27.0 & 237.1 & $-$65.7 & dSph  & MW & 21 & \\
EGB~0427+63 & UGCA~92 & 04 32 01 & +63 36.4 & 144.7 & +10.5 & dIrr & M31 &  23 & \\
{\it LMC}   &          & 05 23 34 & $-$69 45.4 & 280.5 & $-$32.9 & IrrIII-IV   & MW & 11,12,51 & \\
Carina     &          & 06 41 37 & $-$50 58.0 & 260.1 & $-$22.2 & dSph  & MW & 24 &  \\
Leo~A      & DDO~69   & 09 59 24 & +30 44.7 & 196.9 & +52.4 & dIrr    & MW/N3109 & 29,51 & \\
Sextans~B  & DDO~70   & 10 00 00 & +05 19.7 & 233.2 & +43.8 & dIrr        & N3109  & 27,28,51 & \\
NGC~3109   & DDO~236  & 10 03 07 & $-$26 09.5 & 262.1 & +23.1  & IrrIV-V     & N3109 & 29,30,51 & \\
Antlia     &  & 10 04 04 & $-$27 19.8 & 263.1 & +22.3 & dIrr/dSph    & N3109 & 31 & B \\
Leo~I      & DDO~74   & 10 08 27 & +12 18.5 & 226.0 & +49.1 & dSph        &  MW & 32 & \\
Sextans~A  & DDO~75   & 10 11 06 & $-$04 42.5 & 246.2 & +39.9 & dIrr        & N3109  & 33,51 & \\
Sextans    &          & 10 13 03 & $-$01 36.9 & 243.5 & +42.3 & dSph        & MW & 34 & \\
Leo~II     & DDO~93   & 11 13 29 & +22 09.2 & 220.2 & +67.2 & dSph        & MW & 52 & \\
GR~8       & DDO~155  & 12 58 40 & +14 13.0 & 310.7 & +77.0 & dIrr & GR8 & 36,37,54 & \\
Ursa~Minor & DDO~199  & 15 09 11 & +67 12.9 & 105.0 & +44.8 & dSph        & MW & 38 & \\
Draco      & DDO~208  & 17 20 19 & +57 54.8 & \pz86.4  & +34.7 & dSph   & MW & 39 & \\
{\it Milky Way} &     & 17 45 40 & $-$29 00.5 & \pz\pz0.0 & \pmin\pz0.0 & Sbc         & MW & & \\
Sagittarius&          & 18 55 03 & $-$30 28.7 & \pz\pz5.6   & $-$14.1 & dSph-N    & MW & 40 & E \\   
SagDIG     & UKS1927-177 & 19 29 59 & $-$17 40.7 & \pz21.1  & $-$16.3 & dIrr        & LGC & 41,53 \\
NGC~6822   & DDO~209  & 19 44 56 & $-$14 48.1 & \pz25.3  & $-$18.4 & IrrIV-V     & LGC & 42,43,51 \\
DDO~210    & Aquarius & 20 46 46 & $-$12 51.0 & \pz34.0  & $-$31.3 & dIrr/dSph      & LGC & 36,44,54 & C \\
IC~5152    &          & 22 02 42 & $-$51 17.7 & 343.9 & $-$50.2 & dIrr        & LGC & 51 \\
Tucana     &          & 22 41 50 & $-$64 25.2 & 322.9 & $-$47.4 & dSph        & LGC & 47,48 & D \\
UKS2323-326& UGCA~438 & 23 26 27 & $-$32 23.3 & \pz11.9  & $-$70.9 & dIrr        & LGC & 41 \\
Pegasus    & DDO~216  & 23 28 34 & +14 44.8 & \pz94.8  & $-$43.5 & dIrr/dSph & LGC & 25,44,51 \\
\end{tabular}
\end{center}
\end{table}

\clearpage
\hsize=6.5truein
\vsize=8.5truein
\hoffset=0.0truein
\voffset=0.25truein

\noindent EXPLANATION OF COLUMNS OF TABLE 1:

\noindent {\it Column 1:} Galaxy name.  Entries denoted in {\it
italics} refer to the five giant Local Group galaxies that are not
discussed in this review in any detail; {\it Columns 2:} A common
alternative name; {\it Columns 3 and 4:} Right ascension and
declination for epoch J2000.0, respectively; {\it Columns 5 and 6:}
Galactic longitude and latitude, respectively; {\it Column 7:} Galaxy
type following van den Bergh (1994a); {\it Column 8:} Subgroup
membership within the Local Group; {\it Column 9:} References to
optical and 21-cm images (if available); {\it Column 10:} Special
notes.

\vskip2em

\noindent NOTES FOR TABLE 1:
\vskip0.5em

\noindent\hang A. The position listed here is based on an independent
measurement by Paul Hodge (private communication).

\noindent\hang B. Also known as PGC~29194 (Fouqu\'e et al 1990) before
it was re-discovered as a probable Local Group member by Whiting et al
(1997).  A thorough discussion of `pre-discovery' observations of
Antlia is provided by Aparicio et al (1997a).

\noindent\hang C. Marconi et al (1990) claimed that the original position of
DDO~210 (Fisher and Tully 1975) was significantly in error.  However, Lo
et al (1993) noted that the original position seemed to be correct, and
Marconi et al (1995) subsequently agreed.  The
position listed here corresponds to the original one from Fisher and
Tully (1975).

\noindent\hang D. Tucana was known prior to its re-discovery  by Lavery and
Mighell (1992) who thoroughly document all listings of the
galaxy in earlier catalogs.
Lavery and Mighell (1992) did first claim Tucana to be a possible Local
Group member.

\noindent\hang E. The `N' suffix has been added to indicate that these
systems may be nucleated dwarfs.

\vskip2em

\noindent REFERENCES FOR TABLE 1: 
[1] Sandage \&\ Carlson (1985b);
[2] Lautsen et al (1977);
[3] Hummel et al (1986);
[4] Puche et al (1991);
[5] de Vaucouleurs \&\ Ables (1965);
[6] Shostak \&\ Skillman (1989);
[7] Young \&\ Lo (1997a);
[8] Caldwell et al (1992);
[9] Kent (1987);
[10] Hodge (1981);
[11] de Vaucouleurs \&\ Freeman (1972);
[12] Mathewson \&\ Ford (1984);
[13] Carignan et al (1998);
[14] Young \&\ Lo (1997b);
[15] Ables (1971);
[16] Lake \&\ Skillman (1989);
[17] Sandage (1961);
[18] Corbelli et al (1989);
[19] van de Rydt et al (1991);
[20] Carignan et al (1991);
[21] Hodge (1971);
[22] Saha \&\ Hoessel (1991);
[23] Hoessel et al (1988);
[24] Smecker-Hane et al (1994);
[25] Sandage (1986b);
[26] Young \&\ Lo (1996a);
[27] Sandage \&\ Carlson (1985a);
[28] Skillman et al (1988);
[29] Sandage \&\ Carlson (1988);
[30] Jobin \&\ Carignan (1990);
[31] Whiting et al (1997);
[32] Hodge (1963a);
[33] Sandage \&\ Carlson (1982);
[34] Mateo et al (1995a);
[35] Deleted;
[36] Fisher \&\ Tully (1979);
[37] Carignan et al (1990);
[38] van Agt (1967);
[39] Baade \&\ Swope (1961);
[40] Ibata et al (1997);
[41] Longmore et al (1978);
[42] Hodge (1978);
[43] Gottesman \&\ Weliachew (1977);
[44] Lo et al (1993);
[45] Deleted;
[46] Deleted;
[47] Lavery \&\ Mighell (1992);
[48] Oosterloo et al (1996);
[49] van Agt (1978);
[50] Hodge (1992a);
[51] Sandage \&\ Bedke (1994);
[52] Vogt et al (1995);
[53] Cesarsky et al (1977);
[54] Hopp \&\ Schulte-Ladbeck (1995).


\setcounter{table}{1}
\begin{landscape}
\begin{table}
\noindent \ \ Table 2: Distances and Heliocentric Velocities of Local Group Galaxies
\begin{center}
\begin{tabular}{lccccclccll}
\ Galaxy & $l$ & $b$ & E(B$-$V) & (m$-$M)$_0$ & Distance & \ \ \ \ \ Ref & $V_{\odot,opt}$ & $V_{\odot,radio}$ & \ \ \ \ \ \ \ Ref & Notes \\
    & & & mag & mag & kpc & & km sec$^{-1}$ & km sec$^{-1}$ & & \\
\ \ \ (1) & (2) & (3) & (4) & (5) & (6) & \ \ \ \ \ (7) & (8) & (9) & \ \ \ \ \ \ \ (10) & \ \ (11) \\
\\
WLM         & \pz75.9  & $-$73.6 & 0.02$\pm$0.01 & 24.83$\pm$0.08 & \pz925$\pm$40\pz & 1-3,135 & --- & $-$123$\pm$3\pz & 4,5 & D \\
NGC~55       & 332.7 & $-$75.7 & 0.03$\pm$0.02 & 25.85$\pm$0.20 & 1480$\pm$150 & 8,9 & --- & \pmin124$\pm$6\pz & 4,8,136,137 & D \\
IC~10       & 119.0 & \pz$-$3.3  & 0.87$\pm$0.12 & 24.58$\pm$0.12 & \pz825$\pm$50\pz & 10,11  & $-$344$\pm$5 & $-$342$\pm$6\pz &  4,5,12-14 \\
NGC~147     & 119.8 & $-$14.3 & 0.18$\pm$0.03 & 24.30$\pm$0.12 & \pz725$\pm$45\pz & 1,15 & $-$193$\pm$3 & --- &  16,22 \\
And~III     & 119.3 & $-$26.2 & 0.05$\pm$0.02 & 24.40$\pm$0.10 & \pz760$\pm$40\pz & 18 & --- & --- & 4,19 & C \\
NGC~185     & 120.8 & $-$14.5 & 0.19$\pm$0.02 & 23.96$\pm$0.08 & \pz620$\pm$25\pz & 1,20 & $-$210$\pm$7 & $-$204$\pm$4\pz  &  16,21-25 \\
NGC~205     & 120.7 & $-$21.7 & 0.04$\pm$0.02 & 24.56$\pm$0.08 & \pz815$\pm$35\pz& 1,26,27 & $-$242$\pm$3 & $-$229$\pm$5\pz  &  4,16,22,29-31,138 & A \\
M32         & 121.2 & $-$22.0 & 0.08$\pm$0.03 & 24.53$\pm$0.08 & \pz805$\pm$35\pz & 32 & $-$197$\pm$3 & --- &  4,34,157 \\
{\it M31}  & 121.2 & $-$21.6 & 0.08 & 24.43 & 770 & 159 & \multicolumn{2}{c}{$-$297} & 159 & H \\
And~I        & 121.7 & $-$24.9 & 0.04$\pm$0.02 & 24.53$\pm$0.10 & \pz805$\pm$40\pz & 36,37 &--- & --- & 4,19 & C \\
{\it SMC}  & 302.8 & $-$44.3 & 0.08 & 18.82 & \pz58 & 125,159 & \multicolumn{2}{c}{\pmin175} & 159 & H \\
Sculptor    & 287.5 & $-$83.2 & 0.02$\pm$0.02 & 19.54$\pm$0.08 & \pz\pz79$\pm$4\pz\pz & 39,40,145 & \pmin108$\pm$3  & \pmin102$\pm$5\pz &  4,41,42,44,142 & B,C \\
LGS~3       & 126.8 & $-$40.9 & 0.08$\pm$0.03 & 24.54$\pm$0.15 & \pz810$\pm$60\pz & 45,151-153 & $-$282$\pm$4 & $-$272$\pm$6\pz & 4,46,155 \\
IC~1613     & 129.8 & $-$60.6 & 0.03$\pm$0.02 & 24.22$\pm$0.10 & \pz700$\pm$35\pz & 1,27,47,48 & $-$237$\pm$5 & $-$234$\pm$3\pz  &  4,49-51,156 \\
And~II      & 128.9 & $-$29.2 & 0.08$\pm$0.02 & \pz23.6$\pm$0.4\pz & \pz525$\pm$110 & 52 & --- & --- & \\
{\it M33} & 133.6 & $-$31.3 & 0.08 & 24.62 & 840 & 159 & \multicolumn{2}{c}{$-$181} & 159 & H \\
Phoenix     & 272.2 & $-$68.9 & 0.02$\pm$0.01 & 23.24$\pm$0.12 & \pz445$\pm$30\pz & 55 & --- & 56$\pm$3/$-$23$\pm$2 & 54,144 & B \\
Fornax       & 237.1 & $-$65.7 & 0.03$\pm$0.01 & 20.70$\pm$0.12 & \pz138$\pm$8\pz\pz & 56 & \pmin\pz53$\pm$3   & --- &  4,57,142 & C \\
EGB~0427+63 & 144.7 & $-$10.5 & 0.30$\pm$0.15 & \pz25.6$\pm$0.7\pz & 1300$\pm$700 & 61,146,147 & --- & \pz$-$87$\pm$6\pz   & 4 \\
{\it LMC} & 280.5 & $-$32.9 & 0.06 & 18.45 & \pz49 & 125,159 & \multicolumn{2}{c}{\pmin324} & 159 & H \\
Carina       & 260.1 & $-$22.2 & 0.04$\pm$0.02 & 20.03$\pm$0.09 & \pz101$\pm$5\pz\pz & 63-66,148 & \pmin224$\pm$3  & --- & 67-69 & C \\
Leo~A       & 196.9 & +52.4  & 0.01$\pm$0.01 & 24.2$\pm$0.3 & \pz690$\pm$100 & 70,141,161 & --- & \pmin\pz26$\pm$2\pz & 4,5,46,139 & E \\
Sextans~B   & 233.2 & +43.8  & 0.01$\pm$0.02 & 25.64$\pm$0.15 & 1345$\pm$100 & 71 & --- & \pmin303$\pm$2\pz & 4,46,50,152 \\
NGC~3109     & 262.1 & +23.1  & 0.04$\pm$0.02 & 25.48$\pm$0.25 & 1250$\pm$165 & 1,72-75,143 & --- & \pmin404$\pm$2\pz & 4,76,77 & D \\
Antlia      & 263.1 & +22.3  & 0.05$\pm$0.03 & 25.46$\pm$0.10 & 1235$\pm$65\pz & 78,79,158 & --- & \pmin361$\pm$2\pz & 80 \\
Leo~I        & 226.0 & +49.1  & 0.01$\pm$0.01 & 21.99$\pm$0.20 & \pz250$\pm$30\pz & 81,82 & \pmin286$\pm$2  & --- &  4,83,84,142 & C \\
Sextans~A   & 246.2 & +39.9  & 0.03$\pm$0.02 & 25.75$\pm$0.15 & 1440$\pm$110 & 71,86 & \pmin328$\pm$5 & \pmin325$\pm$3\pz &  4,49,87 \\
Sextans     & 243.5 & +42.3  & 0.03$\pm$0.01 & 19.67$\pm$0.08 & \pz\pz86$\pm$4\pz\pz & 88-90 & \pmin227$\pm$3  & --- &  44,91-93 & C \\
Leo~II      & 220.2 & +67.2  & 0.02$\pm$0.01 & 21.63$\pm$0.09 & \pz205$\pm$12\pz & 95-97 & \pmin\pz76$\pm$2 & ---&  4,98,142 & C \\
GR~8   & 310.7 & +77.0 & 0.02$\pm$0.02 & \pz25.9$\pm$0.4\pz & 1590$\pm$600 & 100-102,162 & --- & \pmin215$\pm$3\pz & 4,46,50,154 \\
Ursa~Minor  & 105.0 & +44.8  & 0.03$\pm$0.02 & 19.11$\pm$0.10 & \pz\pz66$\pm$3\pz\pz & 103,104 & $-$248$\pm$2 & --- &  105-106,142 & C \\
Draco       & \pz86.4  & +34.7  & 0.03$\pm$0.01 & 19.58$\pm$0.15 & \pz\pz82$\pm$6\pz\pz & 108,160 & $-$293$\pm$2 & --- &  105,109,142 & C \\
{\it Milky Way} & \pz\pz0.0 & \pmin\pz0.0 & --- & 14.52 & \pz\pz8 & 125 & --- & --- & & H \\
Sagittarius  & \pz\pz5.6   & $-$14.1 & 0.15$\pm$0.03 & 16.90$\pm$0.15 & \pz\pz24$\pm$2\pz\pz & 111-114,149 & \pmin140$\pm$5 & --- &  111,115,116 & C,F \\
SagDIG	    & \pz21.1  & $-$16.3 & 0.22$\pm$0.06 & \pz25.2$\pm$0.3\pz & 1060$\pm$160 & 117-119 & \pz$-$75$\pm$5 & \pz$-$79$\pm$2\pz &  4,119,144 \\
NGC~6822    & \pz25.3  & $-$18.4 & 0.26$\pm$0.04 & 23.45$\pm$0.15 & \pz490$\pm$40\pz & 1,2 & \pz$-$53$\pm$4 & \pz$-$54$\pm$6\pz &  4,49,121-123,136 & D \\
DDO~210     & \pz34.0  & $-$31.3 & 0.06$\pm$0.02 & \pz24.6$\pm$0.5\pz  & \pz800$\pm$250 & 120,124,125 & --- & $-$137$\pm$3\pz & 4,5,46 & E \\
IC~5152     & 343.9 & $-$50.2 & 0.01$\pm$0.02 & 26.01$\pm$0.25 & 1590$\pm$200 & 126,134 & --- & \pmin124$\pm$3\pz & 4,5 \\
Tucana      & 322.9 & $-$47.4 & 0.00$\pm$0.02 & 24.73$\pm$0.08 & \pz880$\pm$40\pz & 127-129 & --- & --- & 132 & G \\
UKS2323$-$326 & \pz11.9  & $-$70.9 & 0.07$\pm$0.03 & \pz25.6$\pm$0.5\pz & 1320$\pm$350 & 118 & --- & \pmin\pz62$\pm$6\pz & 119 \\
Pegasus     & \pz94.8  & $-$43.5 & 0.02$\pm$0.01 & 24.90$\pm$0.10 & \pz955$\pm$50\pz & 130,131,150 & --- & $-$182$\pm$2\pz & 4,5,46,144 & F \\
\\
\end{tabular}
\end{center}
\end{table}
\end{landscape}

\clearpage
\hsize=6.5truein
\vsize=10.5truein
\hoffset=-1.0truein
\voffset=-1.65truein

\noindent EXPLANATION OF COLUMNS OF TABLE 2:

\noindent {\it Column 1:} Galaxy name; names in {\it italics} are
`giant' galaxies included here for convenience but not discussed in
any detail in this review; {\it Columns 2 and 3:} Galactic longitude
and latitude, respectively; {\it Column 4:} Foreground reddening; {\it
Column 5:} True distance modulus; {\it Column 6:} Distance in kpc;
{\it Column 7:} Distance/reddening references; {\it Column 8:}
Heliocentric velocity based on optical measurements; {\it Column 9:}
Heliocentric velocity based on radio measurements; {\it Column 10:}
Velocity references; {\it Column 11:} Special notes.

\vskip1em

\noindent NOTES FOR TABLE 2:
\vskip0.5em

\noindent\hang A. HI, CO and optical velocities differ by small, but possibly
significant amounts.

\noindent\hang B. Ambiguous HI detection; the velocity of the most plausible
component is listed.  For Phoenix, one or both of two HI clouds could 
be associated with the galaxy (Carignan et al 1991).

\noindent\hang C. No HI detected.  See Table 5 for limits.

\noindent\hang D. Photographic spectra were obtained by Humason et al
(1956) from which they measured velocities of $-78$, 210, and 441 km
sec$^{-1}$ for WLM, NGC~55, and NGC~3109, respectively The typical
uncertainty in these results is approximately 50 km sec$^{-1}$.

\noindent\hang E. Large distance discrepancy or uncertainty; see text.

\noindent\hang F. All values refer to the central regions of the galaxy,
taken to coincide with the globular cluster M~54.

\noindent\hang G. A nearby HI cloud ($V_\odot \sim 130$ km sec$^{-1}$)
is observed about 15 arcmin from Tucana which may 
be associated with the Magellanic Stream (Oosterloo et al 1996).

\vskip1em

\noindent REFERENCES FOR TABLE 2: 
[1] Lee et al (1993c); 
[2] Gallart et al (1996c); 
[3] Minniti \&\ Zijlstra (1996); 
[4] Huchtmeier \&\ Richter (1986); 
[5] Richter et al (1987); 
[6] Heisler et al (1997);
[7] C\^ot\'e (1995); 
[8] Puche et al (1991); 
[9] Pritchett et al (1987); 
[10] Wilson et al (1996); 
[11] Saha et al (1996); 
[12] Sait\=o et al (1992); 
[13] Wilson \&\ Reid (1991); 
[14] Shostak \&\ Skillman (1989); 
[15] Han et al (1997); 
[16] Bender et al (1991); 
[17] Deleted in proof;
[18] Armandroff et al (1993);
[19] Thuan \&\ Martin (1979); 
[20] Lee et al (1993b); 
[21] Held et al (1992); 
[22] Young \&\ Lo (1997a); 
[23] Wiklind \&\ Rydbeck (1986); 
[24] Welch et al (1996); 
[25] Ford et al (1977); 
[26] Lee (1996); 
[27] van den Bergh (1995); 
[28] Deleted in proof;
[29] Carter \&\ Sadler (1990); 
[30] Held et al (1990); 
[31] Peterson \&\ Caldwell (1993); 
[32] Grillmair et al (1996); 
[33] Deleted in proof; 
[34] Dressler \&\ Richstone (1988); 
[35] Deleted in proof;
[36] Da Costa et al (1996); 
[37] Mould \&\ Kristian (1990); 
[38] Deleted in proof;
[39] Schweitzer et al (1995); 
[40] Da Costa (1984); 
[41] Queloz et al (1995); 
[42] Armandroff \&\ Da Costa (1986); 
[43] Deleted in proof;
[44] Carignan et al (1998);
[45] Lee (1995b); 
[46] Lo et al (1993); 
[47] Saha et al (1992a); 
[48] Huterer et al (1995); 
[49] Tomita et al (1993); 
[50] Hoffman et al (1996); 
[51] Lake \&\ Skillman (1989); 
[52] K\"onig et al (1993); 
[53] Ortolani \&\ Gratton (1988); 
[54] Carignan et al (1991); 
[55] van de Rydt et al (1991); 
[56] Beauchamp et al (1995); 
[57] Mateo et al (1991b); 
[59] Karachentsev \&\ Tikhonov (1993);
[60] Deleted in proof; 
[61] Hoessel et al (1988); 
[62] Deleted in proof; 
[63] Smecker-Hane et al (1994); 
[64] Hurley-Keller et al (1998); 
[65] Mighell (1997); 
[66] McNamara (1995);
[67] Mateo et al (1993); 
[68] Godwin \&\ Lynden-Bell (1987); 
[69] Mould et al (1990); 
[70] Hoessel et al (1994); 
[71] Piotto et al (1994); 
[72] Lee (1993); 
[73] Capaccioli et al (1992); 
[74] Musella et al (1997); 
[75] Richer \&\ McCall (1992); 
[76] Jobin \&\ Carignan (1990); 
[77] Carignan (1985); 
[78] Whiting et al (1997); 
[79] Aparicio et al (1997c); 
[80] Fouqu\'e et al (1990); 
[81] Lee et al (1993a); 
[82] Demers et al (1994a); 
[83] Mateo et al (1998c); 
[84] Zaritsky et al (1989); 
[85] Deleted in proof;
[86] Sakai et al (1996); 
[87] Skillman et al (1988); 
[88] Mateo et al (1995a); 
[89] Mateo et al (1991a); 
[90] Irwin et al (1990); 
[91] Hargreaves et al (1994a); 
[92] Suntzeff et al (1993); 
[93] Da Costa et al (1991); 
[94] Deleted in proof;
[95] Mighell \&\ Rich (1996); 
[96] Demers \&\ Irwin (1993); 
[97] Lee (1995b); 
[98] Vogt et al (1995); 
[99] Deleted in proof;
[100] Tolstoy et al (1995); 
[101] Wyatt \&\ Dufour (1993); 
[102] Aparicio et al (1988);
[103] Nemec et al (1988); 
[104] Olszewski \&\ Aaronson (1985); 
[105] Armandroff et al (1995); 
[106] Hargreaves et al (1994b); 
[107] Deleted in proof;
[108] Nemec (1985); 
[109] Hargreaves et al (1996b); 
[110] Deleted in proof;
[111] Ibata et al (1994); 
[112] Mateo et al (1995b); 
[113] Alard (1996); 
[114] Alcock et al (1997); 
[115] Ibata et al (1997); 
[116] Koribalski et al (1994);
[117] Longmore et al (1978); 
[118] Deleted in proof;
[119] Longmore et al (1982); 
[120] Marconi et al (1990); 
[121] Gottesman \&\ Wiliachew (1977); 
[122] Israel (1997); 
[123] Wilson (1994b); 
[124] Greggio et al (1993); 
[125] van den Bergh (1994a); 
[126] Caldwell et al (1988); 
[127] Saviane et al (1996); 
[128] Da Costa (1994a); 
[129] Castellani et al (1996); 
[130] Aparicio (1994); 
[131] Hoessel et al (1990); 
[132] Oosterloo et al (1996); 
[133] Richter et al (1991); 
[134] Burstein \&\ Heiles (1982); 
[135] Hodge \&\ Miller (1995); 
[136] Israel et al (1995); 
[137] Hummel et al (1986); 
[138] Sage \&\ Wrobel (1989); 
[139] Young \&\ Lo (1996a); 
[140] Deleted in proof; 
[141] Demers et al (1984);
[142] Knapp et al (1978);
[143] Davidge (1993);
[144] Young \&\ Lo (1997b);
[145] Kaluzny et al (1995);
[146] Karachentseva et al (1996);
[147] Karachentsev et al (1994);
[148] Mateo et al (1998b);
[149] Sarajedini \&\ Layden (1995);
[150] Aparicio et al (1997b);
[151] Aparicio et al (1997a);
[152] Mould (1997);
[153] Tikhonov \&\ Makarova (1996);
[154] Carignan et al (1990);
[155] Cook et al, private communication;
[156] Lu et al (1993);
[157] Nolthenius \&\ Ford (1986);
[158] Sarajedini et al (1997);
[159] Karachentsev \&\ Makarov (1996);
[160] Grillmair et al (1998);
[161] Tolstoy et al (1998);
[162] Dohm-Palmer et al (1998).


\setcounter{table}{2}
\begin{landscape}
\begin{table}
\noindent \ \ Table 3: Integrated Photometric Properties and Structural Parameters of Local Group Dwarf Galaxies
\begin{center}
\begin{tabular}{lcclccccccll}
\ Galaxy & V$_T$ & (B$-$V)$_T$ & \ \ \ \ \ \ Other Colors & $\Sigma_{0}$ & $r$ & $R$ & PA & $e$ & $r_{exp}$ & \ \ \ \ \ \ Ref & Notes \\
       & mag & mag & \ \ \ \ \ \ \ mag [bands] & mag arcsec$^{-2}$ & arcmin & arcmin & deg &  & arcmin & & \\
\ \ \ (1) & (2) & (3) & \ \ \ \ \ \ \ \ \ \ \ \ \ (4) & (5) & (6) & (7) & (8) & (9) & (10) & \ \ \ \ \ \ (11) & \ \ (12) \\
\\
%
WLM         & 10.42$\pm$0.2 & 0.62$\pm$0.12 & $-$0.21 [1] & 20.36$\pm$0.05 & {\it 2.2} & {\it 5.5} & 175$\pm$5\pz & 0.59$\pm$0.04 & 3.3$\pm$0.2 & 57,61 & \\
NGC~55  & \pz7.95$\pm$0.15 & 0.50$\pm$0.04 & 0.87 [3]; 0.77 [4]; 0.42 [5]; & 20.8$\pm$0.5\pz & {\it 4.6} & {\it 20.2} & 105$\pm$5\pz & 0.77 & 3.4$\pm$0.2 & 1,2,4,60,73 & \\
            &         &      & $-$0.05$\pm$0.2 [6] & & & & & & & & \\
IC~10       & 11.55$\pm$0.3\pz & 1.37 & 0.28 [1] & 22.1$\pm$0.4 & --- & 5.0$\pm$2.0 & 140$\pm$20 & 0.3$\pm$0.1 & 1.9$\pm$0.3 & 1,67,70,71,75 & \\
NGC~147     & \pz9.35$\pm$0.15 & 0.92$\pm$0.07 & 0.55 [2]; 1.12 [3] & 21.6$\pm$0.2 & 1.1 & 20 & \pz34$\pm$4\pz & 0.46$\pm$0.03 & 2.1$\pm$0.3 & 1,5-8,11,82 & \\
And~III     & 14.21$\pm$0.15 & --- & \ \ \ \ \ \ \ \ \ \ \ \ \ \ \ \ \ --- & 24.49$\pm$0.05 &  1.27$\pm$0.17 & 6.2$\pm$1.0 & 135$\pm$5\pz & 0.6 & 0.75$\pm$0.01 & 5,8 & \\
NGC~185     & \pz9.09$\pm$0.15 & 0.96$\pm$0.06 & 0.30$\pm$0.05 [1]; 0.60 [2]; & 20.1$\pm$0.4 & 1.0$\pm$0.3 & 16$\pm$2 & \pz41$\pm$5\pz & 0.26 & 1.7$\pm$0.3 & 1,5,6,8-11,74,82 & \\
            &         &      & 1.25 [3]; 2.06 [6] & & & & & & & & \\
NGC~205     & \pz8.05$\pm$0.15 & 0.70$\pm$0.10 & 0.34 [1]; 0.52 [2]; 1.23 [3]; & 20.4$\pm$0.4 & 1.5$\pm$0.5 & 6.2$\pm$2.0 & 165$\pm$15 & 0.46$\pm$0.03 & 1.7 & 1,5,6,8,63-65,79,82 & A,B \\
       &  &  & 3.08 [5] \\
M32         & \pz8.10$\pm$0.15 & 0.99$\pm$0.05 & 0.64 [1]; 0.42 [2]; 1.07 [3]; & $<$11.6 & --- &  9$\pm$2 & 165$\pm$5\pz & 0.18$\pm$0.03 & --- & 1,6,12-14,65,76 & B,C \\
            &         &      & 3.06 [5]; 0.64 [6]; 0.85 [7]  & & & & & & & & \\
And~I       & 12.75$\pm$0.2\pz & 0.75$\pm$0.06 & 0.29 [1] & 24.37$\pm$0.01 & 1.58$\pm$0.08 & 13.4$\pm$1.4 & --- & 0.0 & 1.46$\pm$0.06 & 1,5,47 & \\
Sculptor    & \pz8.5$\pm$0.3 & --- & \ \ \ \ \ \ \ \ \ \ \ \ \ \ \ --- & 23.7$\pm$0.4 & 5.8$\pm$1.6 & 76.5$\pm$5.0 & \pz99$\pm$1\pz & 0.32$\pm$0.03 & 6.7$\pm$0.2 & 5,8,20,77 & \\
LGS~3       & 14.26$\pm$0.15 & 0.74$\pm$0.06 & 0.47 [2]; 1.00 [3] & 24.7$\pm$0.2 & 0.82$\pm$0.05 & 14.5$\pm$4.5 & 175$\pm$5\pz & 0.26$\pm$0.06 & 0.78 & 16-18,78 & \\
IC~1613     & \pz9.59$\pm$0.15 & 0.60$\pm$0.10 & $-$0.25 [1] & 22.8$\pm$0.3 & 3.3$\pm$1.0 & 11$\pm$3 & \pz83$\pm$6\pz & 0.24$\pm$0.06 & 5.4$\pm$3.0 & 48,49,75 & \\
And~II      & 12.7$\pm$0.2 & --- & \ \ \ \ \ \ \ \ \ \ \ \ \ \ \ --- & 24.47$\pm$0.05 & 1.64$\pm$0.08 & 17.2$\pm$1.0 & --- & 0.3 & 1.57$\pm$0.03 & 5,8 & \\
Phoenix     & 13.2$\pm$0.2 & 0.61$\pm$0.05 & $-$0.21 [1] & --- & --- & $>$8.6 & 160$\pm$10 & 0.3$\pm$0.1 & --- & 1,19 & \\
Fornax      & \pz7.6$\pm$0.3 & 0.63$\pm$0.05 & 0.08 [1]; 0.45 [2]; 1.02 [3] & 23.4$\pm$0.3 & 13.8$\pm$0.8 & \pz71$\pm$4\pz & \pz48$\pm$6\pz & 0.31$\pm$0.03 & 10.2 & 5,20-22,24,50 & \\
EGB~0427+63  & 13.88$\pm$0.12 & 1.34$\pm$0.05 & \ \ \ \ \ \ \ \ \ \ \ \ \ \ \ --- & 23.9$\pm$0.3 & {\it 0.5} & {\it 1.0} & 73$\pm$5 & 0.55$\pm$0.05 & --- & 16,72 & D \\
Carina      & 10.85$\pm$0.25 & --- & \ \ \ \ \ \ \ \ \ \ \ \ \ \ \ --- & 25.5$\pm$0.4 & 8.8$\pm$1.2 & 28.8$\pm$3.6 & \pz65$\pm$5 & 0.33$\pm$0.05 & 5.5 & 5,20,26,83 & \\
Leo~A       & 12.8$\pm$0.2 & 0.15$\pm$0.2 & $-$0.2$\pm$0.1 [1] & --- & {\it 2.3} & {\it 3.5} & 94$\pm$5 & 0.36 & --- & 1,2,57,59 & \\
Sextans~B   & 11.43$\pm$0.15 & 0.48 & $-$0.16 [1] & --- & {\it 3.0} & {\it 3.9} & 130$\pm$15 & 0.23 & --- & 2,42,57,58 & \\
NGC~3109    & \pz9.88$\pm$0.15 & 0.52 & $-$0.12 [1] & {\it 23.6$\pm$0.2} & {\it 2.8} & {\it 13.3} & \pz92$\pm$1\pz & 0.80 & 3.1 & 1,53,84 & \\
Antlia      & 14.8$\pm$0.2 & --- & 0.5$\pm$0.1 [2]; 1.2$\pm$0.2 [3] & 24.3$\pm$0.2 & 0.80$\pm$0.05 & \pz5.2$\pm$0.2 & 145$\pm$5\pz & 0.35$\pm$0.03 & 1.1,0.3 & 28,29 & E \\
Leo~I       & 10.1$\pm$0.3 & 0.8$\pm$0.2 & 0.15 [1] & 22.4$\pm$0.3 & 3.3$\pm$0.3 & 12.6$\pm$1.5 & \pz79$\pm$3\pz & 0.21$\pm$0.03 & 1.8$\pm$0.2 & 5,8,20 & \\
Sextans~A   & 11.30$\pm$0.15 & 0.38$\pm$0.07 & $-$0.32 [1] & 23.5$\pm$0.3 & {\it 3.2} & {\it 4.0} & 52$\pm$5 & 0.21 & 1.5$\pm$0.3 & 1,2,75 & \\
Sextans     & 10.3$\pm$0.3 & --- & \ \ \ \ \ \ \ \ \ \ \ \ \ \ \ --- & 26.2$\pm$0.5 & 16.6$\pm$1.2\pz & 160$\pm$50 & \pz56$\pm$5\pz & 0.35$\pm$0.05 & 12.3$\pm$3.0\pz & 5,8,20,30 & \\
Leo~II      & 12.0$\pm$0.2 & 0.65$\pm$0.15 & \ \ \ \ \ \ \ \ \ \ \ \ \ \ \ --- & 24.0$\pm$0.3 & 2.9$\pm$0.6 & 8.7$\pm$0.9 & \pz12$\pm$10 & 0.13$\pm$0.05 & 1.5 & 1,20,32,33 & \\
GR~8   & 14.40$\pm$0.15 & 0.37$\pm$0.05 & $-$0.51 [1]; 0.39 [2] & 22.3$\pm$0.2 & {\it 0.7} & {\it 1.0} & \pz47$\pm$7\pz & 0.31$\pm$0.04 & 0.24$\pm$0.02 & 1,8,35,55-57 & \\
Ursa~Minor  & 10.3$\pm$0.4 & 1.3$\pm$0.3 & $-$0.1$\pm$0.3 [1] & 25.5$\pm$0.5 & 15.8$\pm$1.2\pz & 50.6$\pm$3.6 & \pz53$\pm$5\pz & 0.56$\pm$0.05 & \pz8.0$\pm$2.5\pz & 1,5,20,42,64 & \\
Draco       & 10.9$\pm$0.3 & 0.95$\pm$0.2 & 0.1$\pm$0.3 [1] & 25.3$\pm$0.5 & 9.0$\pm$0.7 & 28.3$\pm$2.4 & \pz82$\pm$1\pz & 0.29$\pm$0.01 & 4.5 & 20,42 & \\
Sagittarius & \pz4.0$\pm$0.5 & --- & \ \ \ \ \ \ \ \ \ \ \ \ \ \ \ --- & 25.4$\pm$0.3 & --- & $>$10$^\circ$ & 120$\pm$10 & 0.80$\pm$0.15 & --- & 38-40,68,69 & \\
SagDIG	    & 13.5$\pm$0.4 & 0.4$\pm$0.2 & \ \ \ \ \ \ \ \ \ \ \ \ \ \ \ --- & 24.4$\pm$0.3 & {\it 0.9} & {\it 1.7} & \pz90$\pm$10 & 0.47$\pm$0.10 & --- & 41,54 & \\
NGC~6822    & \pz9.1$\pm$0.2 & 0.73$\pm$0.05 & 0.04$\pm$0.20 [1] & 21.4$\pm$0.2 & 2.5$\pm$1.0 & 40$\pm$10 & 10$\pm$5 & 0.47$\pm$0.03 & \pz2.4$\pm$0.4 & 1,80-82 & D \\
DDO~210     & 14.71$\pm$0.15 & 0.15 & $-$0.18 [1]; 1.11 [4] & --- & {\it 0.9} & {\it 1.6} & 100$\pm$10 & 0.44 & --- & 1,2,57,58 & \\
IC~5152     & 11.2$\pm$0.2 & 0.34$\pm$0.1 & $-$0.16 [1] & --- & {\it 2.3} & {\it 2.8} & \pz95$\pm$10 & 0.18 & --- & 1,42,62 & \\
Tucana      & 15.15$\pm$0.2\pz & 0.7$\pm$0.1 & 0.55 [2] & 25.05$\pm$0.06 & 0.70$\pm$0.10 & 3.7$\pm$1.2 & \pz97$\pm$2\pz & 0.48$\pm$0.03 & 0.49$\pm$0.08 & 43,44 & \\
UKS2323$-$326 & 13.8$\pm$0.4 & 0.4$\pm$0.2 & \ \ \ \ \ \ \ \ \ \ \ \ \ \ \ --- & 24.6$\pm$0.5 & {\it 1.1} & {\it 1.2} & 135$\pm$25 & 0.05$\pm$0.03 & --- & 41,54 & \\
Pegasus     & 12.04$\pm$0.15 & 0.61 & 0.06$\pm$0.06 [1] & --- & {\it 2.3} & {\it 3.9} & 125$\pm$15 & 0.40 & --- & 1,2,57,58 & \\
\\
\end{tabular}
\end{center}
\end{table}
\end{landscape}

\clearpage
\hsize=6.5truein
\vsize=10.2truein
\hoffset=-1.0truein
\voffset=-0.6truein

\noindent EXPLANATION OF COLUMNS OF TABLE 3:
\vskip0.5em

\noindent {\it Column 1:} Galaxy name; {\it Column 2:} Integrated
apparent V-band magnitude; {\it Column 3:} Integrated (B$-$V) color;
{\it Column 4:} Other integrated colors.  [1] = (U--B), [2] = (V--R),
[3] = (V--I), [4] = (B--R), [5] = (V--K), [6] = (J--H), [7] = (H--K);
{\it Column 5:} V-band central surface brightness; values in {\it
italics} are B-band results; {\it Column 6:} Roman type: core radius
in arcmin; {\it Italic type:} semi-minor axis dimension to the
Holmberg limit of $\Sigma_{0,B} = 26.5$ mag arcsec$^{-2}$; {\it Column
7:} Roman type: tidal radius in arcmin; {\it Italic type:} semi-major
axis dimension to the Holmberg limit as for column 6; {\it Column 8:}
Major-axis position angle, with N = 0$^\circ$ and E = 90$^\circ$; {\it
Column 9:} The ellipticity of the outer parts of the galaxy defined as
$e = (1 - b/a)$, where $b$ = minor axis and $a$ = major axis; {\it
Column 10:} The exponential scale length of the surface-brightness
distribution, typically along the major axis; {\it Column 11:}
References; {\it Column 12:} Special notes.

\vskip2em

\noindent NOTES FOR TABLE 3:
\vskip0.5em

\noindent\hang A. The major axis position angle 
varies significantly outward from the galaxy center.

\noindent\hang B. $\Sigma_0$ is determined from the extrapolation of
the outer surface brightness profile to $r = 0$.  For NGC~205 $r_c$ is
estimated from surface-brightness profile as the location where the
central surface brightness drops a factor of two below the level
listed in this table.  Both M32 and NGC~205 exhibit 
composite surface brightness
profiles; no single King model, exponential profile or power-law
profile can fit the observed profiles at all radii.

\noindent\hang C. The outer isophotes of M32 
are {\it not} truncated according to Kent (1987).

\noindent\hang D. The PA and $e$ values refer to the inner bar structure.

\noindent\hang E. Aparicio et al (1997a) find the radial surface brightness
profile is best fit with a two exponential components.  The values of
$r_{exp}$ refer to the inner and outer components, respectively.  The two
exponential compnents meet at a radial distance of 40 arcsec from the center
of the galaxy.

\vskip2em

\noindent REFERENCES FOR TABLE 3:
[1]  Longo \&\ de Vaucouleurs (1983);  
[2]  de Vaucouleurs et al (1991);  
[3]  Deleted in proof; 
[4]  Pierce \&\ Tully (1992);   
[5]  Caldwell et al (1992);  
[6]  Kent (1987);  
[7]  Hodge (1976);  
[8]  de Vaucouleurs et al (1981);  
[9]  Price (1985);  
[10]  Hodge (1963b);  
[11]  Buta \&\ Williams (1995);  
[12]  Lugger et al (1992);  
[13]  Michard \&\ Nieto (1991);  
[14]  Silva \&\ Elston (1994);  
[15]  Deleted in proof; 
[16]  Karachentseva et al (1996);  
[17]  Lee (1995a);  
[18]  Schild (1980);  
[19]  van de Rydt et al (1991);  
[20]  Irwin \&\ Hatzidimitriou (1995);  
[21]  Hodge \&\ Smith (1974); 
[22]  de Vaucouleurs \&\ Ables (1968);
[23]  Deleted in proof; 
[24]  Poulain \&\ Nieto (1994);  
[25]  Deleted in proof;
[26]  Mateo et al (1993);  
[27]  Deleted in proof; 
[28]  Aparicio et al (1997a);
[29]  Whiting et al (1997);  
[30]  Mateo et al (1991a);  
[31]  Deleted in proof; 
[32]  Hodge (1982);  
[33]  Vogt et al (1995);  
[34]  Deleted in proof; 
[35]  Hopp \&\ Schulte-Ladbeck (1995);  
[36]  Deleted in proof; 
[37]  Deleted in proof; 
[38]  Ibata et al (1994);  
[39]  Mateo et al (1995c);  
[40]  Ibata et al (1997);  
[41]  Longmore et al (1982);  
[42]  Longo \&\ de Vaucouleurs (1985);  
[43]  Saviane et al (1996);  
[44]  Lavery \&\ Mighell (1992);  
[45]  Deleted in proof; 
[46]  Deleted in proof; 
[47]  Mould \&\ Kristian (1990);  
[48]  Hodge et al (1991a);  
[49]  Hodge (1978);  
[50]  Demers et al (1994b);
[51]  Deleted in proof; 
[52]  Deleted in proof; 
[53]  Carignan (1985);  
[54]  Longmore et al (1978);  
[55]  Carignan et al (1990); 
[56]  de Vaucouleurs \&\ Moss (1983);  
[57]  Fisher \&\  Tully (1975);
[58]  Fisher \&\  Tully (1979);
[59]  Allsopp (1978);
[60]  Puche et al (1991);
[61]  Ables \&\  Ables (1977);
[62]  S\'ersic \&\  Cerruti (1979);
[63]  Hodge (1973);
[64]  Lee (1996);
[65]  Peletier (1993);
[66]  Deleted in proof;
[67]  de Vaucouleurs \&\  Freeman (1972);
[68]  Mateo et al (1996);
[69]  Fahlman et al (1996);
[70]  Shostak (1974);
[71]  de Vaucouleurs \&\  Ables (1965);
[72]  Hoessel et al (1988);
[73]  Fitzgibbons (1990);
[74]  Lee (1993);
[75]  Ables (1971);
[76]  Burstein et al (1987);
[77]  Hodge (1966);
[78]  Tikhonov \&\ Makarova (1996);
[79] Price \&\ Grasdalen (1983);
[80] Hodge (1977);
[81] Hodge et al (1991b);
[82] Kodaira et al (1990);
[83] Demers et al (1983);
[84] Jobin \&\ Carignan (1990);
[85] Hodge (1964).

\vskip1em


\setcounter{table}{2}
\begin{landscape}
\begin{table}
\noindent \ \ Table 4: Derived Photometric and Kinematic Properties of Local Group Dwarf Galaxies
\begin{center}
\begin{tabular}{lccccccccccccl}
Galaxy & $M_V$ & $M_B$ & $L_V$ & $R_c$ & $\rho_0$ & $I_0$ & $M_{tot}$ & $(M/L)_{0,V}$ & $(M/L)_{tot,V}$ & $M_{HI}/M_{tot}$ & $M_{HI}/L_B$ & $(v_r/\sigma)^*$ & Notes \\
  & mag & mag & $10^6$\Lsol & pc & \Msol pc$^{-3}$ & \Lsol pc$^{-3}$ & $10^6$\Msol & Solar & Solar &  & \Msol/\Lsol & & \\
\ \ \ (1) & (2) & (3) & (4) & (5) & (6) & (7) & (8) & (9) & (10) & (11) & (12) & (13) & \ (14) \\
\\
WLM & $-$14.5 & $-$13.9 &  \pz\pz50.2\pz &  \pz710 &  (0.46)\pz &  0.19\pz &  \pz\pz150 &  --- &  \pz3.0 &  \pz0.40\pz &  \pz1.2\pz\pz &  (2.8)\pz & D,F \\
NGC~55 & $-$18.0 & $-$17.5 &  1290\pdec\pz &  \pz875 &  (0.26)\pz &  0.10\pz &  15600 &  --- &  12\pdec &  \pz0.09\pz &  \pz0.94\pz &  (10.8)\pz\pz & D,E,F \\
IC~10 & $-$15.7 & $-$15.2 &  \pz160\pdec\pz &  \pz475 &  0.047 &  0.63\pz &  \pz1580 &  \pz0.1 &  \pz9.9 &  \pz0.10\pz &  \pz0.86\pz &  \pz5.8\pz\pz &  \\
NGC~147 & $-$15.5 & $-$14.8 &  \pz131\pdec\pz &  \pz170 &  2.8\pz\pz &  0.39\pz &  \pz\pz110 &  \pz7.1 &  \pz0.8 &  $<$0.001 &  $<$0.001 &  \pz0.32\pz &  \\
And~III & $-$10.3 & \pz$-$9.7 &  \pz\pz\pz1.13 &  \pz180 &  (0.044) &  0.018 &  --- &  --- &  --- &  --- &  $<$0.07\pz &  --- & F \\
NGC~185 & $-$15.5 & $-$14.7 &  \pz125\pdec\pz &  \pz155 &  4.3\pz\pz &   1.76\pz &  \pz\pz130 &  \pz2.5 &  \pz1.0 &  \pz0.001 &  \pz0.001 &  \pz0.08\pz &  \\
NGC~205 & $-$16.6 & $-$16.0 &  \pz366\pdec\pz &  \pz260 &  5.1\pz\pz &  0.43\pz &  \pz\pz740 &  12\pdec &  \pz2.0 &  \pz0.001 &  \pz0.001 &  \pz0.04\pz &  \\
M32 & $-$16.7 & $-$15.8 &  \pz383\pdec\pz &  \pz635 &  1.0\pz\pz &  786 &  \pz2120 &  \pz0.0 &  \pz5.6 &  $<$0.001 &  $<$0.009 &  \pz0.51\pz & A \\
And~I & $-$11.9 & $-$11.2 &  \pz\pz\pz4.71 &  \pz375 &  (0.023) &  0.009 &  --- &  --- &  --- &  --- &  $<$0.02\pz &  --- & F \\
Sculptor & $-$11.1 & $-$10.4 &  \pz\pz\pz2.15 &  \pz110 &  0.60\pz &  0.055 &  \pz\pz6.4 &  11\pdec &  \pz3.0 &  \pz0.004 &  \pz0.01\pz &  --- &  \\
LGS~3 & $-$10.5 & \pz$-$9.9 &  \pz\pz\pz1.33 &  \pz160 &  0.37\pz &  0.018 &  \pz\pz\pz13 &  21\pdec &  \pz9.7 &  \pz0.03\pz &  \pz0.33\pz &  --- &  \\
IC~1613 & $-$14.7 & $-$14.2 &  \pz\pz63.6\pz &  \pz585 &  0.035 &  0.025 &  \pz\pz795 &  \pz1.4 &  12\pdec &  \pz0.07\pz &  \pz0.81\pz &  \pz4.3\pz\pz &  \\
And~II & $-$11.1 & $-$10.5 &  \pz\pz\pz2.35 &  \pz205 &  (0.043) &  0.017 &  --- &  --- &  --- &  --- &  --- &  --- & F \\
Phoenix & $-$10.1 & \pz$-$9.5 &  \pz\pz\pz0.90 &  \pz310 &  0.14\pz &  --- &  \pz\pz\pz33 &  --- &  37\pdec &  \pz0.006 &  \pz0.21\pz &  --- &  \\
Fornax & $-$13.2 & $-$12.6 &  \pz\pz15.5\pz &  \pz460 &  0.086 &  0.018 &  \pz\pz\pz68 &  \pz4.8 &  \pz4.4 &  $<$0.001 &  $<$0.001 &  --- &  \\
EGB~0427+63 & $-$12.6 & $-$11.6 &  \pz\pz\pz9.12 &  \pz\pz85 &  (0.33)\pz &  0.13\pz &  --- &  --- &  --- &  --- &  \pz2.6\pz\pz &  (3.7)\pz & D,F \\
Carina & \pz$-$9.3 & \pz$-$8.6 &  \pz\pz\pz0.43 &  \pz210 &  0.17\pz &  0.006 &  \pz\pz\pz13 &  30\pdec &  31\pdec &  $<$0.001 &  $<$0.002 &  --- &  \\
Leo~A & $-$11.4 & $-$11.3 &  \pz\pz\pz3.03 &  \pz185 &  0.20\pz &  --- &  \pz\pz\pz11 &  --- &  \pz3.5 &  \pz0.72\pz &  \pz1.6\pz\pz &  --- &  \\
Sextans~B & $-$14.2 & $-$13.8 &  \pz\pz40.7\pz &  \pz445 &  0.27\pz &  --- &  \pz\pz885 &  --- &  22\pdec &  \pz0.05\pz &  \pz0.96\pz &  \pz2.1\pz\pz &  \\
NGC~3109 & $-$15.7 & $-$15.2 &  \pz160\pdec\pz &  \pz630 &  0.042 &  0.018 &  \pz6550 &  \pz2.4 &  41\pdec &  \pz0.11\pz &  \pz3.8\pz\pz &  \pz6.8\pz\pz &  \\
Antlia & $-$10.8 & $-$10.2 &  \pz\pz\pz1.73 &  \pz230 &  0.12\pz &  0.016 &  \pz\pz\pz12 &  \pz7.4 &  \pz7.1 &  \pz0.08\pz &  \pz0.58\pz &  --- &  \\
Leo~I & $-$11.9 & $-$11.1 &  \pz\pz\pz4.79 &  \pz215 &  0.28\pz &  0.092 &  \pz\pz\pz22 &  \pz3.1 &  \pz4.6 &  $<$0.001 &  $<$0.007 &  --- &  \\
Sextans~A & $-$14.6 & $-$14.2 &  \pz\pz55.7\pz &  \pz700 &  0.022 &  0.011 &  \pz\pz395 &  \pz2.0 &  \pz7.1 &  \pz0.20\pz &  \pz1.1\pz\pz &  \pz4.1\pz\pz &  \\
Sextans & \pz$-$9.5 & \pz$-$8.8 &  \pz\pz\pz0.50 &  \pz335 &  0.065 &  0.002 &  \pz\pz\pz19 &  34\pdec &  39\pdec &  $<$0.001 &  $<$0.001 &  --- &  \\
Leo~II & \pz$-$9.6 & \pz$-$9.0 &  \pz\pz\pz0.58 &  \pz160 &  0.29\pz &  0.029 &  \pz\pz9.7 &  10\pdec &  17\pdec &  $<$0.001 &  $<$0.02\pz &  --- &  \\
GR~8 & $-$11.6 & $-$11.2 &  \pz\pz\pz3.43 &  \pz110 &  1.7\pz\pz &  0.20\pz &  \pz\pz7.6 &  \pz8.3 &  \pz2.2 &  \pz0.59\pz &  \pz1.0\pz\pz &  \pz1.1\pz\pz &  \\
Ursa~Minor & \pz$-$8.9 & \pz$-$7.6 &  \pz\pz\pz0.29 &  \pz200 &  0.35\pz &  0.006 &  \pz\pz\pz23 &  60\pdec &  79\pdec &  $<$0.002 &  $<$0.25\pz &  \pz0.48\pz &  \\
Draco & \pz$-$8.8 & \pz$-$7.8 &  \pz\pz\pz0.26 &  \pz180 &  0.46\pz &  0.008 &  \pz\pz\pz22 &  58\pdec &  84\pdec &  $<$0.001 &  $<$0.02\pz &  --- &  \\
Sagittarius & $-$13.4 & $-$12.8 &  \pz\pz18.1\pz &  \pz550 &  0.030 &  --- &  --- &  22\pdec &  52\pdec &  $<$0.001 &  $<$0.001 &  $<$0.18 & C \\
SagDIG & $-$12.3 & $-$12.1 &  \pz\pz\pz6.85 &  \pz125 &  0.58\pz &  0.044 &  \pz\pz9.6 &  13\pdec &  \pz1.4 &  \pz9.2\pz\pz &  \pz8.6\pz\pz &  --- & B \\
NGC~6822 & $-$15.2 & $-$14.7 &  \pz\pz94.4\pz &  \pz260 &  (0.97)\pz &  0.39\pz &  \pz1640 &  --- &  17\pdec &  \pz0.08\pz &  \pz1.2\pz\pz &  (6.4)\pz & D,F \\
DDO~210 & $-$10.0 & \pz$-$9.9 &  \pz\pz\pz0.81 &  \pz\pz95 &  0.84\pz &  --- &  \pz\pz5.4 &  --- &  \pz6.7 &  \pz0.35\pz &  \pz1.4\pz\pz &  --- &  \\
IC~5152 & $-$14.8 & $-$14.5 &  \pz\pz70.3\pz &  \pz390 &  0.000 &  --- &  \pz\pz400 &  --- &  \pz5.7 &  \pz0.15\pz &  \pz0.64\pz &  (4.7)\pz & D \\
Tucana & \pz$-$9.6 & \pz$-$8.9 &  \pz\pz\pz0.55 &  \pz130 &  (0.032) &  0.013 &  --- &  --- &  --- &  --- &  $<$0.18\pz &  --- & F \\
UKS2323-326 & $-$12.0 & $-$11.7 &  \pz\pz\pz5.25 &  \pz150 &  (0.051) &  0.020 &  --- &  --- &  --- &  --- &  \pz0.90\pz &  --- & F \\
Pegasus & $-$12.9 & $-$12.3 &  \pz\pz12.0\pz &  \pz280 &  0.16\pz &  --- &  \pz\pz\pz58 &  --- &  \pz4.8 &  \pz0.09\pz &  \pz0.44\pz &  \pz1.7\pz\pz &  \\
\\
\end{tabular}
\end{center}
\end{table}
\end{landscape}

\clearpage
\hsize=6.5truein
\vsize=10.2truein
\hoffset=-1.0truein
\voffset=-0.6truein

\noindent EXPLANATION OF COLUMNS OF TABLE 4:
\vskip0.5em

\noindent {\it Column 1:} Galaxy name; {\it Column 2:} Integrated
V-band absolute magnitude; {\it Column 3:} Integrated B-band absolute
magnitude; {\it Column 4:} Visual luminosity in units of $10^6$\Lsol;
{\it Column 5:} `Core' radius in parsecs, corresponding to the
observed core radius, $r_c$, when available, or 1.25$\times r_{exp}$
if $r_c$ is not measured but the exponential scale length is measured
(compare with Bender et al 1991; the average of $r_c/r_{exp}$ for the
18 galaxies in Table 3 with both radii is 1.27$\pm$0.12), or $a/3$ if
neither $r_c$ nor $r_{exp}$ is available, where $a$ is the observed
Holmberg semi-major axis as defined in Table 3; {\it Column 6:} The
central mass density in \Msol pc$^{-3}$, here approximated as $\rho_0 =
166 \sigma_0^2/R_c$, where $\sigma_0$ is the central velocity
dispersion in km s$^{-1}$, and $R_c$ is in pc; this is computed only for
systems that are pressure supported (see Mateo et al 1991b for
details).  For rotating Local Group dwarfs, the central mass density
is dominated by the visible material and has been approximated as $2.5
I_0$, where $I_0$ is defined in the explanation of column 7 (see Note
F); {\it Column 7:} The central luminosity density in \Lsol pc$^{-3}$,
taken as $I_0 = S_0/2 r_c$, where $S_0$ is the central surface
brightness expressed in units of $L_\odot$ pc$^{-2}$, and $r_c$ is in pc
(see Mateo et al 1991b for details); {\it Column 8:} The total mass;
if a central velocity dispersion is known and exceeds the rotational
velocity, then $M_{tot} = 167 \beta R_c \sigma_0^2$, where $\beta$ is
a scaling factor for King profiles taken to be 8.0 here, appropriate
for low-concentration King models.  If $v_{rot} > \sigma_0$, then
$M_{tot} = R_{rot} v_{rot}^2/G$, where $v_{rot}$ is the rotational
velocity -- corrected for the galaxy inclination -- at the projected
distance $R_{rot}$ from the galaxy center.  {\it Column 9:} The
central V-band mass-to-light ratio, defined as $\rho_0/I_0$, in solar
units; {\it Column 10:} The integrated V-band mass-to-light ratio
defined as $M_{tot}/L_V$, in solar units; {\it Column 11:} The ratio
of integrated H~I mass to the total mass, $M_{tot}$ from column 8;
{\it Column 12:} The ratio of the integrated H~I mass and the {\it
blue} luminosity, in solar B-band units; {\it Column 13:} The
parameter $(v_{rot}/\sigma)^*$ is simply $(v_{rot}/\sigma_0)$ for dIrr
systems, or $(v_{rot}/\sigma_0)(e/(1-e))^{-1/2}$, where $e$ is the
ellipticity, for ellipsoidal systems (Bender et al 1991).  See Note D
for details about values in parentheses; {\it Column 14:} Special
notes.

\vskip2em

\noindent NOTES FOR TABLE 4:
\vskip0.5em

\noindent\hang A. Due to the possible presence of a massive central
black hole, the nuclear kinematics are too complex for the simple
analysis used here.  See Kormendy and Richstone (1995) for details.

\noindent\hang B. $M_{HI}/M_{tot} \geq 1.0$.  This indicates an error
in the distance ($M_{HI}/M_{tot} \propto D$, where $D$ is the
distance), a mis-interpretation of or error in the kinematic data, an
error in the H~I flux measurement, or some combination of these.

\noindent\hang C. Sagittarius probably violates the assumption of
equilibrium that is implicitly adopted in the analysis used here and
its structural parameters remain highly uncertain. The mass and
$M/L$ ratios given here are taken directly from Ibata et al (1997).  
$R_c$ and $\rho_0$ are also taken from this source.

\noindent\hang D. The ISM velocity dispersion of these galaxies is assumed
to be 8 km s$^{-1}$ (see Table 7).

\noindent\hang E. The results given here for NGC~55 are especially
uncertain because of internal reddening which affects the inferred
total luminosity and central luminosity density.  See Puche et al (1991)
for a complete discussion and for more precise estimates of the
luminosity and mass of this galaxy.  Note that these authors assumed a
distance of 1.6 Mpc for NGC~55, in contrast to the distance of
1.48$\pm$0.15 Mpc used here (Table~2).

\noindent\hang F. The central mass density has not been determined
kinematically for these galaxies.  Instead, we assume $\rho_0 = 2.5 I_0$.



\setcounter{table}{0}
\begin{table}
\noindent \ \ Table 6: Heavy Element Abundances of Local Group Dwarf Galaxies
\begin{center}
\begin{tabular}{lcclccll}
\\
Galaxy & [Fe/H] & $\sigma_{\rm [Fe/H]}$ & \ \ Ref & $12 + \log ({\rm O/H})$ & [N/O] & \ Ref & Notes \\
 & dex & dex & & dex & dex & \\
\ \ \ (1) & (2) & (3) & \ \ (4) & (5) & (6) & \ (7) & \ \ (8) \\
\\
WLM        & $-$1.5$\pm$0.2 & --- & 1,2 & 7.75$\pm$0.2 & $-$1.46$\pm$0.15 & 3,4 & \\
NGC~55     &  --- & --- & --- & 8.32$\pm$0.15 & $-$1.44$\pm$0.15 & 67,75,76 & \\
IC~10      &  --- & --- & --- & 8.19$\pm$0.15 & $-$1.37$\pm$0.12 & 5,74 & A \\
NGC~147    & $-$1.1$\pm$0.2 & 0.4$\pm$0.1 & 7-9 & --- & --- & --- & B,C,H \\
And III    & $-$2.0$\pm$0.2 & $\leq$0.2$\pm$0.04 & 10 & --- & --- & --- & \\
NGC~185    & $-$1.22$\pm$0.15 & 0.4$\pm$0.1 & 11 & 8.2$\pm$0.2 & --- & 6 & F,H \\
NGC~205    & $-$0.8$\pm$0.1 & 0.5$\pm$0.1 & 12-14 & 8.6$\pm$0.2 & --- & 6 & D,F,H \\
M32        & $-$1.1$\pm$0.2 & 0.7$\pm$0.2 & 15,16 & --- & --- & --- & D \\
And~I      & $-$1.5$\pm$0.2 & 0.3$\pm$0.1 & 17,71 & --- & --- & --- & \\
Sculptor   & $-$1.8$\pm$0.1 & 0.3$\pm$0.05 & 18,19 & --- & --- & --- & \\
LGS~3      & $-$1.8$\pm$0.3 & 0.3$\pm$0.2 & 20,65 & --- & --- & \\
IC~1613    & $-$1.3$\pm$0.2 & --- & 2 & 7.8$\pm$0.2 & --- & 67 & \\
And~II     & $-$1.6$\pm$0.3 & 0.5$\pm$0.1 & 21 & --- & --- & --- & \\
Phoenix    & $-$1.9$\pm$0.1 & 0.5$\pm$0.1 & 22,23 & --- & --- & --- & \\
Fornax     & $-$1.3$\pm$0.2 & 0.6$\pm$0.1 & 24,50 & 7.98$\pm$0.4 & --- & 6,72 & F,H \\
EGB~0427+63 & --- & --- & --- & 7.62$\pm$0.1 & $\leq-$1.5 & 4 & \\
Carina     & $-$2.0$\pm$0.2 & $<$0.1 & 51,52 & --- & --- & --- &  \\
Leo~A      & --- & --- & --- & 7.3$\pm$0.1 & --- & 25 & \\
Sextans~B  & $\sim -1.2$ & --- & 60 & 7.84$\pm$0.3 & --- & 25,26 & E \\
NGC~3109   & $-$1.5$\pm$0.3 & $<$0.3 & 2,28-30 & 8.06$\pm$0.2 & --- & 6 & \\
Antlia     & $-$1.8$\pm$0.25 & 0.3$\pm$0.1 & 48,49,77 & --- & --- & --- & \\
Leo~I      & $-$1.5$\pm$0.4 & 0.3$\pm$0.1 & 31,32,69,70 & --- & --- & --- & \\
Sextans~A  & $-$1.9$\pm$0.3 & --- & 61 & 7.49$\pm$0.2 & --- & 25 & \\
Sextans    & $-$1.7$\pm$0.2 & 0.2$\pm$0.05 & 33-36 & --- & --- & --- & \\
Leo~II     & $-$1.9$\pm$0.1 & 0.3$\pm$0.1 & 31,37,38 & --- & --- & --- & \\
GR~8       & --- & --- & --- & 7.62$\pm$0.1 & --- & 26 & E \\
Ursa~Minor & $-$2.2$\pm$0.1 & $\lsim$0.2 & 39 & --- & --- & --- & \\
Draco      & $-$2.0$\pm$0.15 & 0.5$\pm$0.1 & 40,41 & --- & --- & --- & \\
Sagittarius & $-$1.0$\pm$0.2 & 0.5$\pm$0.1 & 54-57,63,64 & 8.30$\pm$0.08 & $-$1.0$\pm$0.3 & 62 & F,G,H  \\
SagDIG     & --- & --- & --- & 7.42$\pm$0.3 & --- & 3 \\
NGC~6822   & $-$1.2$\pm$0.3 & 0.5$\pm$0.1 & 2,43,44 & 8.2$\pm$0.2 & $-$1.7$\pm$0.1 & 3,59,68 & E,F \\
DDO~210    & $<-$1.0 & --- & 30 & --- & --- & --- & I \\
IC~5152    & --- & ---  & --- & 8.36$\pm$0.2 & --- & 67 & \\
Tucana     & $-$1.7$\pm$0.15 & 0.3$\pm$0.2 & 45,46 & --- & --- & --- & \\
UKS2323-326 & --- & --- & --- & --- & --- & --- \\
Pegasus    & $-$1.0$\pm$0.3 & --- & 47,49,66 & 7.93$\pm$0.14 & $-$1.24$\pm$0.15 & 73 & \\
\end{tabular}
\end{center}
\end{table}

\clearpage
\hsize=6.5truein
\vsize=8.5truein
\hoffset=0.0truein
\voffset=0.25truein

\noindent EXPLANATION OF COLUMNS OF TABLE 6:
\vskip0.5em

\noindent {\it Column 1:} Galaxy name; {\it Column 2:} The mean iron
abundance for the old and intermediate-age stellar populations where
[Fe/H] $\equiv \log (Z/Z_\odot)$; {\it Column 3:} The intrinsic
dispersion in [Fe/H].  If the reference gave a full metallicity range
instead of a dispersion, I assumed $\sigma = 0.5 \Delta$[Fe/H].  This
reasonably approximates the scaling between these quantities for the
few galaxies that have independent estimates of both $\sigma_{\rm
[Fe/H]}$ and $\Delta$[Fe/H].  {\it Column 4:} References for the
stellar abundances; {\it Column 5:} The oxygen abundance defined as
$12 + \log(O/H)$, where (O/H) is the number ratio of oxygen to
hydrogen atoms; {\it Column 6:} The nitrogen to oxygen ratio defined
as [N/O] = $\log$(N/O), where (N/O) is the number ratio of
nitrogen to oxygen atoms; {\it Column 7:} References for the oxygen
and nitrogen abundances; {\it Column 8:} Special notes.

\vskip2em

\noindent NOTES FOR TABLE 6:
\vskip0.5em

\noindent\hang A. The sulphur abundance was also determined:
$12 + \log$(S/H) = 6.77 (Richer \&\ McCall 1995).

\noindent\hang B. A weak abundance gradient is seen with [Fe/H]
increases with radius from the galaxy center (Han et al 1997).

\noindent\hang C. The stellar metallicity dispersion is higher in the
center of the galaxy than in the outer region (Han et al 1997).

\noindent\hang D. The metallicity distribution seems to be skewed towards
positive values of [Fe/H] (NGC~205: Mould et al 1984; M32: Grillmair et
al 1996).

\noindent\hang E. The helium abundance was also determined (N(He)/N(H)):
0.098 (GR~8; Moles et al 1990); 0.097 (Sextans~B; Moles et al 1990);
0.074 (NGC~6822; Pagel et al 1980).

\noindent\hang F. The oxygen abundance was derived, at least in part,
from one or more planetary nebula.

\noindent\hang G. The associated globular clusters of this galaxy
exhibit a considerably larger spread in [Fe/H] than the field stars.

\noindent\hang H. The globular clusters of this galaxy are more metal poor on 
average than the field stars.

\noindent\hang I.\ \ 
At the closer distance adopted in Table 2, the stellar iron 
abundance is likely to be considerably lower than found by Greggio et
al (1993) for their assumed distance of over 2.5 Mpc.

\vskip2em

\noindent REFERENCES FOR TABLE 6:
[1] Minniti \&\ Zijlstra (1996);
[2] Lee et al (1993c);
[3] Skillman et al (1989b);
[4] Hodge \&\ Miller (1995);
[5] Garnett (1990);
[6] Richer \&\ McCall (1995);
[7] Davidge (1994);
[8] Mould et al (1983);
[9] Han et al (1997);
[10] Armandroff et al (1993);
[11] Lee et al (1993b);
[12] Mould et al (1984);
[13] Lee (1996);
[14] Richer et al (1984);
[15] Davidge \&\ Jones (1992);
[16] Grillmair et al (1996);
[17] Da Costa et al (1996);
[18] Kaluzny et al (1995);
[19] Da Costa (1984);
[20] Lee (1995a);
[21] K\"onig et al (1993);
[22] van de Rydt et al (1991);
[23] Ortolani \&\ Gratton (1988);
[24] Beauchamp et al (1995);
[25] Skillman et al (1989a);
[26] Moles et al (1990);
[27] Deleted in proof; 
[28] Lee (1993);
[29] Davidge (1993);
[30] Greggio et al (1993);
[31] Suntzeff et al (1986);
[32] Reid \&\ Mould (1991);
[33] Suntzeff et al (1993);
[34] Mateo et al (1995a);
[35] Mateo et al (1991a);
[36] Da Costa et al (1991);
[37] Demers \&\ Irwin (1993);
[38] Lee (1995b);
[39] Olszewski \&\ Aaronson (1985);
[40] Lehnert et al (1992);
[41] Carney \&\ Seitzer (1986);
[42] Deleted in proof; 
[43] Gallart et al (1996a);
[44] Gallart et al (1996b);
[45] Saviane et al (1996);
[46] Castellani et al (1996);
[47] Aparicio \&\ Gallart (1995);
[48] Whiting et al (1997);
[49] Aparicio et al (1997b);
[50] Buonanno et al (1985);
[51] Smecker-Hane et al (1994);
[52] Mould \&\ Aaronson (1983);
[53] Hurley-Keller et al (1998);
[54] Mateo et al (1995c);
[55] Ibata et al (1997);
[56] Sarajedini \&\ Layden (1995);
[57] Da Costa \&\ Armandroff (1995);
[58] Deleted in proof; 
[59] Pagel et al (1980);
[60] Tosi et al (1991);
[61] Dohm-Palmer et al (1997);
[62] Walsh et al (1997);
[63] Whitelock et al (1996);
[64] Marconi et al (1998);
[65] Aparicio et al (1997c);
[66] Deleted in proof; 
[67] Talent (1980);
[68] Dufour \&\ Talent (1980);
[69] Lee et al (1993a);
[70] Demers et al (1994a);
[71] Mould \&\ Kristian (1990);
[72] Maran et al (1984);
[73] Skillman et al (1997);
[74] Lequeux et al (1979);
[75] Stasinska et al (1986);
[76] Webster \&\ Smith (1983);
[77] Sarajedini et al (1997).



\setcounter{table}{0}
\begin{table}
\noindent \ \ Table 7: Internal Kinematic Properties of Local Group Dwarf Galaxies
\begin{center}
\begin{tabular}{lcclccccll}
\\
\ Galaxy & $\sigma_*$ & $v_{rot,*}$ & \ \ \ \ \ \ Ref & $\sigma_{ISM}$ & $v_{rot,ISM}$ & $R_{rot}$ & $i$ & \ \ Ref & Notes \\
  & km s$^{-1}$ & km s$^{-1}$ & & km s$^{-1}$ & km s$^{-1}$ & arcmin & deg & & \\
\ \ \ (1) & (2) & (3) & \ \ \ \ \ \ (4) & (5) & (6) & (7) & (8) & \ \ (9) & \ \ (10) \\
\\
WLM   & --- & --- & & (8) & 21$\pm$2 & 4.8 & 69$\pm$5 & 1 & A,G \\
NGC~55 & --- & --- & & (8) & 86$\pm$3 & 21\pdec & 88$\pm$2 & 1,2,49 & A,D,G \\
IC~10      &  --- & --- & & 8$\pm$2 & 30$\pm$3 & 13\pdec & 40$\pm$5 & 1,3 & D \\
NGC~147    & 22$\pm$4 & 6.5$\pm$1.1 & 4 & --- & --- & --- & --- & & \\
And III    & --- & --- & & --- & --- & --- & --- & & \\
NGC~185    & 25$\pm$4 & 1.2$\pm$1.1 & 4-8 & --- & --- & --- & --- & & B \\
NGC~205 (Outer) & 46$\pm$8 & 1.5$\pm$0.8 & 4,9-12 & 16$\pm$4 & --- & --- & --- & 7 & B \\
NGC~205 (Inner) & 21$\pm$6 & --- & 4,9-12 & --- & --- & --- & --- & & \\
M32        & 50$\pm$10 & 12$\pm$3 & 13,14,47,48,50 & --- & --- & --- & --- & & C \\
And~I      & --- & --- & &  --- & --- & --- & --- & & \\
Sculptor   & 6.6$\pm$0.7 & $<$2.0 & 15-18 & --- & --- & --- & --- & & \\
LGS~3      & 6.5$\pm$3.0 & --- & 43 & 9$\pm$3 & $<$2$\pm$2 & --- & 50$\pm$5 & 1,19,20 & \\
IC~1613    & --- & --- & & 8.5$\pm$1.0 & 21$\pm$2 & 12.5 & 35$\pm$3 & 1,21,22 & D \\
And~II     & --- & --- & & --- & --- & --- & --- & & \\
Phoenix    & --- & --- & & 8.9$\pm$1.5 & $<$2.0 & --- & 55$\pm$4 & 23 & \\
Fornax     & 10.5$\pm$1.5 & $<$2.0 & 24,25 & --- & --- & --- & --- & & \\
EGB~0427+63 & --- & --- & & (8) & 33$\pm$10 & --- & --- & 26 & G \\
Carina     & 6.8$\pm$1.6 & --- & 27 & --- & --- & --- & --- & & \\
Leo~A      & --- & --- & & 9.3/3.5 & $<$3.0 & --- & 45$\pm$5 & 1,19,28 & \\
Sextans~B  & --- & --- & & 18 & 22$\pm$3 & 6.6 & 35$\pm$15 & 22,51 & D,G \\
NGC~3109   & --- & --- & & 10$\pm$2 & 67$\pm$4 & 17.0 & 83$\pm$6 & 1,31,45 & D \\
Antlia     & --- & --- & & 6.3$\pm$1.7 & --- & --- & --- & 32 & \\
Leo~I      & 8.8$\pm$0.9 & --- & 44 & --- & --- & --- & --- & & \\
Sextans~A  & --- & --- & & 8$\pm$3 & 19$\pm$2 & 3.7 & 35$\pm$3 & 1,30 & A \\
Sextans    & 6.6$\pm$0.7 & --- & 33,42 & --- & --- & --- & --- & & \\
Leo~II     & 6.7$\pm$1.1 & --- & 34 & --- & --- & --- & --- & & \\
GR~8       & --- & --- & & 11$\pm$3 & 7$\pm$3 & 0.5 & 48$\pm$3 & 19,22,35 & E \\
Ursa~Minor & 9.3$\pm$1.8 & 5.0$\pm$2.0 & 36-38 & --- & --- & --- & --- & & \\
Draco      & 9.5$\pm$1.6 & $<$2.0 & 37-39 & --- & --- & --- & --- & & \\
Sagittarius & 11.4$\pm$0.7 & $\lsim$4.0 & 46 & --- & --- & --- & --- & & \\
SagDIG     & --- & --- & & 7.5$\pm$2.0 & $<$2.0 & --- & 60$\pm$10 & 1,19 & F \\
NGC~6822   & --- & --- & & (8) & 47$\pm$3 & 19\pdec & 67$\pm$3 & 1,41 & D \\
DDO~210    & --- & --- & & 6.6$\pm$1.8 & $<$5.0 & --- & --- & 19 & \\
IC~5152    & --- & --- & & (8) & 31$\pm$3 & 2.6 & 55$\pm$5 & 1 & \\
Tucana     & --- & --- & & --- & --- & --- & --- & & \\
UKS2323-326 & --- & --- & & --- & --- & --- & --- & & F \\
Pegasus    & --- & --- & & 8.6$\pm$1.4 & 10$\pm$5 & 4.0 & 42$\pm$10 & 1,19,22 & \\
\\
\end{tabular}
\end{center}
\end{table}

\clearpage
\hsize=6.5truein
\vsize=8.5truein
\hoffset=0.0truein
\voffset=0.25truein

\noindent EXPLANATION OF COLUMNS OF TABLE 7:
\vskip0.5em

\noindent {\it Column 1:} Galaxy name; {\it Column 2:} The stellar
central velocity dispersion; {\it Column 3:} The rotation velocity
from stellar velocity measurements; {\it Column 4:} Optical kinematic
references; {\it Column 5:} The velocity dispersion measured from the
ISM (HI line widths, HI clump-clump dispersions, and CO cloud-cloud
dispersions).  See Notes A, F, and G for additional details; 
{\it Column 6:} The peak or outermost observed rotation
velocity of the ISM; {\it Column 7:} The angular distance from the
kinematic center of the galaxy to where the rotation velocity listed
in column 6 is observed; {\it Column 8:} The inclination of the disk
of the galaxy from the plane of the sky; {\it Column 9:} Radio/ISM
kinematic references; {\it Column 10:} Special notes.

\vskip2em

\noindent NOTES FOR TABLE 7:
\vskip0.5em

\noindent\hang A. Used the formulae given by Huchtmeier \&\ Richter (1988) to 
convert the full width at 20\%\ intensity to FWHM.  The rotation velocity
was taken to be FWHM/2.0.  If the inferred rotation was found
to be less than 20 km s$^{-1}$, then the line width was instead 
interpreted as a measure of the 
internal velocity dispersion such that $\sigma_0 =$ FWHM/2.35.

\noindent\hang B. Streaming motions that cannot be attributed to 
rotation are observed in the HI velocity maps.

\noindent\hang C. The nuclear kinematics of M32 are complex, possibly
due to the 
presence of a massive central black hole (Kormendy \&\ Richstone 1995).  The
dispersion and rotational velocity quoted here refer to regions far outside
the nucleus ($r > 10$ arcsec).

\noindent\hang D. The higher velocity and spatial resolution study is
used here to obtain the rotation velocity.

\noindent\hang E. The inner parts appear to follow a solid-body
rotation curve, but the outer regions do not exhibit organized
rotation and appear to be pressure supported (Carignan et al 1990).

\noindent\hang F. Both SagDIG and UKS2323-326 were observed at 21cm by
Longmore et al (1982).  For SagDIG, their estimate of the velocity
dispersion (derived here using the methods described in note A above)
is highly discrepant compared to the two other independent
measurements of the galaxy's dispersion.  One of these references (Lo
et al 1993) finds no evidence for rotation, so this discrepancy is
probably not merely a mis-interpretation of the cause for the
line broadening.
I have chosen to disregard the Longmore et al (1982) HI
velocity widths in this table for both SagDIG (for which other values
are available) and UKS2323-326 (for which no other kinematic data are
published).

\noindent\hang G. $\sigma_{ISM}$ is assumed to be 8 km s$^{-1}$.  No
correction has been applied to $v_{rot}$ for WLM or IC~5152 to account
for this (small) dispersion.  For the other galaxies, detailed
corrections for asymmetric drift have been applied to the velocity
curves as necessary in the cited studies.

\vskip2em

\noindent REFERENCES FOR TABLE 7:
[1] Huchtmeier \&\ Richter (1988);
[2] Hummel et al (1986);
[3] Shostak \&\ Skillman (1989);
[4] Bender et al (1991);
[5] Held et al (1992);
[6] Welch et al (1996);
[7] Young \&\ Lo (1997a);
[8] Wiklund \&\ Rydbeck (1986);
[9] Carter \&\ Sadler (1990);
[10] Held et al (1990);
[11] Ford et al (1987); 
[12] Peterson \&\ Caldwell (1993);
[13] Bender \&\ Nieto (1990);
[14] Dressler \&\ Richstone (1988);
[15] Da Costa (1992);
[16] Armandroff \&\ Da Costa (1986);
[17] Queloz et al (1995);
[18] Da Costa (1994a);
[19] Lo et al (1993);
[20] Thuan \&\ Martin (1979);
[21] Lake \&\ Skillman (1989);
[22] Hoffmann et al (1996);
[23] Carignan et al (1991);
[24] Mateo et al (1991b);
[25] Paltaglou \&\ Freeman (1987);
[26] Huchtmeier \&\ Richter (1986);
[27] Mateo et al (1993);
[28] Young \&\ Lo (1996a);
[29] Deleted in proof; 
[30] Skillman et al (1988);
[31] Jobin \&\ Carignan (1990);
[32] Fouqu\'e et al (1990);
[33] Suntzeff et al (1993);
[34] Vogt et al (1995);
[35] Carignan et al (1990);
[36] Hargreaves et al (1994b);
[37] Olszewski et al (1995);
[38] Armandroff et al (1995);
[39] Hargreaves et al (1996b);
[40] Deleted in proof; 
[41] Gottesman \&\ Weliachew (1977);
[42] Hargreaves et al (1994a);
[43] Cook et al, private communication (1998);
[44] Mateo et al (1998c);
[45] Carignan (1985);
[46] Ibata et al (1997);
[47] Nolthenius \&\ Ford (1986);
[48] Tonry (1984);
[49] Puche et al (1991);
[50] Carter \&\ Jenkins (1993);
[51] Hewitt et al (1983).


\setcounter{table}{2}
\begin{landscape}
\begin{table}
\noindent \ \ Table 5: Integrated and Derived Properties of the ISM of Local Group Dwarf Galaxies
\begin{center}
{\small 
\begin{tabular}{lcclcclcclccccll}
\ \ \ Galaxy & S(HI) & M(HI) & \ \ \ Ref & S(CO) & M$_{mol}$ & \ \ Ref & $\log(F_{\alpha})$ & SFR & \ \ Ref & $f_{60}$ & $f_{100}$ & M$_d$(FIR) & Dust & \ \ \ Ref & Notes \\
       & Jy km s$^{-1}$ & $10^6$ M$_\odot$ & & Jy km s$^{-1}$ & $10^6$ M$_\odot$ & & & M$_\odot$ yr$^{-1}$ & & mJy & mJy & $10^2$ M$_\odot$ & & & \\
\ \ \ \ \ \ (1) & (2) & (3) & \ \ \ (4) & (5) & (6) & \ \ (7) & (8) & (9) & \ \ (10) & (11) & (12) & (13) & (14) & \ \ \ (15) & \ (16) \\
\\
WLM &   300$\pm$25 & 61$\pm$6 & 1 &     $<$70 & $<$1900 & 80 &   6.38 & 0.003 & 2,3 &   \pz\pz320 &  \pz\pz1040 & 4.7$\pm$0.3   & Yes? & 13,64 & \\
NGC~55 &  2680$\pm$200 & 1390$\pm$224 & 1,4 & 43$\pm$8 & 99$\pm$23 & 5,6 &   7.85 & 0.18 & 7,8 & 77000 & 174090 & 1050$\pm$150 & Yes & 13,76 & \\
IC~10 &   950$\pm$140 & 153$\pm$35 & 1 & 80$\pm$25 & 57$\pm$20 & 9,80 & 8.11 & (0.71) & 2,10 & 31230 & \pz71250 & 136$\pm$12 & Yes & 13-51,81 & H \\
NGC~147 &     $<$ 0.04 & $<$0.005 & 1,11,82 &     --- & --- & & --- & --- &  &     \pz$<$56 & \pz$<$160 & 0.34$\pm$0.03   & No & 11,12 & \\
And~III &     $<$0.6 & $<$0.08 & 1,14 & --- & --- & &   --- & --- & &  --- &  --- & --- & --- & & \\
NGC~185 &     1.4$\pm$0.2 & 0.13$\pm$0.02 & 1,11,82 & 69$\pm$15 & 28$\pm$6 & 11,15-17,84 &   4.47 & 0.0 & 11 &   \pz\pz440 &  \pz\pz1720 & 5.0$\pm$0.3   & Yes & 12,56 & A \\
NGC~205 &     2.4$\pm$0.2 & 0.38$\pm$0.04 & 1,11 & 62$\pm$17 & 43$\pm$12 & 11,17,18,83 &   $<$1.88 & 0.0 & 11 &   \pz\pz570 &  \pz\pz3130 & 33$\pm$2   & Yes & 12,68-70 & \\
M~32 &    $<$17.8 & $<$2.7 & 1 &     --- & --- & & --- & --- & &  \pz$<$40 & \pz$<$144 & 0.60$\pm$0.04  & No & 12,19,79 & \\
And~I &     $<$0.65 & $<$0.1 & 1,14 &     --- & --- & & --- & --- & & --- & --- & --- & --- & & \\
Sculptor &    17.3$\pm$1.5 & 0.026$\pm$0.003 & 20,46 & $<$2.8 & $<$0.05 & 46 &   --- & --- & &  --- & --- & --- & --- & & B \\
LGS~3 &     2.7$\pm$0.1 & 0.42$\pm$0.05 & 14,21,22 & 28$\pm$5 & 19$\pm$4 & 23,31 &  $<$2.10 & 0.0 & 2,3 &   \pz$<$75 & \pz$<$109 & $<$0.1 & --- & 13 & C \\
IC~1613 &   480$\pm$40 & 54$\pm$11 & 58,60 & $<$3.0 & $<$4.6 & 24,80 &   6.73 & 0.003 & 2,25,26 &  \pz1420 &  \pz\pz3690 & 6.3$\pm$0.4  & Yes & 12,13,63 & \\
And~II &  --- & --- & &     --- & --- & & --- & --- & & --- & --- & --- & --- & & \\
Phoenix (+55) &     2.0$\pm$0.4 & 0.08$\pm$0.02 &  21,27 & --- & --- & & --- & ---& & --- & --- & --- & --- & & D \\
Phoenix ($-$23) &     2.6$\pm$0.1 & 0.11$\pm$0.01 & 21,27 &  --- & --- & & --- & --- & &  --- &  --- & --- & --- & & D \\
Fornax &     $<$1.05 & $<$0.005 & 1,20 &     --- & --- & &   --- & --- & & \pz$<$17 & \pz\pz$<$86 & 0.0 & --- & 12 & \\
EGB~0427+63 &   105$\pm$2 & 16$\pm$6 & 1 &     --- & --- & &   5.40 & 0.0004 & 3 &  --- &  --- & --- & --- & & \\
Carina &   $<$0.29 & $<$0.0007 & 29 &     --- & --- & &   --- & --- & &  --- &  --- & --- & --- & & \\
Leo~A &    68$\pm$3 & 80$\pm$8 & 1,22 &  $<$1.6 & $<$2.4 & 31,80,85,86 &   4.77 & 0.0003 & 32 & \pz$<$90 & \pz$<$270 & $<$6 & --- & 13 & \\
Sextans~B &   106$\pm$10 & 45$\pm$6 & 1,60 & $<$70 & $<$400 & 80 &   5.58 & 0.0008 & 2,32 &   \pz\pz246 &   \pz\pz\pz689 & 5.0$\pm$0.5   & --- & 13 & \\
NGC3109 &  1880$\pm$110 & 690$\pm$140 & 61,62 & $<$1.0 & $<$4.9 & 33,80 &   7.08 & 0.02 & 2,33,57 &  \pz3410 &  \pz\pz7970 & 38$\pm$5   & Yes & 13,52 & \\
Antlia &     2.7$\pm$0.5 & 0.97$\pm$0.19 & 34 & --- & --- & &   --- & --- & &  --- &  --- & --- & Yes? &  & G \\
Leo~I &     $<$2.1 & $<$0.03 & 1,20 &     --- & --- & &   --- & --- & & \pz$<$33 & \pz\pz$<$72 & 0.0   & No? & 12,66 & \\
Sextans~A &   160$\pm$20 & 78$\pm$13 & 1,35 & $<$4.1 & $<$27 & 24,80,85 & 5.98 & 0.002 & 2,36,78 &   \pz\pz503 &   \pz\pz\pz849 & 3.0$\pm$0.3  & --- & 13 & \\
Sextans &     $<$0.08 & $<$0.0001 & 46 &     --- & --- & &   --- & --- & &  --- &  --- & --- & --- & & \\
Leo~II &     $<$1.05 & $<$0.01 & 1,20 & --- & --- & &   --- & --- & &  --- &  --- & --- & --- & & \\
GR~8 &     8.4$\pm$0.6 & 4.5$\pm$1.4 & 22,59,60 & $<$1.6 & $<$12 & 24,31,67 & 5.47 & 0.0007 & 37 &    \pz\pz\pz20 &   \pz\pz\pz143 & 9.2$\pm$2.7 & --- & 13 & \\
Ursa~Minor &    $<$42 & $<$0.04 & 20 &     --- & --- & &   --- & --- & & \pz$<$27 & \pz\pz$<$73 & $<$0.1 & --- & 12 & \\
Draco &    $<$18 & $<$0.003 & 20 &     --- & --- & &   --- & --- & &  --- &  --- & --- & --- & & \\
Sagittarius &     $<$0.56 & $<$0.0001 & 38 &     --- & --- & & --- & --- & & ---& --- & --- & --- & & E \\
SagDIG &    33$\pm$2 & 8.8$\pm$1.9 & 21,22,39 & --- & --- & & 4.72 & 0.0001 & 32 &     \pz$<$94 & \pz$<$204 & $<$0.6 & & 13 & \\
NGC6822 &  2370$\pm$150 & 134$\pm$18 & 1 & 91$\pm$17 & 23$\pm$5 & 24,41,42,47 &   8.11 & (0.06) & 2,43 & 47630 & \pz95420 & 52$\pm$4  & Yes & 13,72-75 & H \\
DDO210 &    12.8$\pm$1.4 & 1.9$\pm$0.8 & 1,22 &  $<$2.0 & $<$4.0 & 23,31,85 &   --- & --- & &  \pz\pz139 & \pz$<$449 & $<$4.8   & --- & 13 & \\
IC5152 &    98$\pm$8 & 59$\pm$11 & 1 &  $<$70 & $<$560 & 80 & --- & --- & &  \pz2461 &  \pz\pz6861 & 69$\pm$12   & --- & 13 & \\
Tucana &     $<$0.48 & $<$0.09 & 45 &     --- & --- & &   --- & --- & &  --- &  --- & --- & --- & & \\
UKS2323-326 &    15$\pm$3 & 6.2$\pm$2.6 & 1,39 &     --- & --- & &   --- & --- & &  --- &  --- & --- & --- & 13 & \\
Pegasus &    25$\pm$2 & 5.4$\pm$0.6 & 22,60 & $<$2.1 & $<$6.0 & 23,31,80,85 &   3.80 & 0.0 & 2,77 &   \pz$<$55 & \pz$<$531 & $<$34 & --- & 13 & \\
\\
\end{tabular}
}
\end{center}
\end{table}
\end{landscape}

\clearpage
\hsize=6.5truein
\vsize=10.2truein
\hoffset=-1.0truein
\voffset=-1.3truein

\noindent EXPLANATION OF COLUMNS OF TABLE 5:
\vskip0.5em

\noindent {\it Column 1:} Galaxy name; {\it Column 2:} Integrated 21
cm flux; {\it Column 3:} Total HI mass in solar units: M$_{HI} = 2.36
\times 10^{11} {\rm F(HI)} d^2$, where $d$ is the galaxy distance in
Mpc, and F(HI) in Jy km s$^{-1}$; {\it Column 4:} HI References; {\it
Column 5:} The integrated CO flux, S(CO), in Jy~km s$^{-1}$.  If the
CO intensity, I(CO), is given, I derived S(CO) from the relation ${\rm
S(CO)} = g {\rm I(CO)}$, where I(CO) is in K~km s$^{-1}$, and $g$ is a
factor (in Jy K$^{-1}$) that accounts for the beam response of a point
source. I adopt $g$ from Roberts et al (1991), or use $g$ = 14.2 or
1.7 for SEST, and the Nobemaya 50m, respectively.  When necessary, I
assume $S_{2-1}/S_{1-0}$ = 0.75.  Upper limits are 3$\sigma$ values
and assume a velocity dispersion of 12 km s$^{-1}$; {\it Column 6:}
Total molecular mass computed as M$_{CO} = 1.61 \times 10^4 d^2 {\rm
S(CO)}$ in solar units for $d$ in Mpc (Roberts et al 1991; Wilson
1995); {\it Column 7:} CO References ; {\it Column 8:} The $\log_{10}$
of the integrated H$\alpha$ flux in units of $10^{-18}$ ergs s$^{-1}$
cm$^{-2}$; {\it Column 9:} The current star formation rate in
M$_\odot$ yr$^{-1}$ using the extinction-corrected H$\alpha$ fluxes
(Kennicutt 1983); {\it Column 10:} H$\alpha$ references; {\it Columns
11 and 12:} The IRAS 60$\mu$m and 100$\mu$m integrated fluxes,
respectively; {\it Column 13:} The total mass of cool dust in solar
units: M$_d = 0.00478 f_{100} d^2 [\exp(2.94 R^{0.4}) -1]$, where $R =
f_{100}/f_{60}$ and the distance $d$ is in Mpc.  If only the 100$\mu$m
fluxes are available, M$_d = 2.6 f_{100} d^2$. Fluxes are in mJy and
$d$ is in Mpc (Roberts et al 1991); {\it Column 14:} A flag indicating
whether there are optical indicators of dust in the galaxy either
through detection of internal reddening or direct observation of
opaque dust clouds; {\it Column 15:} Far-IR and dust references; {\it
Column 15:} Special notes.

\vskip1.2em

\noindent NOTES FOR TABLE 5:
\vskip0.5em

\noindent\hang A. Young and Lo (1997a) argue that the H$\alpha$ emission is
unlikely to be related to star formation processes, but may instead arise
from shock-excited gas in an old supernova remnant.  Ho et al (1997)
classify NGC~185 as a Seyfert 2!

\noindent\hang B. HI emission extended on a scale comparable to the beam
size; the HI flux and mass may be significantly underestimated
(Carignan et al 1998).

\noindent\hang C. The extended HI distribution is much larger than the
beam sizes of the earliest observations; the largest reported
HI flux is listed here.

\noindent\hang D. Young and Lo (1997b) 
report two distinct HI clouds in the vicinity of Phoenix, one at
$v_\odot = +55$, the other at $v_\odot = -23$.  Results for each cloud
are listed separately.  The stellar velocity is not known.

\noindent\hang E. HI observations obtained only in fields near the
center of the galaxy.

\noindent\hang F. A nearby neutral-hydrogen
cloud has F(HI) = 7.7 Jy km s$^{-1}$ and $V_\odot = +130\pm$2
km s$^{-1}$.  Oosterloo et al (1996) favor an association of this cloud with
the Magellanic Stream rather than Tucana.

\noindent\hang G. Aparicio et al (1997a) note a possible small HII
region near the center of Antlia.  This HII region is visible in the
color photograph of Whiting et al (1997).

\noindent\hang H. Internal reddening is probably significant.  The
star-formation rate estimated from the H$\alpha$ flux (along with most
other photometric parameters) may be greatly underestimated.

\vskip1.2em

\noindent REFERENCES FOR TABLE 5:
[1]   Huchtmeier \&\ Richter (1986);
[2]   Hunter et al (1993);
[3]   Hodge \&\ Miller (1995);
[4]   Hummel et al (1986);
[5]   Israel et al (1995);
[6]   Dettmar \&\ Heithausen (1989);
[7]   Hoopes et al (1996);
[8]   Ferguson et al (1996);
[9]   Wilson (1995);
[10]   Hodge \&\ Lee (1990);
[11]   Young \&\ Lo (1997a);
[12]   Knapp et al (1985);
[13]   Melisse \&\ Israel (1994a,b);
[14]   Thuan \&\ Martin (1979);
[15]   Sofue \&\ Wakamatsu (1993);
[16]   Welch et al (1996);
[17]   Roberts et al (1991);
[18]   Sage \&\ Wrobel (1989);
[19]   van Dokkum \&\ Franx (1995);
[20]   Knapp et al (1978);
[21]   Young \&\ Lo (1997b);
[22]   Lo et al (1993);
[23]   Young et al (1995);
[24]   Ohta et al (1993);
[25]   Hodge et al (1990);
[26]   Price et al (1990);
[27]   Carignan et al (1991);
[28] Deleted in proof; 
[29]   Mould et al (1990);
[30] Deleted in proof; 
[31]   Tacconi \&\ Young (1987);
[32]   Strobel et al (1991);
[33]   Bresolin et al (1993);
[34]   Fouqu\'e et al (1990);
[35]   Skillman et al (1988);
[36]   Hodge et al (1994);
[37]   Hodge et al (1989);
[38]   Koribalski et al (1994);
[39]   Longmore et al (1982);
[40]  Deleted in proof;  
[41]   Israel (1997);
[42]   Wilson (1994b);
[43]   Collier \&\ Hodge (1994);
[44] Deleted in proof; 
[45]   Oosterloo et al (1996);
[46]   Carignan et al (1998);
[47]   Wilson (1992b);
[48] Deleted in proof; 
[49]   Ohta et al (1988);
[50]   de Vaucouleurs \&\ Ables (1965);
[51]   Klein \&\ Gr\"ave (1986);
[52]  Davidge (1993);
[53] Deleted in proof;
[54] Deleted in proof;
[55] Deleted in proof;
[56] Hodge (1963b);
[57] Hodge (1969a);
[58] Lake \&\ Skillman (1989);
[59] Carignan et al (1990);
[60] Hoffman et al (1996);
[61] Jobin \&\ Carignan (1990);
[62] Carignan (1985);
[63] Hodge (1978);
[64] Ables \&\ Ables (1977);
[65] Gottesman \&\ Weliachew (1977);
[66] Bowen et al (1997);
[67] Verter \&\ Hodge (1995);
[68] Hodge (1973);
[69] Price \&\ Grasdalen (1983);
[70] Lee (1996);
[71] Hodge (1976);
[72] Hodge et al (1991b);
[73] Hodge (1977);
[74] Wilson (1992a);
[75] Gallart et al (1996b);
[76] Fitzgibbons (1990);
[77] Aparicio \&\ Gallart (1995);
[78] Aparicio \&\ Rodr\'\i guez-Ulloa (1992);
[79] Bendinelli et al (1992);
[80] Rowan-Robinson et al (1980);
[81] Yang \&\ Skillman (1993);
[82] Johnson \&\ Gottesman (1983);
[83] Young \&\ Lo (1996b);
[84] Wiklund \&\ Rydbeck (1986);
[85] Taylor, private communication (1998);
[86] Young \&\ Lo (1996a).

\vskip1em


\setcounter{table}{1}
\begin{landscape}
\begin{table}
\noindent \ \ Table 8: Summary of the Contents of Local Group Dwarf Galaxies
\begin{center}
\begin{tabular}{lcccclcccclcccll}
\ Galaxy & $N_{RR}$ & $N_{Ceph}$ & $N_{Mira}$ & $N_{AC}$ & \ \ \ \ Refs & $N_{OBA}$ & $N_{WR}$ & $N_{HII}$ & $N_{Dust}$ & \ \ \ \ \ Refs & $N_{AGB}$ & $N_{PN}$ & $N_{GC}$ & \ \ \ \ Refs & Notes \\
\ \ \ (1) & (2) & (3) & (4) & (5) & \ \ \ \ \ (6) & (7) & (8) & (9) & (10) & \ \ \ \ \ (11) & (12) & (13) & (14) & \ \ \ \ (15) & \ (16) \\
\\
WLM         & --- & 15 & --- & --- & 64 & --- & --- & 21 & --- & 37 & 10 & 2 & 1 & 2,18,32,98 & E \\
NGC~55      & --- & --- & --- & --- & & --- & --- & 13 & --- & 95 & 14+ & --- & 5 & 55,94,96 & \\
IC~10       & --- & 5-9 & --- & --- & 61,97 & --- & 16 & 144 & --- & 34,51,76 & --- & 1 & --- & 98 & \\
NGC~147     & 32 & --- & --- & --- & 60,99 & --- & --- & 0 & --- & 100 & Yes  & 5  & 3  & 23,28,32,101 & \\
And~III     & --- & --- & --- & --- & & --- & --- & --- & --- & & Yes?  & --- & --- & 5 & \\
NGC~185     & 151  & --- & --- & --- & 59 & --- & --- & 1 & 2 & 42,100,102 & Yes  & 5  & 9  & 16,23,32,49 & \\
NGC~205     & 30  & --- & 12  & $\sim$9  & 58 & Yes? & --- & --- & 12 & 92 &  7+  & 28  & 12  & 16,32,47,56,92,93 & A,D \\
M32         & --- & --- & --- & --- & & --- & --- & 0 & 0 & 103-105 & Yes  & 30  & 0  & 16,26,32,72,73,144 & \\
And~I       & 4+ & --- & --- & --- & 91 & --- & --- & --- & --- & & Yes  & --- & --- & 5,6 & \\
Sculptor    & 226+  & --- & 1  & 11? & 43,52,66,82,140 & --- & --- & --- & --- & & 8  & --- & 0 & 8,9,106-108,143 & A \\
LGS~3       & --- & --- & --- & --- & & --- & --- & 0  & --- & 37,109 & Yes  & --- & --- & 46 & \\
IC~1613     & 15 & 77+ & --- & --- & 13,92,110,111 & 20  & 1  & 77  & 11 & 7,31,35,54,112 & 15  & 1 & 0  & 18,32,40,71,113  & \\
And~II      & --- & --- & --- & --- & & --- & --- & --- & --- & & --- & --- & --- & & \\
Phoenix     & --- & 60?  & --- & --- & 69 & --- & --- & --- & --- & & Yes & --- & --- & 114 & \\
Fornax      & 400+  & 1  & 30  & 1 & 52,70,75,136 & --- & --- & --- & --- & & 82 &  1 & 5  & 10,32,106,115,116,143 & F,G \\
EGB~0427+63 & --- & --- & --- & --- & & --- & --- & 25  & --- & 37 & --- & --- & --- & & \\
Carina      & 69  & --- & --- & 9 & 43,50,52,74,83 & --- & --- & --- & --- & & 9  & --- & --- & 8,9,143 & A \\
Leo~A       & --- & 5?  & --- & --- & 38,39 & --- & --- & 6-10 & --- & 109,117,118 & Yes & 2-4 & --- & 98,117,120 & \\
Sextans~B   & 36  & 7  & 1  & 6 & 13,53,63 & --- & --- & 12 & --- & 118,121 & --- & 0 & --- & 98 & \\
NGC~3109    & --- & 29  & --- & --- & 13,14,22,65,80 & 18  & --- & 39 & Yes & 12,14,29 & Yes & 7  & 0-10 & 22,57,98,122 & \\
Antlia      & --- & --- & --- & --- & & --- & --- & 1  & Yes? & 78,124 & Yes & --- & --- & 78,123,124 & \\
Leo~I       & Yes? & --- & --- & 15 & 19,50,52,85,125 & --- & --- & --- & --- & & 19  & --- & --- & 8,9,10,48,143 & A \\
Sextans~A   & --- & 10  & --- & --- & 13,53,62,126 & 1+ & --- & 25  & --- & 4,33,109,127 & Yes & 1 & --- & 98,129 & \\
Sextans     & 36 & --- & 1 & 6 & 50 & --- & --- & --- & --- & & 0 & --- & --- & 143 & \\
Leo~II      & 152  & --- & --- & 4 & 21,50,52,81 & --- & --- & --- & --- & & 8 & --- & --- & 8,9,10,20,45,130,143 & A \\
GR~8        & --- & 1? & 5?  & --- & 67 & --- & 0 & 32 & --- & 36,132 & Yes & 0 & --- & 96,133 & \\
Ursa~Minor  & 82  & --- & --- & 7 & 32,52 & --- & --- & --- & --- & & 1  & --- & --- & 9,10  & B \\
Draco       & 280+ & --- & --- & 5+ & 52,85,134,139 & --- & --- & --- & --- & & 3-4  & --- & --- & 9,10 & A \\
Sagittarius & 313+ & --- & 4  & --- & 1,84,135 & --- & --- & --- & --- & & 4+  & 2 & 4 & 41,84,87-90 & \\
SagDIG	    & --- & --- & --- & --- & & --- & --- & 3 & --- & 118 & Yes & --- & --- & 137 & \\
NGC~6822    & --- & 13  & --- & --- & 13,25,44 & 16  & 4  & 157 & 11 & 7,17,30,68,131 & 40 & 8+ & 1  & 18,24,32,138  & C,E \\
DDO~210     & --- & 0  & --- & --- & 69 & --- & --- & --- & --- & & Yes? & --- & 1  & 27 & \\
IC~5152     & --- & Yes? & --- & --- & 69 & --- & --- & --- & --- & & --- & --- & --- &  & \\
Tucana      & Yes & --- & 3?  & --- & 15,142 & --- & --- & --- & --- & & No? & --- & --- & 15,141,142 & \\
UKS2323$-$326 & --- & --- & --- & --- & & --- & --- & --- & --- & & --- & --- & --- & & \\
Pegasus     & --- & 7-10 & --- & --- & 38 & --- & --- & 1  & --- & 3 & Yes  & 1 & --- & 3,98 & \\
\\
\end{tabular}
\end{center}
\end{table}
\end{landscape}

\clearpage
\hsize=6.5truein
\vsize=9.8truein
\hoffset=-1.0truein
\voffset=-1.2truein

\noindent EXPLANATION OF COLUMNS OF TABLE 8:

\noindent {\it Column 1:} Galaxy name; {\it Columns 2-5:} The census
of known populations of variable stars: RR~Lyr stars, Cepheids, Mira
(long-period) variables, and Anomalous Cepheids; {\it Column 6:}
References for variable stars; {\it Columns 7-10:} The census of
tracers of young populations: OB associations, Wolf-Rayet stars, H~II
regions, and discrete dust clouds; {\it Column 11:} References for the
young population tracers; {\it Columns 12-14:} The census of known
intermediate-age and old-age population tracers: asymptotic giant
branchstars, planetary nebulae, and globular clusters; {\it Column
15:} References for the intermediate-age and old-age tracers; {\it
Column 16:} Special Notes.

\vskip2em

\noindent NOTES FOR TABLE 8:
\vskip0.5em

\noindent\hang A. The AGB stars noted are principally C stars.

\noindent\hang B. Additional low-luminosity CH stars are also known
on or near the AGB (Aaronson et al 1983).

\noindent\hang C. The numbers of Cepheids are reported by Capaccioli et al
(1992); see that paper for the original sources.

\noindent\hang D. The numbers of Planetary Nebulae are reported by 
Ciardullo et al (1989); see that paper for the original sources.

\noindent\hang E. The AGB census is for stars with $M_I \leq -4.4$.

\noindent\hang F. Approximately 21 RR~Lyr stars may be associated with
Fornax cluster 1 (Smith et al 1996).

\noindent\hang G. The Cepheid is most likely a W~Vir or Pop~II Cepheid
(Light et al 1986).

\vskip2em

\noindent REFERENCES FOR TABLE 8:
[1] Alard (1996);
[2] Ables \&\ Ables (1977);
[3] Aparicio (1994);
[4] Aparicio \&\ Rodri\'guez-Ulloa (1992);
[5] Armandroff et al (1993);
[6] Mould \&\ Kristian (1990);
[7] Armandroff \&\ Massey (1991);
[8] Azzopardi et al (1985);
[9] Azzopardi et al (1986);
[10] Aaronson et al (1983);
[11] Deleted in proof;
[12] Bresolin et al (1993);
[13] Capaccioli et al (1992);
[14] Sandage \&\ Carlson (1988);
[15] Castellani et al (1996);
[16] Ciardullo et al (1989);
[17] Collier \&\ Hodge (1994);
[18] Cook et al (1986);
[19] Demers et al (1994a);
[20] Demers \&\ Irwin (1993);
[21] Swope (1967);
[22] Demers et al (1985);
[23] Ford et al (1977);
[24] Gallart et al (1996a); 
[25] Gallart et al (1996c); 
[26] Grillmair et al (1996);
[27] Greggio et al (1993);
[28] Han et al (1997);
[29] Hodge (1969);
[30] Hodge (1977);
[31] Hodge (1978);
[32] Hodge (1988);
[33] Hodge et al (1994);
[34] Hodge \&\ Lee (1990);
[35] Hodge et al (1990);
[36] Hodge et al (1989);
[37] Hodge \&\ Miller (1995);
[38] Hoessel et al (1990);
[39] Hoessel et al (1994);
[40] Freedman (1988a);
[41] Ibata et al (1994);
[42] Hodge (1963b);
[43] Kaluzny et al (1995);
[44] Kayser (1967);
[45] Lee (1995b);
[46] Lee (1995a);
[47] Lee (1996);
[48] Lee et al (1993a);
[49] Lee et al (1993c);
[50] Mateo et al (1995a);
[51] Massey et al (1992);
[52] Nemec et al (1988);
[53] Piotto et al (1994);
[54] Price et al (1990);
[55] Pritchet et al (1987);
[56] Richer et al (1984);
[57] Richer \&\ McCall (1992);
[58] Saha et al (1992b);
[59] Saha \&\ Hoessel (1990);
[60] Saha et al (1990);
[61] Saha et al (1996);
[62] Sandage \&\ Carlson (1982);
[63] Sandage \&\ Carlson (1985a);
[64] Sandage \&\ Carlson (1985b);
[65] Sandage \&\ Carlson (1988);
[66] Schweitzer et al (1995);
[67] Tolstoy et al (1995);
[68] Wilson (1992a);
[69] Caldwell et al (1988);
[70] Demers \&\ Irwin (1987);
[71] Deleted in proof; 
[72] Freedman (1992);
[73] Elston \&\ Silva (1992);
[74] Kuhn et al (1996);
[75] Light et al (1986);
[76] Massey et al (1992);
[77] Saha et al (1992a);
[78] Aparicio et al (1997c);
[80] Musella et al (1997);
[81] van Agt (1973);
[82] van Agt (1978);
[83] Saha et al (1986);
[84] Whitelock et al (1996);
[85] Hodge \&\ Wright (1978);
[86] Deleted in proof; 
[87] Minniti \&\ Zijlstra (1996);
[88] Walsh et al (1997); 
[89] Da Costa \&\ Armandroff (1995);
[90] Zijlstra \&\ Walsh (1996); 
[91] Da Costa et al (1996);
[92] Hodge (1973);
[93] Battistini et al (1987);
[94] Da Costa \&\ Graham (1982);
[95] Ferguson et al (1996);
[96] Liller \&\ Alcaino (1983); 
[97] Wilson et al (1996);
[98] Jacoby \&\ Lesser (1981); 
[99] Saha \&\ Hoessel (1987). 
[100] Young \&\ Lo (1997a);
[101] Hodge (1976);
[102] Gallagher \&\ Hunter (1981); 
[103] Ford \&\ Jenner (1976); 
[104] Ford et al (1978); 
[105] van Dokkum \&\ Franx (1995);
[106] Richer \&\ Westerlund (1983); 
[107] Hodge (1966);
[108] Eskridge (1988b); 
[109] Hunter et al (1993);
[110] Carlson \&\ Sandage (1990); 
[111] Freedman (1988b); 
[112] Hodge (1980);
[113] Lequeux et al (1987); 
[114] van de Rydt et al (1991);
[115] Maran et al (1984);
[116] Danzinger et al (1978); 
[117] Tolstoy (1996);
[118] Strobel et al (1991);
[119] Deleted in proof;
[120] Tolstoy et al (1998); 
[121] Hodge (1974);
[122] Davidge (1993);
[123] Whiting et al (1997);
[124] Sarajedini et al (1997); 
[125] Mateo et al (1998d);
[126] Sakai et al (1996);
[127] Hunter \&\ Gallagher (1990); 
[128] Deleted in proof; 
[129] Dohm-Palmer et al (1997);
[130] Mighell \&\ Rich (1996);
[131] Hodge et al (1988); 
[132] Drissen et al (1993); 
[133] Dohm-Palmer et al (1998);  
[134] Harris et al (1997). 
[135] Mateo et al (1995b);
[136] Mateo et al (1998a);  
[137] Cook (1987);  
[138] Killen \&\ Dufour (1982); 
[139] Baade \&\ Swope (1961);
[140] van Agt (1978);
[141] Saviane et al (1996);
[142] Lavery et al (1996);  
[143] Azzopardi (1994);
[144] Davidge \&\ Nieto (1992).